%% file: Bottomonia_Review.tex
\RequirePackage{lineno}
\documentclass[review]{elsarticle}
\usepackage{verbatim}
\usepackage{graphicx}
\usepackage[utf8]{inputenc}
\usepackage[usenames,dvipsnames,svgnames,table]{xcolor}
\usepackage[breaklinks=true,colorlinks=true,linkcolor=blue,urlcolor=blue,citecolor=blue]{hyperref}
\usepackage{rotating}
\usepackage{graphicx}
\usepackage{dcolumn}
\usepackage{bm}
\usepackage{epsfig}
\usepackage{hyperref}
\usepackage{ulem}
\usepackage{appendix}
\usepackage{mwe}
\usepackage{subfig}
\usepackage{lineno}
\expandafter\let\csname equation*\endcsname\relax
\expandafter\let\csname endequation*\endcsname\relax
\usepackage{amsmath}
\usepackage{geometry}
\newgeometry{vmargin={20mm}, hmargin={30mm,22mm}}
\usepackage{times}  
\newcommand{\Jpsi}{\ensuremath{J\hspace{-.08em}/\hspace{-.14em}\psi}\xspace} 

\newcommand{\QQbar}{\ensuremath{Q \overline{Q}\xspace}\xspace}

\newcommand{\sNN}{\sqrt{s_{NN}}}

\newcommand{\jpsi}{\ensuremath{\mathrm{J}/\psi}\hspace {0.05in}} 

\newcommand{\upsa}{\ensuremath{\Upsilon\mathrm{(1S)}}\hspace {0.05in}}
\newcommand{\upsb}{\ensuremath{\Upsilon\mathrm{(2S)}}\hspace {0.05in}}
\newcommand{\upsc}{\ensuremath{\Upsilon\mathrm{(3S)}}\hspace {0.05in}}

\newcommand{\psiP}{\ensuremath{\psi\mathrm{(2S)}}\hspace {0.05in}}
\newcommand{\chic}{\ensuremath{\chi_c}\hspace {0.05in}}
\newcommand{\chib}{\ensuremath{\chi_b}\hspace {0.05in}}

\newcommand{\be}{\begin{equation}}
\newcommand{\ee}{\end{equation}}
\newcommand{\bea}{\begin{eqnarray}}
\newcommand{\eea}{\end{eqnarray}}

\begin{document}
  \title{Production of bottomonia states in proton+proton and heavy-ion collisions}
  \author[NPD]{Vineet Kumar}
  \author[NPD,HBNI]{Prashant Shukla\corref{mycorrespondingauthor}}
  \ead{pshukla@barc.gov.in}
  \author[UOC]{Abhijit Bhattacharyya\corref{mycorrespondingauthor}}
  \cortext[mycorrespondingauthor]{Corresponding authors}
  \ead{abhattacharyyacu@gmail.com}
  \address[NPD]{Nuclear Physics Division, Bhabha Atomic Research Centre, Mumbai 400085, India}
  \address[HBNI]{Homi Bhabha National Institute, Anushakti Nagar, Mumbai 400094, India}
  \address[UOC]{Department of Physics, University of Calcutta, 92, A. P. C. Road Kolkata-700009, India}
  \date{\today}
\fontsize{11}{15}
\selectfont
  
\begin{abstract}
    
 In this work, we review the experimental and theoretical developments of bottomonia
production in proton+proton and heavy-ion collisions. The bottomonia production
process is proving to be one of the most robust processes to investigate the
fundamental aspects of Quantum Chromodynamics at both low and high temperatures.  
The LHC experiments in the last decade have produced large statistics of
bottomonia states in wide kinematic ranges in various collision
systems. The bottomonia have three $\Upsilon$ S-states which are
reconstructed in dilepton invariant mass channel with high mass resolution by
LHC detectors and P-states are measured via their decay to S-states. 
We start with the details of measurements in proton+proton collisions and their
understanding in terms of various effective theoretical models. Here we cover both the
Tevatron and LHC measurements with $\sqrt{s}$ spanning from 1.8 TeV to 13 TeV. 
  The bottomonia states have particularly been very good probes to understand
strongly interacting matter produced in heavy-ion collisions. The Pb+Pb
collisions have been performed at   $\sqrt{s_{NN}}$ = 2.76 TeV and 5.02 TeV at LHC. This led to the
detailed study of the modification of bottomonia yields as a function of various observables
and collision energy. At the same time, the improved results of bottomonia
production became available from RHIC experiments which have proven to be useful
for a quantitative comparison. 
 A systematic study of bottomonia production in p+p,
p+Pb and Pb+Pb has been very useful to understand the medium effects
in these collision systems. We review some of the (if not all the) models of bottomonia
evolution due to various processes in a large dynamically evolving medium and
discuss these in comparison with the measurements. 
\end{abstract}

  
 
\maketitle
\newpage
\tableofcontents
\input{Introduction}

\input{BottPP}
\input{BottAAexp}

\input{BottAAthe}

\input{Conclusions}

\section*{Acknowledgement}
AB thanks Partha Pratim Bhaduri for helpful discussions. 

\bibliography{Bottomonia_Review.bib}
\end{document}

%% file: Introduction.tex
\newpage
\section{Introduction}
\label{sec:Introduction}

The strong interaction among quarks and gluons is described by
Quantum Chromodynamics (QCD) that has two regimes; asymptotic freedom at short
distances and color confinement at long distances.
At short distances, the interaction can be well described using perturbative methods. 
However, confinement, which is a non-perturbative phenomenon, is not 
very well understood. The study of quarkonia ($Q\bar{Q}$) involves
both the perturbative and non-perturbative aspects of QCD. 
The quarkonia are composed of heavy constituents (charm and bottom quarks) and their
velocity can be considered to be small allowing one to use the non-relativistic
formalism~\cite{Povh:1995mua,Ikhdair:2005jf} in the study. 
In a simple picture,  a quarkonium can be understood as a heavy quark pair ($Q\bar{Q}$)
bound in a color singlet state by some effective potential interaction.
In this bound state, the constituents are separated by distances much smaller
than $1/\Lambda_{\rm QCD}$ where $\Lambda_{\rm QCD}$ is the QCD scale.

It is expected that the dynamics of strongly interacting matter changes
at temperatures and/or densities which are similar to or larger than
the typical hadronic scale.
It can be argued that under such extreme conditions, 
one should have the onset of deconfinement of quarks and gluons  and thus the 
strongly-interacting matter could then be described in terms of these
elementary degrees of freedom.
This new form of matter is known as 
quark-gluon plasma~\cite{Shuryak:1980tp,Satz:2011wf}, or QGP.
The support for the existence of such a transition has indeed been demonstrated 
from first principles using QCD simulation on lattice.
The heavy-ion collisions provide experimental means to study the deconfinement transition 
and the properties of hot and dense strongly-interacting matter~\cite{Satz:2000bn}.
Significant parts of different experimental heavy-ion programs are dedicated to studying
quarkonium yields. Such studies are 
motivated by the suggestion of  Matsui and Satz
that quarkonium suppression could be a signature of 
deconfinement~\cite{Matsui:1986dk}.
In fact, the observation of anomalous suppression of J/$\psi$ at SPS energies
was considered to be a key signature of deconfinement~\cite{Kluberg:2005yh}.

The $\Upsilon^{'}$s having three states with different binding
energies are far richer probes of the QCD dynamics in p+p and Pb+Pb collisions than
the charmonia states.
It is therefore important to achieve a good understanding of their
production mechanism in the vacuum as well as of how the nuclear effects in proton-nucleus
collisions affect them.
  At Large Hadron Collider (LHC) energy, the cross section of bottomonia production
is large and also the detector technologies enabled the study of various bottomonia 
states separately both in p+p and heavy-ion collisions.
As proposed by different theories, bottomonium is an important and clean probe 
of hadronic collisions for at least two reasons. 
First, the effective field theory approach, which provides a link to first 
principles QCD, is more suitable for bottomonium due to better separation of 
scales and higher binding energies. Second, the statistical recombination effects 
are less important due to the higher mass of bottom quarks.  
Experimentally, the bottomonia are detected via their decay in the dimuon channel.
Bottomonia can be reconstructed with better mass resolution and smaller
combinatorial background due to higher mass as compared to other resonances. 
 All these properties make bottomonium a good probe of QGP formation in heavy-ion collisions.

The detailed experimental study of the bottomonia states in p$+\overline{\rm p}$ collisions 
was carried out at Fermilab at $\sqrt{s}$ = 1.8 and 1.96 TeV~\cite{CDF:1995gwi,CDF:2001fdy,D0:2005klj}.
At the LHC, the experimental collaborations carried out 
the bottomonia study in p+p collisions at
$\sqrt{s}=$ 7 TeV~\cite{CMS:2010wld,CMS:2015xqv,ATLAS:2012lmu} and
13 TeV~\cite{CMS:2017dju,LHCb:2018yzj}.
There have been immense
experimental~\cite{Sirunyan:2017isk,Sirunyan:2017lzi,CMS:2018zza,Acharya:2019iur,ALICE:2018wzm}
and theoretical works~\cite{Strickland:2011mw,Song:2011nu,Kumar:2014kfa,Kumar:2019xdj} on
quarkonia modifications in Pb+Pb collisions.
The bottomonia states in heavy-ion collisions are suppressed with respect to their yields
in proton-proton collisions scaled by the number of binary nucleon-nucleon
(NN) collisions.
The amount of quarkonia suppression is expected to be sequentially ordered by the binding
energies of the quarkonia states.
 As the screening depends on the  binding energy, the bottomonium states ($\Upsilon$(1S), $\Upsilon$(2S),
$\Upsilon$(3S), $\chi_{b}$, etc.) are extremely useful probes to understand the color screening
properties of the QGP.
The sequential suppression of the yields of $\Upsilon$(nS) states was first observed by
CMS at $\sNN =$ 2.76 TeV~\cite{Chatrchyan:2011pe,Chatrchyan:2012lxa}. Later, the results with
improved statistical precision have been reported by both the ALICE~\cite{ALICE:2018wzm}
and CMS Collaborations ~\cite{Sirunyan:2017lzi,CMS:2018zza} at $\sNN =$ 5.02 TeV.
The suppression of the $\Upsilon$(1S)
meson has also been studied at $\sNN =$ 200 GeV at Relativistic Heavy Ion Collider
(RHIC)~\cite{STAR:2013kwk}, although the 
bottomonia production cross section is small at lower energies.
There have been many reviews written on experimental and theoretical developments on Quarkonia and
their modifications in heavy-ion collisions~\cite{Brambilla:2010cs,Andronic:2015wma,Rothkopf:2019ipj}.
The future measurement prospects with high luminosity LHC can be found in Ref.~\cite{Chapon:2020heu}.

In this writeup, we review experimental and theoretical aspects of bottomonia production in p+p, p+A
and A+A collisions at RHIC and LHC energies.

%% file: BottPP.tex
\section{Bottomonia production in p+p collisions: Experimental overview}
\label{Sec:BottPP}

The $\Upsilon$ meson was discovered by E288 collaboration at Fermilab in the collision of
a beam of 400 GeV protons with nuclei in 1977~\cite{PhysRevLett.39.252}.
Detailed measurements of all the states of $\Upsilon$ were also done at Fermilab.
The Collider Detector at Fermilab (CDF) measured $\Upsilon$(1S), $\Upsilon$(2S) and $\Upsilon$(3S) 
differential ($d^{2}\sigma/dp_{T}dy$) and integrated cross sections in p+$\overline{\rm p}$
collisions at $\sqrt{s}$ = 1.8 TeV~\cite{CDF:1995gwi} at Tevatron.
The three states were reconstructed via their decays 
to $\mu^{+}$ and $\mu^{-}$. The differential ($d^{2}\sigma/dp_{T}dy$) and integrated
cross sections have been measured for  $\Upsilon$(1S) in the transverse momentum range
0$\textless p_{T} \textless$16 GeV/c and for $\Upsilon$(2S) and $\Upsilon$(3S)
in the range 0$\textless p_{T}\textless$10 GeV/c.
  In 2002, CDF measured both the cross sections and polarizations of $\Upsilon$
for $|y|\textless$ 0.4 in p+$\overline{\rm p}$ collisions at $\sqrt{s}$ = 1.8 TeV with
an integrated luminosity of 77 pb$^{-1}$~\cite{CDF:2001fdy} and the cross sections are
listed in Table~\ref{Tab:YCrossCDF02}. These studies helped to understand the
relative production of $\Upsilon$ states in hadronic collisions.


\begin{table}
  \begin{center}
    \caption[]{The cross section of $\Upsilon$(nS) at midrapidity
($|y|\textless$ 0.4) in p+$\overline{\rm p}$ collisions at $\sqrt{s}$ = 1.8 TeV with
an integrated luminosity of 77 pb$^{-1}$ as measured by CDF~\cite{CDF:2001fdy}}
\label{Tab:YCrossCDF02}
\begin{tabular}{cl} 
\hline 
\hline
$\Upsilon$(nS) state             &$\frac{d\sigma(\Upsilon(nS))}{dy}\times B(\Upsilon(nS)\rightarrow\mu^{+}\mu^{-})$ (pb)    \\              
\hline
$\Upsilon$(1S)                   &680$\pm$15(stat.)$\pm$18(syst.)$\pm$26(lumi.)\\
$\Upsilon$(2S)                   &175$\pm$9(stat.)$\pm$8(syst.)\\
$\Upsilon$(3S)                   &97$\pm$8(stat.)$\pm$5(syst.)\\   
\hline
\hline
\end{tabular}
\end{center}
\end{table}

The Run II of Tevatron was carried out at p+$\overline{\rm p}$ collisions at
$\sqrt{s}$ = 1.96 TeV, and D0 experiment published results from these. 
The D0 experiment measured $\Upsilon$(1S) cross section in different 
rapidity ranges in p+$\overline{\rm p}$ collisions at $\sqrt{s}$  = 1.96 TeV 
at luminosity of 185 pb$^{-1}$~\cite{D0:2005klj}.
Table~\ref{Tab:YCrossD0RunII} summarizes the D0 $\Upsilon$ cross section
measurements.

\begin{table}
 \begin{center}
   \caption[]{ $\Upsilon$(1S) cross sections in different rapidity ranges in p+$\overline{\rm p}$
     collisions at $\sqrt{s}$ = 1.96 TeV measured in Tevatron Run II with an integrated luminosity
     of 185 pb$^{-1}$~\cite{D0:2005klj}. }
\label{Tab:YCrossD0RunII}
\begin{tabular}{cc} 
\hline 
\hline
rapidity range             &$\frac{d\sigma(\Upsilon(1S))}{dy}\times B(\Upsilon(nS)\rightarrow\mu^{+}\mu^{-})$ (pb)    \\              
\hline
0.0-0.6                   &628$\pm$16(stat.)$\pm$63(syst.)$\pm$38(lumi.)\\
0.6-1.2                   &654$\pm$17(stat.)$\pm$65(syst.)$\pm$40(lumi.)\\
1.2-1.8                   &515$\pm$16(stat.)$\pm$46(syst.)$\pm$31(lumi.)\\
0.0-1.8                   &597$\pm$12(stat.)$\pm$58(syst.)$\pm$36(lumi.)\\
\hline
\hline
\end{tabular}
\end{center}
\end{table}

The measurements of the production of $\Upsilon$(1S,2S,3S) in p+p collisions at the
unprecedented center of mass energies of 2.76, 5.02, 7, 8, and 13 TeV have been undertaken,
within various rapidity windows and in the dimuon momentum range of
$p_{T}<$ 100 GeV/c at LHC by
ATLAS~\cite{ATLAS:2011nal,ATLAS:2012lmu},
CMS~\cite{CMS:2013qur,CMS:2017dju} and LHCb collaborations \cite{LHCb:2018yzj}.

 The Large Hadron Collider (LHC) performed $\Upsilon$ measurements at 
$\sqrt{s}$ = 7 TeV in p+p collisions that is roughly four times the energy of the Tevatron. 
CMS measured the $\Upsilon$ cross section in 2011 in the kinematic range 
$|y|\textless$ 2, and $p_{T} \textless$ 30 GeV~\cite{CMS:2010wld} 
with an integrated luminosity of 3.1 pb$^{-1}$.
  In 2013, CMS measured the $\Upsilon$ cross section in p+p collisions at $\sqrt{s}$ = 7 TeV
with the increased integrated luminosity of 35.8 pb$^{-1}$ and in the kinematic range
$|y|\textless$ 2.4 and $p_{T}\textless$ 50 GeV~\cite{CMS:2015xqv} as shown in
Table~\ref{Tab:CMSYCrossPLB}.


\begin{table}
  \begin{center}
    \caption[]{ $\Upsilon$(nS) cross sections
  in kinematic range $|y|\textless$ 2.0 and $p_{T}\textless$ 50 GeV 
      measured by CMS in p+p collisions at $\sqrt{s}$ = 7 TeV.
  for integrated luminosity of 35.8 pb$^{-1}$~\cite{CMS:2015xqv}.}
\label{Tab:CMSYCrossPLB}
\begin{tabular}{cl} 
\hline 
\hline
$\Upsilon$(nS) state             &$ \sigma(p+p \rightarrow \Upsilon(nS)X) \times B(\Upsilon(nS)\rightarrow\mu^{+}\mu^{-})$ (nb)    \\              
\hline
$\Upsilon$(1S)                   &8.55$\pm$0.05(stat.)$^{+0.56}_{-0.50}$(syst.)$\pm$0.34(lumi.)\\
$\Upsilon$(2S)                   &2.21$\pm$0.03(stat.)$^{+0.16}_{-0.14}$(syst.)$\pm$0.09(lumi.)\\
$\Upsilon$(3S)                   &1.11$\pm$0.02(stat.)$^{+0.10}_{-0.08}$(syst.)$\pm$0.04(lumi.)\\
\hline
\hline
\end{tabular}
\end{center}
\end{table}

The ATLAS experiment measured the $\Upsilon$(nS) production cross section
in p+p collisions at $\sqrt{s}$ = 7 TeV in kinematic range $|y|\textless$ 2.25,
and $p_{T}\textless$ 70 GeV~\cite{ATLAS:2012lmu}.  
The results are shown in Table~\ref{Tab:ATLASYCross}.

\begin{table}
  \begin{center}
    \caption[]{ATLAS measurement of $\Upsilon$(nS) cross section in $|y|\textless$ 2.25 and $p_{T}\textless$ 70 GeV
      at $\sqrt{s}$ = 7 TeV~\cite{ATLAS:2012lmu}.}
\label{Tab:ATLASYCross}
\begin{tabular}{cl} 
\hline 
\hline
$\Upsilon$(nS) state             &$ \sigma(p+p \rightarrow \Upsilon(nS)X) \times B(\Upsilon(nS)\rightarrow\mu^{+}\mu^{-})$ (nb)    \\              
\hline
$\Upsilon$(1S)                   &8.01$\pm$0.02(stat.)$\pm$0.36(syst.)$\pm$0.31(lumi.)\\
$\Upsilon$(2S)                   &2.05$\pm$0.01(stat.)$\pm$0.12(syst.)$\pm$0.08(lumi.)\\
$\Upsilon$(3S)                   &0.92$\pm$0.01(stat.)$\pm$0.07(syst.)$\pm$0.04(lumi.)\\
\hline
\hline
\end{tabular}
\end{center}
\end{table}

Several $\Upsilon$ polarization measurements have also been made. With an integrated luminosity
of 77 pb$^{-1}$, CDF measured the $\Upsilon$(1S) polarization in 2002 in 
p+$\overline{\rm p}$ collisions at $\sqrt{s}$ = 1.8 TeV in knematic range
$|y|\textless$ 0.4 and found that $\Upsilon$(1S) is mostly unpolarized~\cite{CDF:2001fdy}.
At $\sqrt{s}$  = 1.96 TeV, the D0 experiment measured the $\Upsilon$(1S) and
$\Upsilon$(2S) polarizations in 2008 in p+$\overline{\rm p}$ data
with an integrated luminosity of 1.3 fb$^{-1}$~\cite{D0:2008yos}. The measurement done by D0 found
longitudinal polarization for the $\Upsilon$(1S)~\cite{D0:2008yos}.
However, these first polarization measurements were only done in one reference frame.
They were sensitive to the bias resulted due to the choice of the reference frame
and also to the acceptance of the detector. Later, measurements were done in multiple
reference frames which enabled the calculation of the frame invariant parameter to
prevent bias from the choice of the reference frame and detector acceptance~\cite{CMS:2012bpf}.

The first full polarizations for all $\Upsilon$(nS) states were measured in 2012 by
CDF in Tevatron Run II at $\sqrt{s}$  = 1.96 TeV~\cite{CDF:2011ag}.
This CDF Run II measurement with an integrated luminosity of 6.7 fb$^{-1}$
with $|y|\textless$ 0.6 and $p_{T}\textless$ 40 GeV found no evidence for
polarized production of $\Upsilon$(nS) states~\cite{CDF:2011ag}.
CMS measured the $\Upsilon$(nS) polarization in 2003 in p+p collisions
at $\sqrt{s}$  = 7 TeV with an integrated luminosity 4.9 fb$^{-1}$~\cite{CMS:2012bpf}.
In these measurements, the angular distribution of the muons produced in
the $\Upsilon$(1S, 2S, 3S) decays was analyzed in different reference frames to determine the
polarization parameters. The CMS measurement found no evidence of large transverse or
longitudinal polarizations in the explored kinematic region~\cite{CMS:2012bpf}.
CMS measured the differential cross section of $\Upsilon$(nS)
in p+p collisions at $\sqrt{s}$ = 13 TeV in central rapidity $|y|<2.4$,
as a function of transverse momentum in high momentum range~\cite{CMS:2017dju}.
These are used to constrain the theoretical calculations in the high
momentum range as shown in the next section.

The $\Upsilon$(nS) cross sections are measured by LHCb~\cite{LHCb:2018yzj}
in p+p collisions at $\sqrt{s}$ = 13 TeV, in transverse momentum
range $0 < p_T < 15$ GeV/$c$ and forward rapidity region $2 < y < 4.5$.
The cross sections are listed here. \\
B($\Upsilon$(1S) $\rightarrow$ $\mu^+\mu^-$) $\times$ $\sigma$($\Upsilon$(1S)) = (4687 $\pm$ 10 (stat) $\pm$ 294 (sys)) pb,\\
B($\Upsilon$(2S) $\rightarrow$ $\mu^+\mu^-$) $\times$ $\sigma$($\Upsilon$(2S)) = (1134 $\pm$ 6 (stat) $\pm$ 71 (sys)) pb, \\
B($\Upsilon$(3S) $\rightarrow$ $\mu^+\mu^-$) $\times$ $\sigma$($\Upsilon$(3S)) = (561 $\pm$ 4 (stat) $\pm$ 36 (sys)) pb.

The measurements of the cross sections and polarizations have shed light on the
$\Upsilon$(1S, 2S, 3S) production mechanisms in p+p collisions.
LHC data \cite{CMS:2010wld,CMS:2015xqv,ATLAS:2012lmu,CMS:2017dju,LHCb:2018yzj}
  has substantially extended the reach of the kinematics to test the
Non-Relativistic QCD (NRQCD) and other models with
higher-order corrections which becomes more distinguishable with the increase of $p_{T}$.

\section{Bottomonia production mechanism in p+p collisions}
\label{sec:Bottomonia_pp_th}

In general, one can subdivide the quarkonia production process into two major parts;
the production of a heavy quark pair in hard collisions and the formation of quarkonia
out of the two heavy quarks.
  The massive quarks (with $m_c\sim 1.6$ GeV/$c^2$, $m_b\sim 4.5$ GeV/$c^2$) are produced
in the initial stages of hadronic collision with high momentum transfer and 
can be treated perturbatively~\cite{Nason:1989zy}. The formation of quarkonia
from the two massive quarks is a non-perturbative phenomenon that is
described using different effective models~\cite{Bodwin:1994jh,Brambilla:2014jmp}.
The Color Singlet Model (CSM)~\cite{Einhorn:1975ua,Berger:1980ni},
the Color Evaporation Model (CEM)~\cite{Fritzsch:1977ay,Amundson:1995em}, the Fragmentation
Scheme and the NRQCD factorisation formalism are some of the popular
approaches used for the description of quarkonia production.

The hadronic cross section in p+p collisions at center of mass energy
$\sqrt{s}$ can be written as
\begin{eqnarray}
\sigma_{\rm pp}(s,m^2) & = & \sum_{i,j = q, \overline q, g} 
\int dx_1 \, dx_2 \, 
f_i^p (x_1,\mu_F^2) \,
f_j^p(x_2,\mu_F^2) \, \widehat{\sigma}_{ij}(\hat{s},m^2,\mu_F^2,\mu_R^2).
\label{sigpp}
\end{eqnarray}
Here, $f_i^p$ are the parton (${i = q, \overline q, g}$) densities of the proton,
$x_1$ and $x_2$ are the fractional momenta carried by the colliding
partons and $\mu_F$ and $\mu_R$ are, respectively, fragmentation and renormalization scales. 
The total partonic cross section has been calculated up to next-to-leading order (NLO)
\cite{Nason:1989zy,Nason:1987xz} and can be expressed as
\begin{eqnarray}
\widehat{\sigma}_{ij}(\hat{s},m,\mu_F^2,\mu_R^2) & = & 
\frac{\alpha_s^2(\mu_R^2)}{m^2}
\left\{ f^{(0,0)}_{ij}(\rho) \right. \nonumber \\
 & + & \left. 4\pi \alpha_s(\mu_R^2) \left[f^{(1,0)}_{ij}(\rho) + 
f^{(1,1)}_{ij}(\rho)\ln\bigg(\frac{\mu_F^2}{m^2} \bigg) \right] 
+ {\cal O}(\alpha_s^2) \right\},
\,\, 
\label{sigpart}
\end{eqnarray}
where $\rho = 4m^2/\hat{s}$ and 
$f_{ij}^{(k,l)}$ are the scaling functions upto NLO \cite{Nason:1989zy,Nason:1987xz}. 
At small $\rho$, the ${\cal O}(\alpha_s^2)$ and ${\cal O}(\alpha_s^3)$
$q \overline q$ and the ${\cal O}(\alpha_s^2)$ $gg$ scaling functions 
become small while the ${\cal O}(\alpha_s^3)$ $gg$ and $qg$ scaling functions
plateau at finite values.  Thus, at collider energies, the total cross sections
are primarily dependent on the small $x$ parton densities and phase space.
The total cross section does not depend on any kinematic variables. It depends  
only on the quark mass, $m$, and the renormalization and factorization scales with central
values $\mu_{R,F} =\mu_0 = m$.

The formation of quarkonium through a non-perturbative evolution of the $Q\bar Q$ pair 
has been discussed extensively using different models and also in the framework of the 
  effective field theories of QCD
\cite{Bodwin:1994jh,Brambilla:2004wf}. Different
treatments of this evolution have led to various theoretical models for
inclusive quarkonium production. Most notable among these are the color-singlet
model (CSM), the color-evaporation model (CEM) and the non-relativistic QCD
(NRQCD) factorization approach. In this review, we will describe all these models and will compare 
the findings of the CEM and the NRQCD 
approaches with the experimental results. 

\subsection{The color singlet model}

The color singlet model (CSM) was proposed shortly after the discovery of the 
J/$\psi$~\cite{Einhorn:1975ua,Berger:1980ni,Ellis:1976fj,Carlson:1976cd}.
In the CSM model, it is assumed that the $Q\bar Q$ pair which evolves into
the quarkonium is in a color-singlet state and also the pair has the same spin
and angular-momentum quantum numbers as the final quarkonium.
In the CSM, the production rate for the quarkonium state is obtained in terms of 
the absolute values of the color-singlet $Q\bar Q$ wave function and its
derivatives, evaluated at small $Q\bar Q$ separation.
These quantities can be extracted by comparing theoretical 
calculations in the CSM with experimental measurements. Once these quantities 
are extracted, the CSM  can predict the cross sections at any given collision energy.
At low energies, the CSM can successfully  predict the quarkonium
production cross sections~\cite{Schuler:1994hy}.
Nevertheless, at high energies, very large corrections to the CSM appear at next-to-leading
order (NLO) and next-to-next-to-leading order (NNLO) in $\alpha_s$
\cite{Artoisenet:2007xi,Campbell:2007ws,Artoisenet:2008fc}.
It thus indicates that there may be some additional production mechanism emerging at
higher energy. However, given the very large corrections at
NLO and NNLO, it is not clear that the perturbative expansion in
$\alpha_s$ is convergent. 

\subsection{The color evaporation model}  
\label{prod_sec:CEM}

The color evaporation model (CEM)~\cite{Fritzsch:1977ay,Amundson:1995em,Amundson:1996qr}
is motivated by the principle of quark-hadron duality. In the CEM, it
is assumed that every produced $\QQbar$ pair can evolve into a quarkonium
if it has an invariant mass that is less than the threshold for
producing a pair of open-flavor heavy mesons. It is further assumed that
the non-perturbative probability for the $\QQbar$ pair to evolve into a
quarkonium state $H$ is given by a constant $F_H$ that is independent of 
energy-momentum and process. Once $F_H$ has been fixed by
comparison with the measured total cross section for the production of
the quarkonium $H$, the CEM has no more free parameters and
can predict, the momentum distribution of the quarkonium production rate
at any collision energy.
The CEM predictions provide good descriptions of the CDF data for $\Jpsi$,
$\psi(2S)$ and $\chi_{c}$ production at $\sqrt{s} = $ 1.8~TeV
\cite{Amundson:1996qr}.

The quarkonium production cross sections are calculated in the
color evaporation model with normalizations determined by fitting the
scale parameter to the shape of the energy-dependent cross sections in Ref.~\cite{Nelson:2012bc}.
The production cross sections for heavy flavor and quarkonia at $\sNN$ = 5.02 
TeV~\cite{Kumar:2012qx} calculated using CEM are given in Table~\ref{NLOcros}.
The bottom quark production cross section is calculated to NLO in pQCD  
using the CT10 parton densities \cite{Lai:2010vv}.
The quark mass and scale parameters are $m_b = (4.65 \pm 0.09)$ GeV,
$\mu_F/m_{T\, b} = 1.40^{+0.75}_{-0.47}$, and $\mu_R/m_{T\, b} = 1.10^{+0.26}_{-0.19}$.
The central EPS09 NLO parameter set~\cite{Eskola:2009uj} is used to 
calculate the modifications of the parton distribution functions (nPDF) in 
Pb+Pb collisions, referred to as cold-nuclear matter (CNM) effects.
The yields in a minimum bias 
Pb+Pb event is obtained from the per nucleon cross
section, $\sigma_{\rm PbPb}$, in Table~\ref{NLOcros}, as
\begin{eqnarray}
N = {A^2 \sigma_{\rm PbPb} \over  
\sigma_{\rm PbPb}^{\rm tot}} \, \, .
\end{eqnarray}
Here $A$ is the mass number of the two colliding nuclei.  At 5.02 TeV, the total Pb+Pb cross section, $\sigma_{\rm PbPb}^{\rm tot}$, 
is 7.7 b~\cite{Loizides:2017ack}.

%
%

\begin{table}
  \begin{center}
\caption[]{Heavy quark and quarkonia production  cross sections at
$\sNN = $ 5.02 TeV~\cite{Kumar:2012qx}. The cross sections are given per nucleon pair while
$N^{\rm PbPb}$ gives the initial number of heavy quark pair/quarkonia per minimum bias Pb+Pb event.}
\label{NLOcros}
\begin{tabular}{l|l|l} 
\hline 
\hline
                        & $ b \overline b$                    & $\Upsilon$   \\              
\hline
$\sigma_{\rm pp}$            & $210.3^{+70.8}_{-77.6}~\mu$b            & $0.42^{+0.14}_{-0.16}~\mu$b  \\

$\sigma_{\rm PbPb}$        & $179.3^{+60.3}_{-66.2}~\mu$b             & 0.359$^{+0.121}_{-0.132}~\mu$b  \\

$N^{\rm PbPb}$              & $1.007^{+0.339}_{-0.372}$               & $0.0020^{+0.0007}_{-0.0007}$   \\

\hline
\hline
\end{tabular}
\end{center}
\end{table}

Recently, work in Ref.~\cite{Cheung:2018upe} presents an Improved Color Evaporation Model
(ICEM). They obtained bottomonium production cross sections as a function of
transverse momentum and rapidity and calculate the polarization of prompt
$\Upsilon$($n$S) production at leading order employing the $k_T$-factorization approach.
We reproduce here some of the representative calculations using ICEM.

\begin{figure*}
\centering
\includegraphics[width=0.60\textwidth]{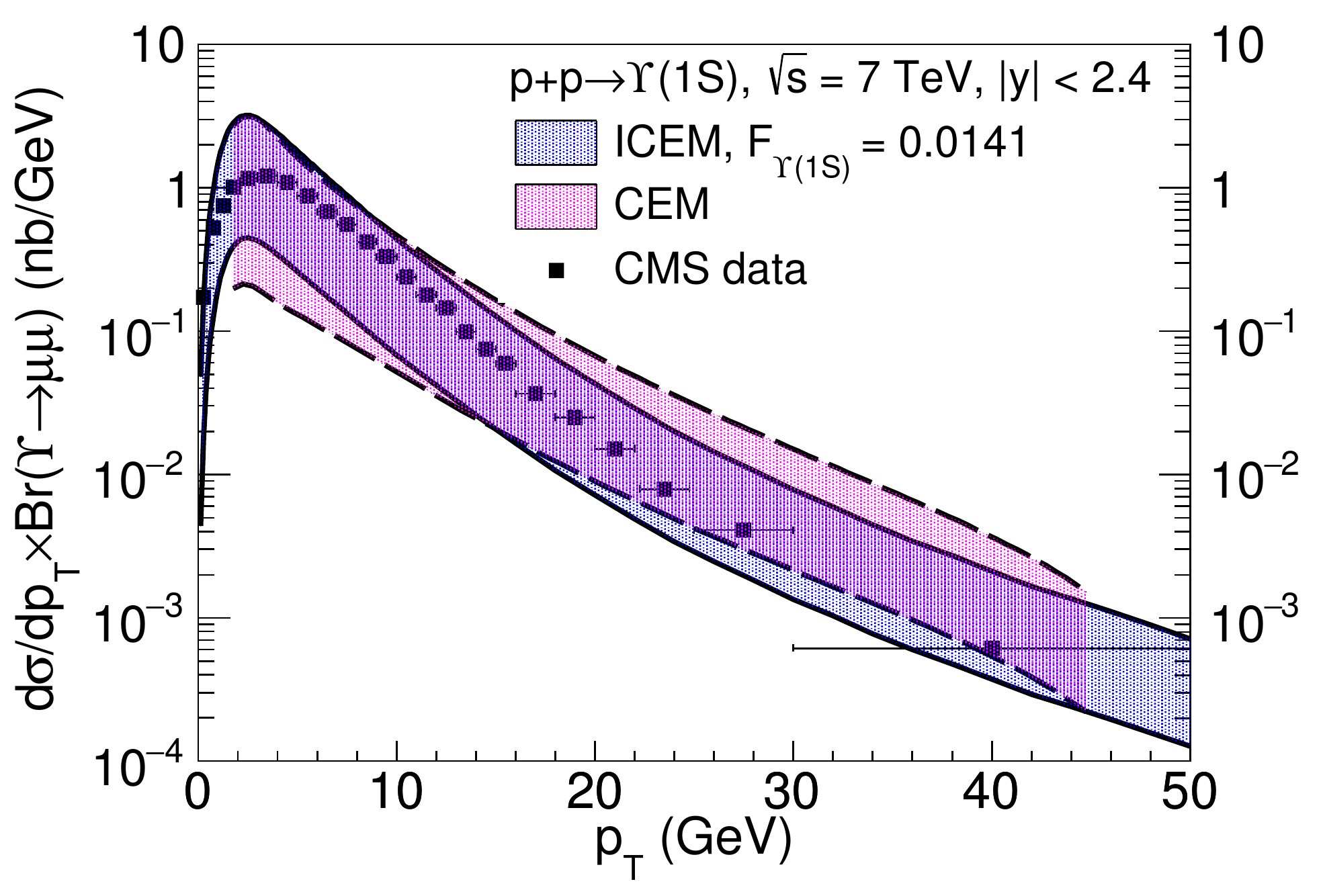}
\caption{(Color online) The differential cross sections for $\Upsilon$(1S) production as a function
  of $p_T$ in p+p collisions at $\sqrt{s} = $7~TeV in midrapidity $|y|<2.4$ calcuated using
ICEM~\cite{Cheung:2018upe} with combined mass and renormalization scale
uncertainties (blue).  Also shown the are calculations with CEM using collinear
factorization approach (magenta).
 The calculations are compared with the CMS midrapidity data \cite{CMS:2013qur}.}
\label{CMS_1S_pt}
\end{figure*}

Figure~\ref{CMS_1S_pt} shows the
differential cross sections for $\Upsilon$(1S) production as a function
  of $p_T$ in p+p collisions at $\sqrt{s} =$ 7~TeV in midrapidity, $|y|<2.4$, calculated using
ICEM~\cite{Cheung:2018upe} with combined mass and renormalization scale
uncertainties (blue).  Also shown are the calculations with CEM using a collinear
factorization approach (magenta). The calculations are compared with the CMS
midrapidity data \cite{CMS:2013qur}.

\begin{figure*}
\centering
\includegraphics[width=0.48\textwidth]{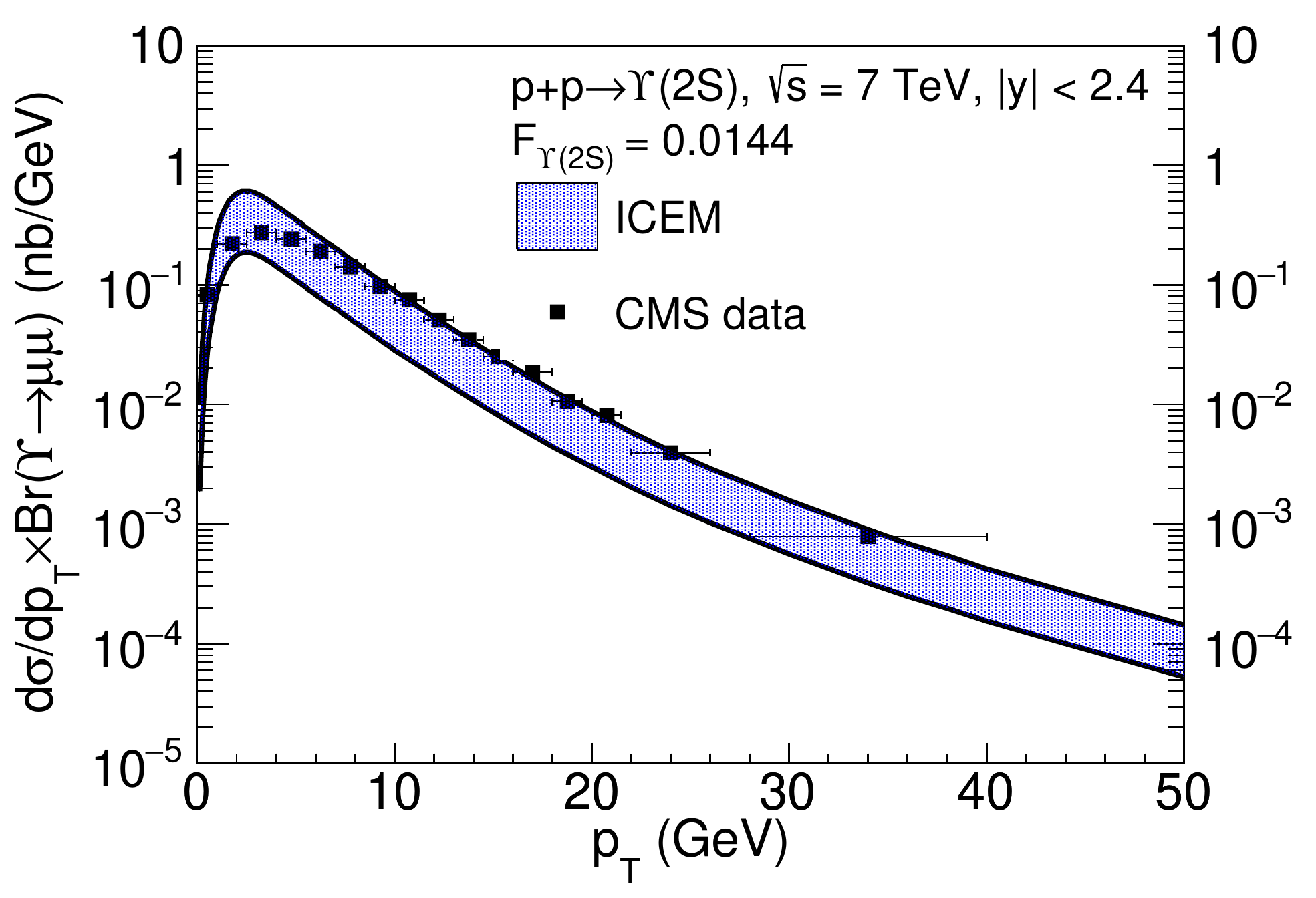}
\includegraphics[width=0.48\textwidth]{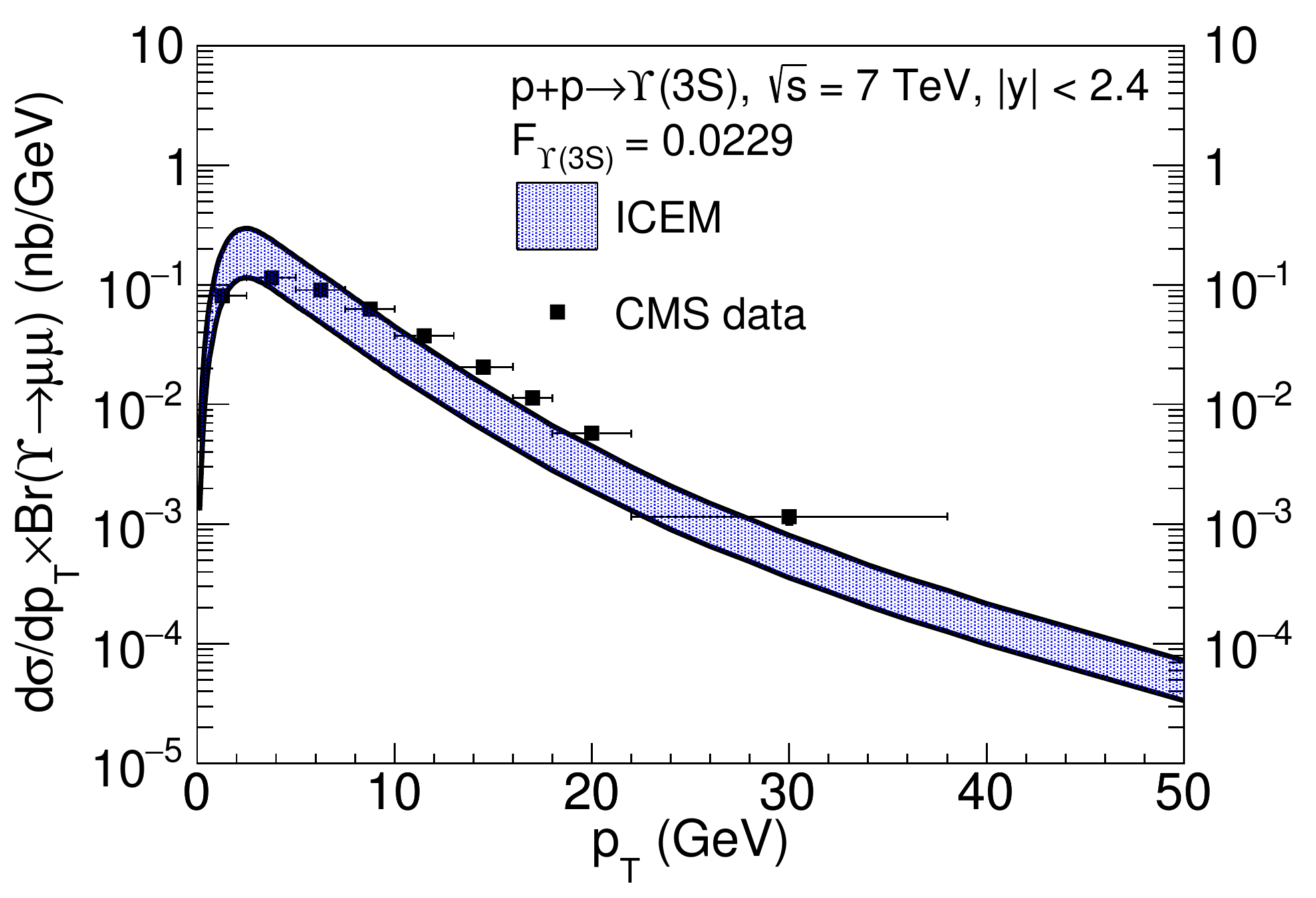}
\caption{(Color online) The differential production cross sections of prompt $\Upsilon$(2S) (left)
  and prompt $\Upsilon$(3S) (right) as a function
  of $p_T$ in p+p collisions at $\sqrt{s} =$ 7~TeV in midrapidity, $|y|<2.4$
  calculated using ICEM~\cite{Cheung:2018upe} with combined mass and renormalization
  scale uncertainties compared with the CMS midrapidity data \cite{CMS:2013qur}.}
\label{CMS_2S_3S_pt}
\end{figure*}

Figure~\ref{CMS_2S_3S_pt} shows the
differential production cross sections of prompt $\Upsilon$(2S) (left)
and prompt $\Upsilon$(3S) (right) as a function
of $p_T$ in p+p collisions at $\sqrt{s} = $ 7~TeV in midrapidity, $|y|<2.4$, 
calculated using ICEM~\cite{Cheung:2018upe} with combined mass and renormalization
scale uncertainties compared with the CMS midrapidity data \cite{CMS:2013qur}.

These results show that the $p_{\rm T}$ dependence of the cross sections of all three states
is explained by ICEM within the model uncertainties. The model gives 
the probability $F_H$ for the $\QQbar$ pair to evolve into a 
quarkonium state $H$ which is given in the corresponding figures. 
  
\subsection{The NRQCD factorization approach}

The $Q\bar{Q}$ pair, evolving into the quarkonium,
is assumed to have the same spin and angular momentum same as that of quarkonium.
NRQCD approach incorporates both the color singlet as
well as the Color Octet (CO) states of quarkonium.
In the formalism of the NRQCD factorisation approach, the evolution
probability of $Q\bar{Q}$ pair into a state of quarkonium is expressed as matrix
elements of NRQCD operators.
These operators are expanded
in terms of the velocity $v$ (for $v\ll$1) of heavy quarks~\cite{Bodwin:1994jh}.
The production cross sections and decay rates of quarkonia states are then
calculated using factorisation formulae.
The full structure of the $Q\bar{Q}$ Fock space
is spanned by $n$=$^{2s+1}L_J^{[a]}$ states where $L$ is the orbital angular momentum, $s$
is the spin, $J$ is the total angular momentum
and $a$ (color multiplicity) = 1 for CS and 8 for CO states. 
The produced CO states of $Q\bar{Q}$ pair at short distances finally emerge as 
CS quarkonia by emitting soft gluons non-perturbatively.

There have been several studies on bottomonia production based on
NRQCD formalism \cite{Domenech:1999qg,Domenech:2000ri,Braaten:2000cm,Gong:2010bk,Sharma:2012dy}.
Both the production and polarisation of $\Upsilon$(nS) at NLO have been discussed in 
Ref.~\cite{Gong:2013qka} within the framework of NRQCD.
The CO matrix elements are obtained by fitting the calculations with experimental data.
The study is updated in Ref.~\cite{Feng:2015wka} by considering
feed down from $\chi_{bJ}$(mP) states in $\Upsilon$(nS) production.
The yields and polarisations of $\Upsilon$(nS) measured at Tevatron and
LHC are well explained by this work.
The NLO study in Ref.~\cite{Han:2014kxa} includes feed down contributions
from higher states and describes the cross sections and
polarisations of $\Upsilon$(nS) at LHC energy.
In Ref.~\cite{Yu:2017pot}, production cross section for $\Upsilon$(nS),
$\chi_{bJ}$, $\eta_b$ and $h_b$ have been calculated using NRQCD, as produced
in hard photo production and fragmentation processes at LHC energies. 
In Ref.~\cite{Kumar:2021sek} it is shown that there is a large difference among the
Long Distance Matrix Elements (LDME) obtained by different analyses at NLO.

In Ref.~\cite{Kumar:2021sek} the LO NRQCD calculations for the differential production
cross sections of $\Upsilon$ states in p+p collisions have been discussed. 
This work uses a large set of data from Tevatron~\cite{Acosta:2001gv} and
LHC~\cite{CMS:2013qur,LHCb:2012aa,CMS:2015xqv,ATLAS:2012lmu,CMS:2017dju} 
to extract the LDMEs required for the $\Upsilon$ production.
It is to be noted that an LO NRQCD analyses is straightforward and has excellent
predictability power for unknown cross sections.


The processes that govern the production of heavy mesons like bottomonium,
can be denoted generically by 
$i+j\rightarrow \Upsilon +X$, where $i$ and $j$ are the incident light partons,
$\Upsilon$ is the heavy meson and $X$ is final state light parton.
The double differential cross section as a function of $p_T$ and $y$ of 
the heavy meson can be written as~\cite{Kumar:2016ojy},
\begin{eqnarray}
  E\,\frac{d^3\sigma^{\Upsilon} }{d^3p} &=& \sum_{i,j=q,\overline{q},g} \int dx_1 dx_2 f_{i}^p(x_1,\mu_F^2)
  f_{i}^p(x_2,\mu_F^2) \delta(s+u+t-m^2) {\hat{s} \over \pi} \, \frac{d\sigma}{d\hat{t}}.
  \label{eq4}
\end{eqnarray}
where, $f_{i}^p$($f_{j}^p$) are distribution functions of the colliding parton $i(j)$ in
the incident protons as a function of $x_1$($x_2$); the fractions of the total momentum
carried by the incident partons and the scale of factorisation $\mu_F$.
Here $\sqrt{s}$ is the total center of mass energy of the p+p system and $m_T~(=\mu_F)$ stands for
the transverse mass, $m_T^2=p_T^2 + M^2$ of the quarkonium.
The ${d\sigma}/{d\hat{t}}$ in Eq.~\ref{eq4} is the parton level cross section and is
defined as~\cite{Bodwin:1994jh},
\begin{equation}
  \frac{d\sigma}{d\hat{t}} = \frac{d\sigma}{d\hat{t}}(ab\rightarrow Q\bar{Q}(^{2s+1}L_J)+X)
  M_L(Q\bar{Q}(^{2s+1}L_J)\rightarrow \Upsilon)
  \label{eq6}
\end{equation}
The first term in righ hand side (RHS) is the short distance contribution, which corresponds to the $Q\bar{Q}$
pair production in a specific color and spin configuration and is calculable using 
perturbative QCD (pQCD)~\cite{Braaten:2000cm,Baier:1983va,Humpert:1986cy,Gastmans:1987be,Cho:1995vh,Cho:1995ce}.
The other term in the RHS of Eq. (\ref{eq6}) is the LDME 
and gives the probability of the $Q\bar{Q}$ converting into a quarkonium state.
They are determined by comparing the calculations with the measurements.

The quarkonium yield depends on the $^3S_1^{[1]}$ 
and $^3P_J^{[1]}$(J=0,1,2) CS states and $^1S_0^{[8]}$, $^3S_1^{[8]}$ and $^3P_J^{[8]}$
CO states in the limit $v\ll 1$.
The superscripts in square brackets represent the color structure of the bound state,
1 for the CS and 8 for the CO.

\begin{figure}
  \centering
  \includegraphics[width=0.99\textwidth]{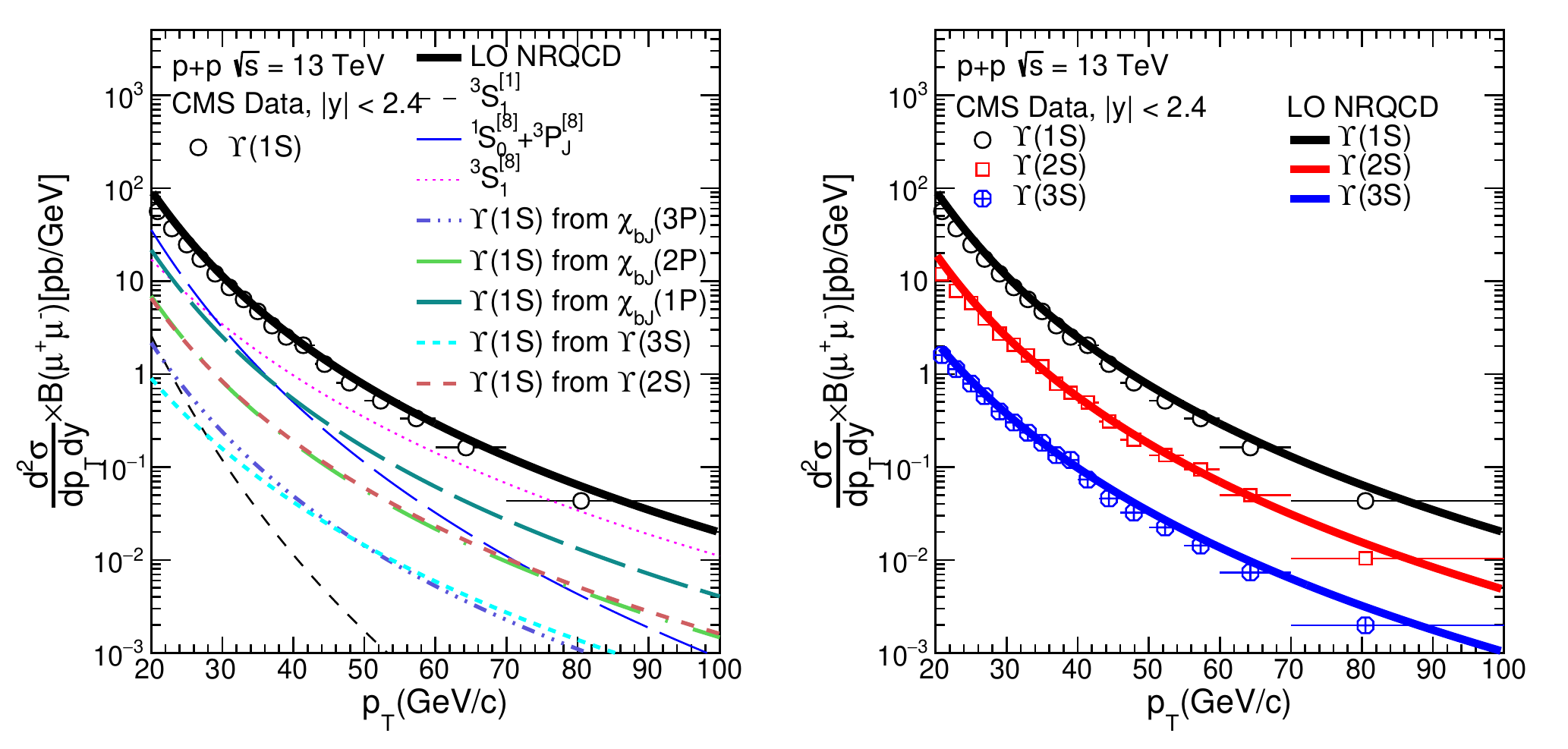}
  \caption{\small{(Color online) The NRQCD calculations~\cite{Kumar:2021sek} of the production cross section of $\Upsilon$(nS)
      in p+p collisions at $\sqrt{s}$ = 13 TeV in central rapidities, as a function of
      transverse momentum, compared with the measured data at CMS~\cite{CMS:2017dju}
      experiment. The left figure shows relative contributions in $\Upsilon$(1S) from
      singlet and octet states as well as from feed down. The right figure shows the sum
      of all contributions for all the 3 states where the results for $\Upsilon$(1S) and
      $\Upsilon$(2S) are shifted vertically by factors of 10 and 5, respectively
      for better visibility.}}
  \label{Fig:SigmaYnSCMS13TeV}
\end{figure}

\begin{figure}
  \centering
  \includegraphics[width=0.99\textwidth]{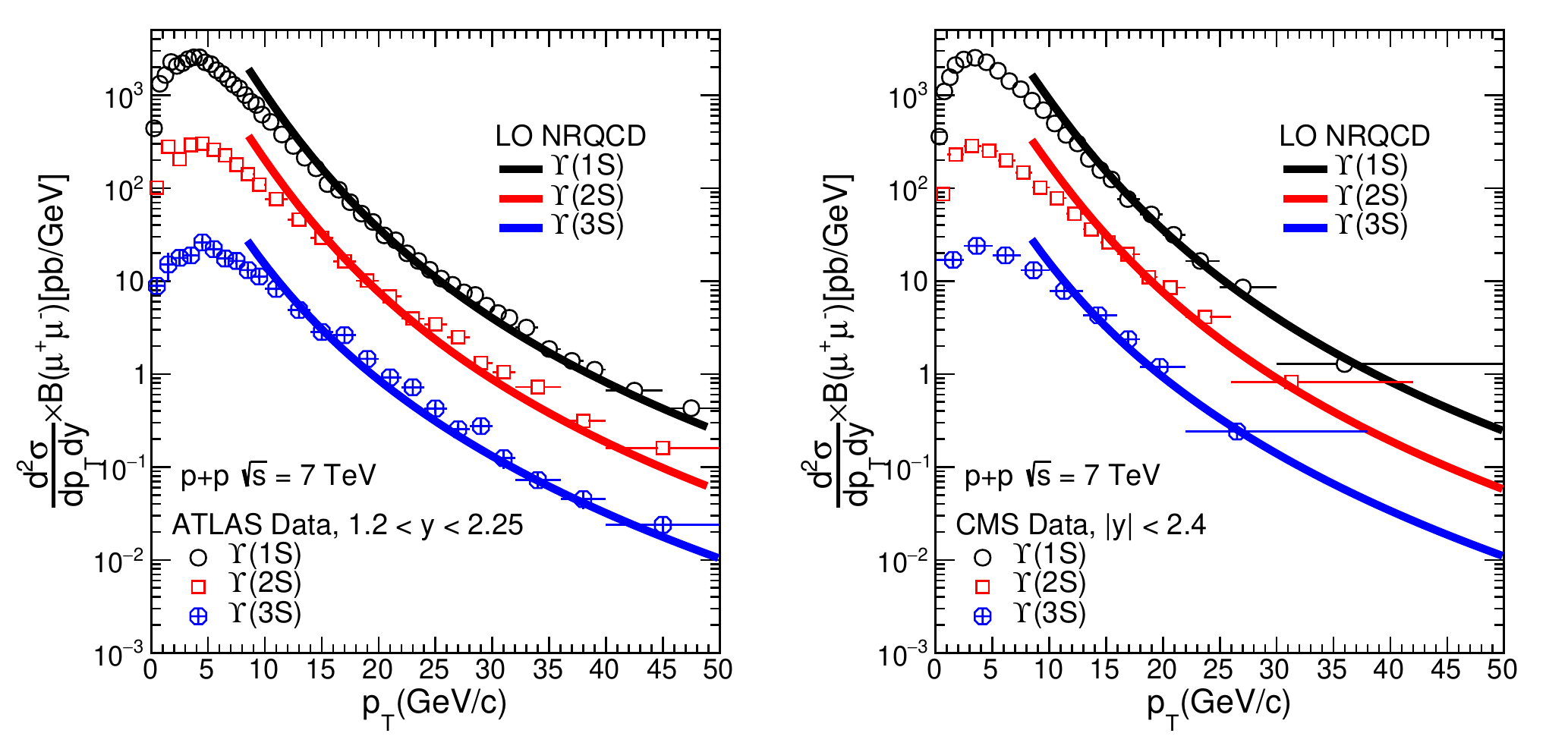}
  \caption{\small{(Color online) The NRQCD calculations~\cite{Kumar:2021sek} of the production cross section of $\Upsilon$(nS) in
      p+p collisions at $\sqrt{s}$ = 7 TeV, as a function of transverse momentum, compared with
      the measured data by ATLAS~\cite{ATLAS:2012lmu} in the left figure and CMS~\cite{CMS:2013qur}
      in the right figure. The cross section of $\Upsilon$(1S) and $\Upsilon$(2S) as well as
      calculations are shifted vertically by factors of 10 and 5, respectively for better visibility.}}
  \label{Fig:SigmaYnSCMS7TeV}
\end{figure}

One requires both CS and CO matrix elements in order to get theoretical
predictions for the production of bottomonia at the Tevatron and LHC energies.
The corresponding expressions and
numerical values for CS states are obtained from Ref.~\cite{Braaten:2000cm}.
The CO states are obtained using experimentally measured data sets 
as in Refs.~\cite{Braaten:2000cm,Cho:1995vh,Cho:1995ce}. 
The CO elements related to p-wave states, needed as the 
feed down contributions, are obtained by Ref.~\cite{Sharma:2012dy,Feng:2015wka}.

Here we present the NRQCD results obtained in Ref.~\cite{Kumar:2021sek}.
The calculations use CT18NLO parametrisation~\cite{Hou:2019efy} for parton distribution
functions and the bottom quark mass $m_b$ is taken to be 4.88 GeV.
Measured transverse momentum distributions of $\Upsilon$(3S), 
$\Upsilon$(2S) and $\Upsilon$(1S) in p +{$\bar {\rm p}$} collisions at
$\sqrt{s} = $ 1.8 TeV and in p+p collisions at 7 TeV and 13 TeV are used to constrain
the LDMEs.

\begin{table*}
  \centering
  \caption{Comparison of color singlet (CS) elements and color octet (CO) LDMEs extracted from fitting with experimental data
    using NRQCD formalism for $\Upsilon$(1S).}
  \footnotesize
  \begin{tabular*}{\textwidth}{@{\extracolsep{\fill}}lrrrrrl@{}}
    \hline
    \hline
    Ref. (LO/NLO) & PDF & $m_b$ & $M_L(b\bar{b}([^3S_1]_1$ & $M_L(b\bar{b}([^3S_1]_8$ & 
    $M_L(b\bar{b}([^1S_0]_8$, & $p_T$-cut \\
    & & & $\rightarrow\Upsilon(1S)$ & $\rightarrow\Upsilon(1S)$ & $[^3P_0]_8\rightarrow\Upsilon(1S)$ & \\
    & & (GeV) & $({\rm GeV^3})$ & $({\rm GeV^3})$ & $({\rm GeV^3})$ & GeV/$c$ \\
    \hline
    \hline
    & & & & & & \\
    \cite{Kumar:2021sek} (LO) & CT18 &4.88 &10.9 &0.0601$\pm$0.0017 & 0.0647$\pm$0.0016 & 8   \\
    & & & & & & \\
    \cite{Domenech:2000ri} (LO) & CTEQ4L & 4.88 & 11.1 & 0.077$\pm$0.017 & 0 & 2 \\
    & & & & 0.087$\pm$0.016 & 0 & 4 \\
    & & & & 0.106$\pm$0.013 & 0 & 8 \\
    & & & & & & \\
    \cite{Braaten:2000cm} (LO) & CTEQ5L & 4.77 & 12.8$\pm$1.6 & 0.116$\pm$0.027 & 0.109$\pm$0.062 & 8 \\
    & & & & 0.124$\pm$0.025 & 0.111$\pm$0.065 & \\
    & & & & & & \\
    & MRSTLO & 4.77 & 12.8$\pm$1.6 & 0.117$\pm$0.030 & 0.181$\pm$0.072 & 8 \\
    & & & & 0.130$\pm$0.028 & 0.186$\pm$0.075 & \\
    & & & & & & \\
    \cite{Sharma:2012dy} (LO) & MSTW08LO & 4.88 & 10.9 & 0.0477$\pm$0.0334 & 0.0121$\pm$0.0400 & -  \\
    & & & & & & \\
    \cite{Gong:2013qka} (NLO) & CTEQ6M & 4.75 & 9.282 & -0.0041$\pm$0.0024 & 0.0780$\pm$0.0043 & 8 \\
    & & & & & & \\
    \cite{Feng:2015wka} (NLO) & CTEQ6M & PDG & 9.282 & 0.0061$\pm$0.0024 & 0.0895$\pm$0.0248 & 8 \\
    \hline
    \hline
  \end{tabular*}
  \label{LDMEsY1S}
\end{table*}

Figure~\ref{Fig:SigmaYnSCMS13TeV} shows the NRQCD calculations of production cross section
of $\Upsilon$(nS) in p+p collisions at $\sqrt{s}$ = 13 TeV in central rapidities, as a function of
transverse momentum compared with the measured data at CMS~\cite{CMS:2017dju}
experiment. The left figure shows relative contributions in $\Upsilon$(1S) from
singlet and octet states as well as from feeddown. The right figure shows the sum
of all contributions for all the 3 states where the results for $\Upsilon$(1S) and
$\Upsilon$(2S) are shifted vertically by factors  of 10 and 5, respectively
for better visibility.

Figure~\ref{Fig:SigmaYnSCMS7TeV} shows the NRQCD calculations of the production cross section
of $\Upsilon$(nS) in p+p collisions at $\sqrt{s}$ = 7 TeV, as a function of transverse momentum compared with
the measured data by ATLAS~\cite{ATLAS:2012lmu} in the left figure and CMS~\cite{CMS:2013qur}
in the right figure. The cross section of $\Upsilon$(1S) and $\Upsilon$(2S) as well as
calculations are shifted vertically by factors  of 10 and 5, respectively for better visibility.
The NRQCD calculations have steeper slopes in comparison to the data at lower $p_T$ because the quarkonia 
production cross section saturates at lower $p_T$ due to the saturation of gluon distribution function inside protons.
This effect can be explained if we use the color glass condensate (CGC) model along with NRQCD~\cite{Ma:2014mri}. 

The calculations for  $\Upsilon$(3S), $\Upsilon$(2S) and $\Upsilon$(1S) are compared with 
the measured data at Tevatron and LHC~\cite{Kumar:2021sek}. The NRQCD formalism provides good description
of the data in large transverse momentum ranges at different collision energies. 
At high $p_T$, the color singlet contribution is very small and the LHC data in large $p_{\rm T}$ range 
help to constrain the relative contributions of different color octet contributions.
Table~\ref{LDMEsY1S} summarizes the LDME values for $\Upsilon$(1S) obtained by 
different groups.

\subsection{Additional methods}

In this section we, very briefly, touch upon two specialized processes namely
i) Fragmentation and ii) $k_T$ factorisation. 

\paragraph{Fragmentation}

In heavy-ion collisions at high energies, the produced partons
carry large transverse momentum.
When such a parton  with large transverse momentum $(k_T)$ decays into the final
hadronic state (quarkonium state here)~\cite{frag} then the process of production is called
fragmentation. At large enough $k_T$, quarkonium production is dominated by
fragmentation instead of the short distance mechanism which 
is suppressed by powers of $m_Q/k_T$ even though fragmentation is of higher
order in $\alpha_s$~\cite{frag}. 
It was first shown by Braaten and Yuan~\cite{frag,frag1} that fragmentation of
gluons and heavy quarks
 could be an important source of large-$k_T$ quarkonia production.
 A process like $A  B \rightarrow H  X$ (where $A,B$ are  hadrons)
is factorised  into a part containing the hard-scattering cross section which produces
 a gluon or a heavy quark 
and a part which takes care of the fragmentation of the gluon or the heavy quark into the
relevant quarkonia state. One may write 
\begin{equation}
d\sigma (A  B \rightarrow H X) = \sum_i \int_0^1 dz \ d\sigma  (A  B \rightarrow i  X) \ D_{i \rightarrow H} (x,\mu).
\end{equation}
In the above equation, $i$ is either a gluon or a heavy quark. The term $D_{i \rightarrow H} (x,\mu)$ 
is called the fragmentation function which depends on the fraction $(x)$ of momentum of the parent
parton carried by the quarkonia state and the scale $\mu$ which is of the order of $k_T$. 

\paragraph{$k_T$ factorisation}

Another approach to quarkonium production is the $k_T$ factorisation method~\cite{kt1,kt2}.
In the  standard collinear approach, it is assumed that the momentum of all partons is
in the same direction as the initial particle and thus the  transverse momentum $(k_T)$ is
considered to be zero. But, at large collision energies, the transverse
momentum $(k_T)$ is not negligible at all. 
 
In the $k_T$ factorisation approach, the quarkonium cross section is factorised
into two parts,  a cross section ${\hat \sigma} (x, k_T, \mu)$ and a parton
density function $f(x, k_T, \mu)$, where both depend on the transverse
momentum $k_T$~\cite{kt3}.  The quarkonium cross section is given by 
 
\begin{eqnarray}
   \sigma &=& \sum_{i,j} \int \frac {dx_1}{x_1} \frac {dx_2}{x_2} \
            f_i (x_1, k_{T,1}^2, \mu) \  f_j (x_2, k_{T,2}^2, \mu) \nonumber \\
    && \times \ {\hat \sigma}_{i+j \rightarrow H} (k_{T,1}, k_{T,2}, x_1, x_2, s) \ dk^2_{T,1} \ dk^2_{T,2}.
\end{eqnarray}
where $i$ and $j$ are initial partons, $H$ is the final state,
 and ${\hat \sigma}_{i+j \rightarrow H}$ is the parton cross
section giving the probability that initial partons $i$ and $j$ will form final state $H$.

%% file: BottAAexp.tex
\section{Experimental overview of Bottomonia results in heavy-ion collisions}
\label{Sec:BottAAexp}
\subsection{$\Upsilon$(nS) Nuclear Modification Factor $R_{AA}$}
A large set of heavy-ion collision data is available at both RHIC and LHC energies.
RHIC at BNL is designed for Au+Au collisions
upto $\sNN$ = 200 GeV and can accelerate ions up to Uranium. Both PHENIX and
STAR experiments at RHIC can measure quarkonia in the dimuon channel. 
LHC runs part of the time for heavy-ion program and it can perform Pb+Pb collision
up to $\sNN$ = 5.5 TeV. In addition, d+Au collisions
are performed at RHIC and p+Pb collisions are performed at LHC to study the 
intermediate system. The CMS, ATLAS and ALICE detectors at LHC have obtained large
amounts of $\Upsilon$ data in different kinematic ranges.

To quantify the effect of the medium in the quarkonia production scenario, one takes
recourse to a quantity called the nuclear modification factor ($R_{AA})$. This quantity
is defined as the ratio of the quarkonium yield in the A+A collisions to that
in p+p collisions scaled by the average number of collisions $\langle N_{\rm coll} \rangle$: 
 \begin{equation}
 R_{AA} = \frac{1}{\langle N_{\rm coll} \rangle} \ \frac {N^{Q{\bar Q}}_{AA}} {N^{Q{\bar Q}}_{pp}}.
 \end{equation}
 The ratio will be unity if the physics of the A+A collisions is simply the sum of
 a scaled number of p+p collisions. The effect of the medium should make it vary from unity. 
In this section, we review the current status of the experimental measurement
of the nuclear modification factor of the $\Upsilon$ states. The results from
different experiments are compared to understand the effects of the medium and their
dependence on the collision energy and kinematic ranges. 

 The cross sections of bottomonia at LHC are large and hence all three 
bottomonia states are measured at the LHC with very good statistical
precision~\cite{Chatrchyan:2011pe,Chatrchyan:2012lxa,Abelev:2014nua,Khachatryan:2016xxp}.
 Pb+Pb collisions at LHC were performed at $\sNN$ = 2.76 TeV starting from 
the year 2010. p+p collisions were performed at the same energy and p+Pb collisions were
performed at $\sNN$ = 5.02 TeV. 
During the second LHC run, Pb+Pb collisions were performed at $\sNN$ = 5.02 TeV 
and p+Pb collisions were performed at $\sNN$ = 8 TeV. 
The CMS experiment can reconstruct all three states of $\Upsilon$ starting from
$p_T$=0 covering a large central rapidity region with $|y| < 2.4$.
The ALICE experiment has reconstructed $\Upsilon$ in the forward rapidity range
$2.5 < y < 4.0$ in muon arm. The reach of ATLAS experiment is within $|y| < 1$ but
it can measure up to very high $p_T$. At lower energy, the STAR experiment can reconstruct in mid-rapidity range
$|y| < 1.0$ from zero $p_{\rm T}$ onwards. 

\begin{figure}
  \includegraphics[width=0.99\textwidth]{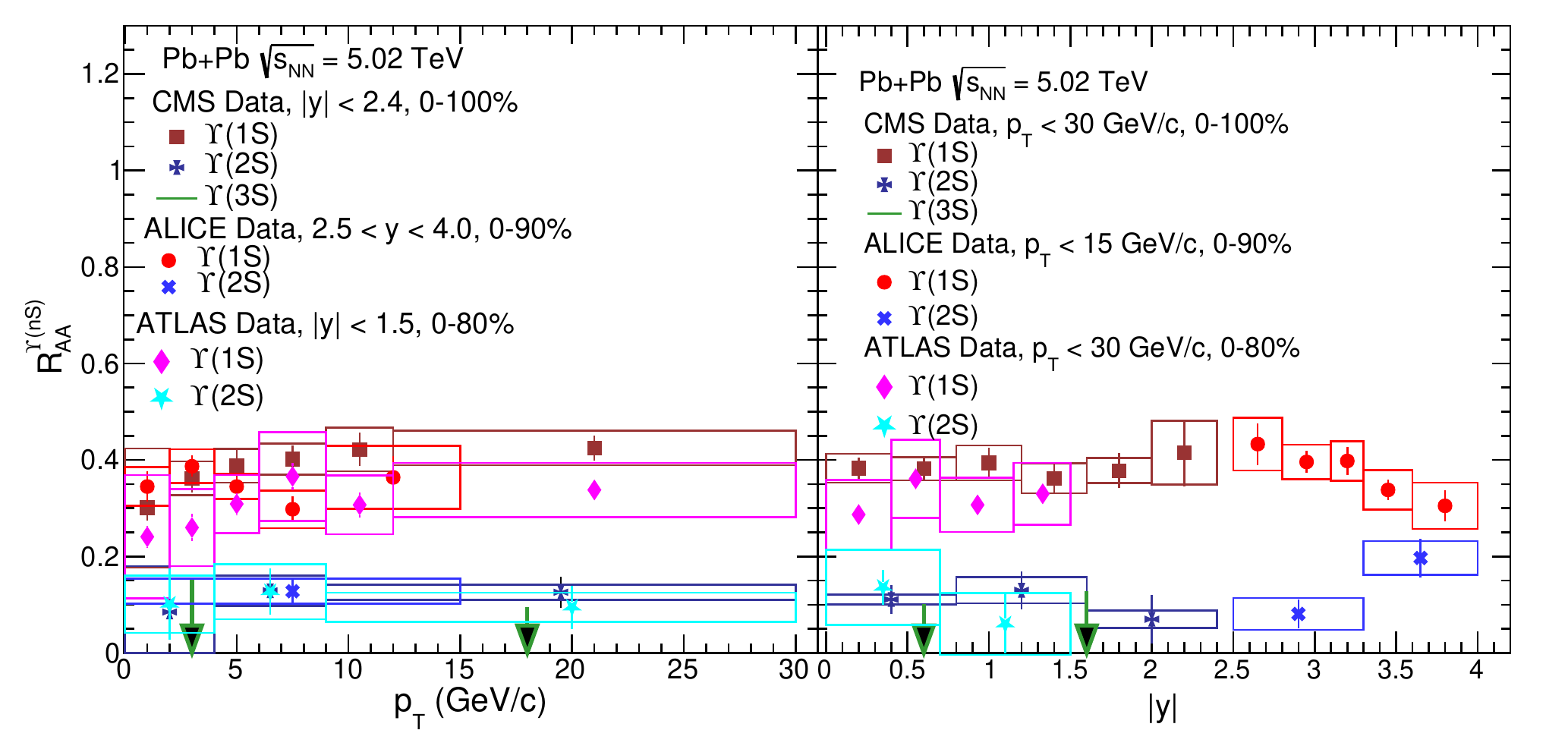}
  \caption{(Color online) The $\Upsilon$(nS) nuclear modification factor, $R_{AA}$
in Pb+Pb collisions at $\sNN$ = 5.02 TeV, (left) as a function of transverse momentum $p_{T}$
    and (right) as a function of rapidity measured by CMS~\cite{CMS:2018zza}, ALICE~\cite{ALICE:2020wwx}
    and ATLAS experiments~\cite{ALICE:2020wwx}.
    The vertical bars denote statistical uncertainties, and the rectangular boxes
    show the total systematic uncertainties.
  }
  \label{fig:LHCYnSRAAPtRap}
\end{figure}

\begin{figure}
  \begin{center}
  \includegraphics[width=0.6\textwidth]{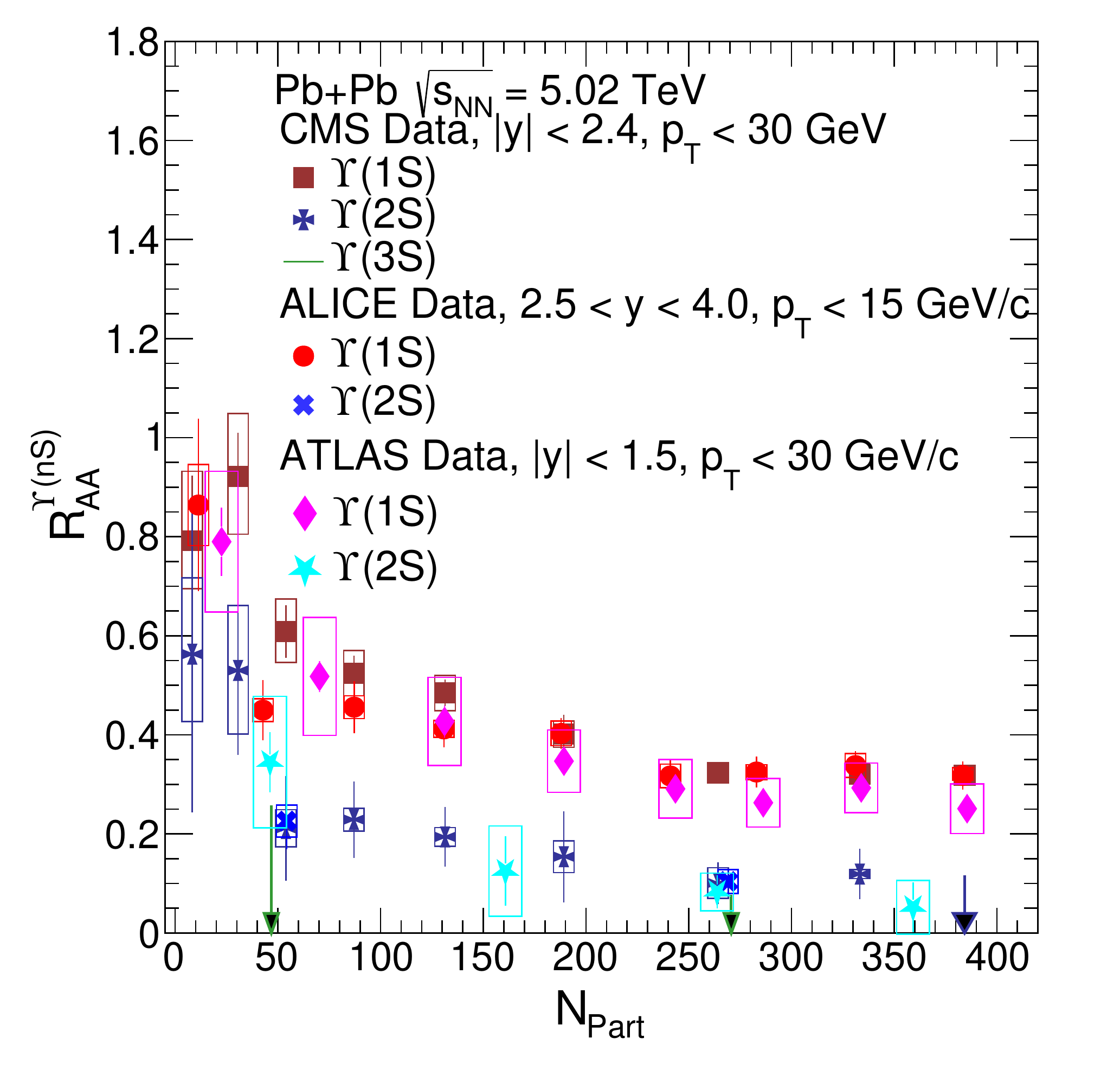}
  \caption{(Color online) The $\Upsilon$(nS) nuclear modification factor, $R_{AA}$
   in Pb+Pb collisions at $\sNN$ = 5.02 TeV as a function of $N_{\rm Part}$
   measured by CMS~\cite{CMS:2018zza}, ALICE experiments~\cite{ALICE:2020wwx} and
   ATLAS experiments~\cite{ALICE:2020wwx}.
   The vertical bars denote statistical uncertainties and the rectangular boxes show
   the total systematic uncertainties.
  }
  \label{fig:LHCYnSRAANPart}
  \end{center}
\end{figure}

\begin{figure}
  \includegraphics[width=0.99\textwidth]{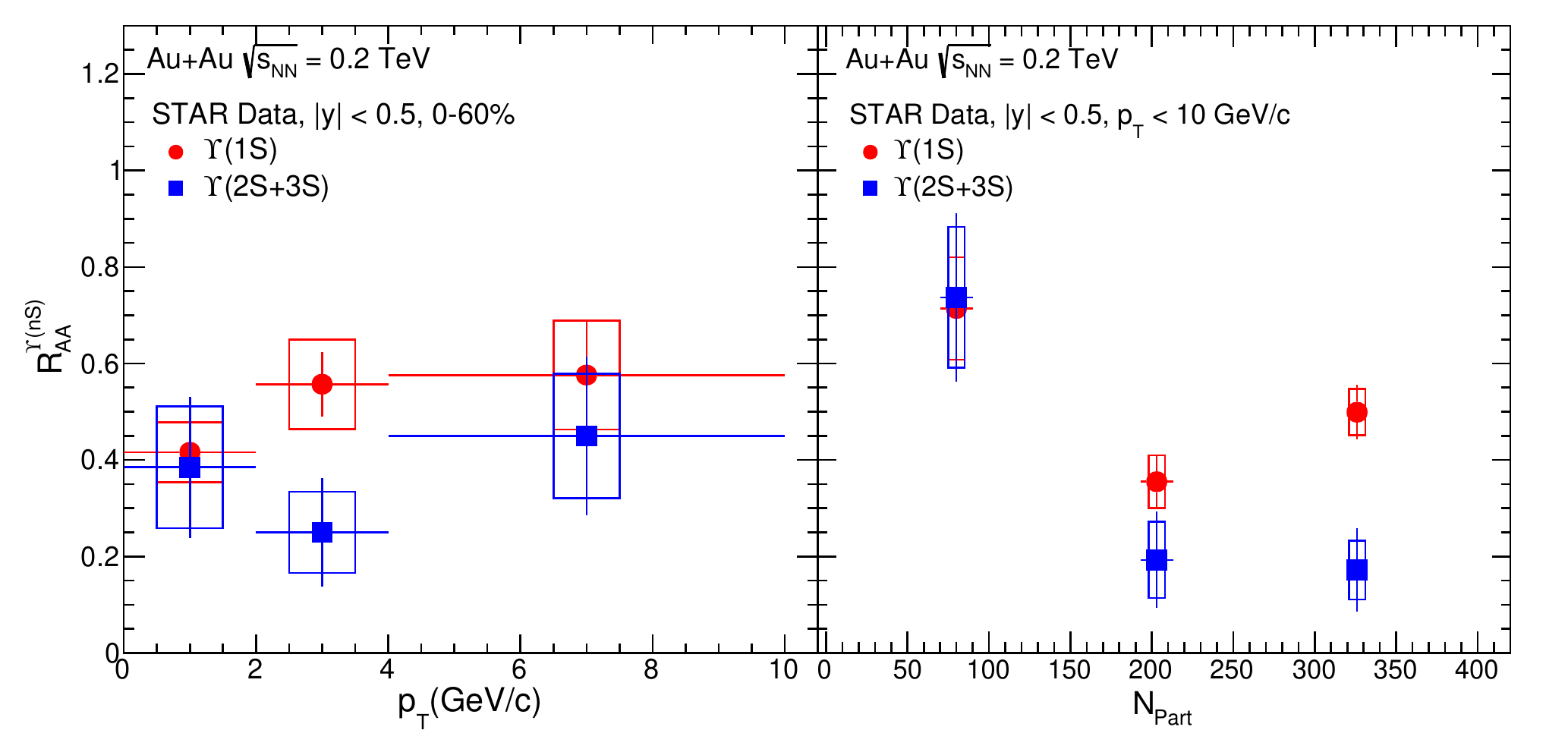}
  \caption{(Color online) The $\Upsilon$(nS) nuclear modification factor, $R_{AA}$
in Au+Au collisions at $\sNN$ = 200 GeV,
     (left) as a function of $p_{T}$
    and (right) as a function of $N_{\rm Part}$ measured by the STAR experiments~\cite{Wang:2019vau}. The vertical bars denote
    statistical uncertainties, and the rectangular boxes show the total systematic uncertainties.
  }
  \label{fig:RHICYnSRAAPt}
\end{figure}

\begin{figure}
  \includegraphics[width=0.99\textwidth]{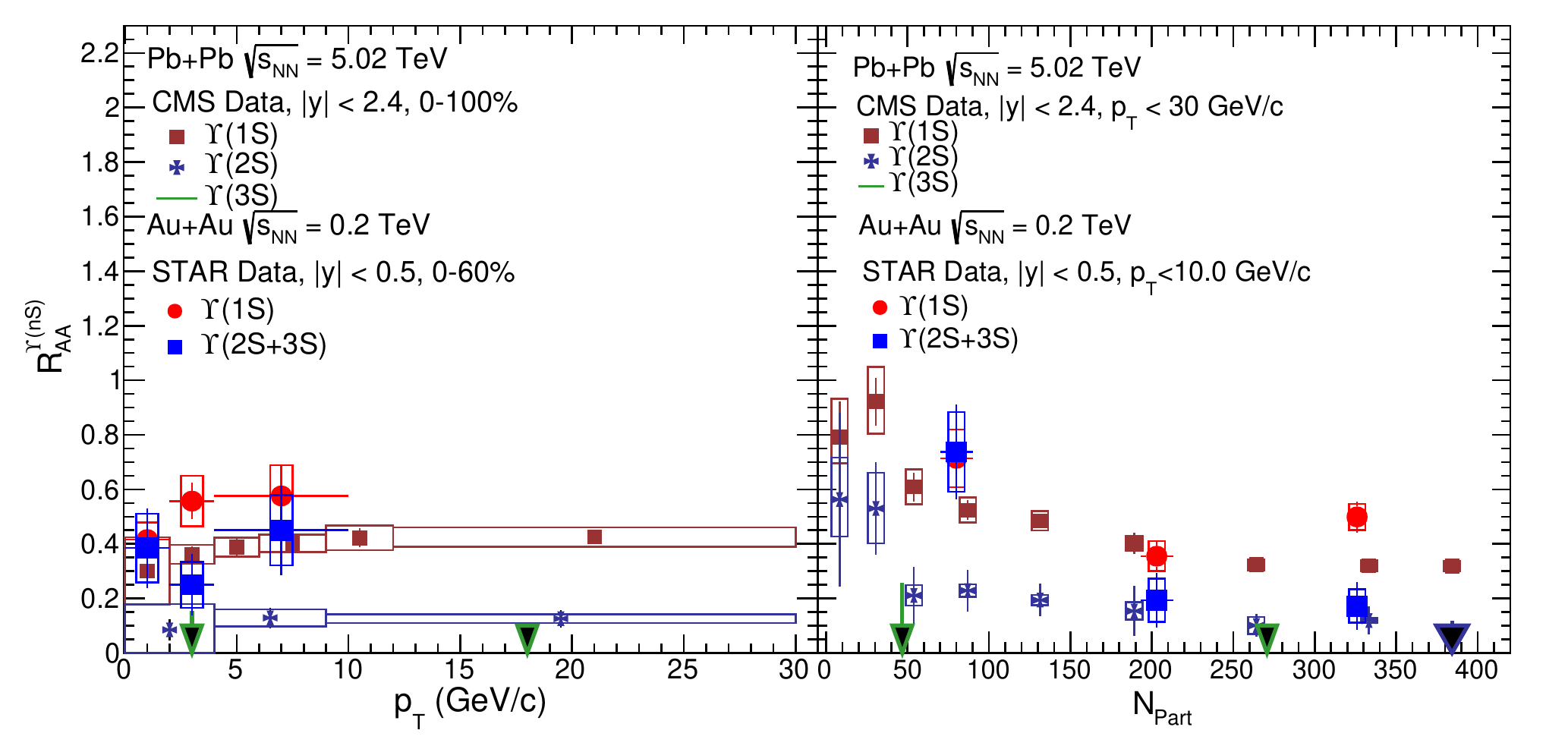}
  \caption{(Color online) The $\Upsilon$(nS) nuclear modification factor, $R_{AA}$, (left) as a function of $p_{T}$
    and (right) as a function of $N_{\rm Part}$ measured by the STAR experiments~\cite{Wang:2019vau} at 0.2 TeV and CMS experiment~\cite{CMS:2018zza} at 5.5 TeV.
    The vertical bars denote statistical uncertainties, and the rectangular boxes show the total systematic uncertainties.
  }
  \label{fig:RHICYnSRAANPart}
\end{figure}

Figure~\ref{fig:LHCYnSRAAPtRap} shows the $\Upsilon$(nS) nuclear modification factor, $R_{AA}$
in Pb+Pb collisions at $\sNN$ = 5.02 TeV, (left) as a function of $p_{T}$
and (right) as a function of rapidity measured by CMS~\cite{CMS:2018zza}, ALICE~\cite{ALICE:2020wwx}
and ATLAS experiments~\cite{ALICE:2020wwx}.
The vertical bars denote statistical uncertainties and the rectangular boxes
show the total systematic uncertainties. From these figures, it is clear 
that the individual $\Upsilon$ states are suppressed in
the Pb+Pb collisions relative to the production in the p+p collisions.
One can also notice that $\Upsilon$(2S) and $\Upsilon$(3S) are 
more suppressed relative to the ground state $\Upsilon$(1S) and there is sequential
suppression pattern as per the binding energies of the states.
$\Upsilon$(3S) almost disappears in Pb+Pb collisions. 
The suppression of $\Upsilon$ states increases with $p_{\rm T}$ and $R_{AA}$ looks
to be flat at high $p_T$ although more precise measurements (at high $p_{\rm T}$) are required to ascertain
this behavior. 
With increasing rapidity, the suppression remains the same and decreases
slightly but only at larger rapidities.
The forward rapidity ($2.5 \leq y^{\Upsilon} \leq 4.0$) measurement of the $\Upsilon$ suppression at 
ALICE is found to be consistent with the mid-rapidity ($|y^{\Upsilon}|\,\leq 2.4$)
measurement of the $\Upsilon$ suppression at the CMS which again shows the weak dependence
of suppression on rapidity.

Figure~\ref{fig:LHCYnSRAANPart} shows
the $\Upsilon$(nS) nuclear modification factor, $R_{AA}$
in Pb+Pb collisions at $\sNN$ = 5.02 TeV, as a function of $N_{\rm Part}$
measured by CMS~\cite{CMS:2018zza}, ALICE~\cite{ALICE:2020wwx}
and ATLAS experiments~\cite{ALICE:2020wwx}.
 The vertical bars denote statistical uncertainties and the rectangular
boxes show the total systematic uncertainties. The $\Upsilon$ nuclear modification
factor, $R_{AA}$, shows a strong dependence on collision centrality and the
suppression of all the states increases as the collisions become more central
corresponding to bigger system size. The results from different experiments
seem to agree with each other although the measurements of different experiments
correspond to different rapidity ranges.

Figure~\ref{fig:RHICYnSRAAPt} shows the $\Upsilon$(nS) nuclear modification factor, $R_{AA}$
in Au+Au collisions at $\sNN$ = 200 GeV,
 (left) as a function of $p_{T}$
and (right) as a function of $N_{\rm Part}$ measured by the STAR
experiment~\cite{Wang:2019vau}. The vertical bars denote
statistical uncertainties, and the rectangular boxes show the total systematic
uncertainties. At RHIC energy, there is a substantial suppression of $\Upsilon$ states.
Moreover, the suppression pattern of $\Upsilon$ states at RHIC 
looks similar as we discussed for LHC; Heavier states are more suppressed,
the suppression has weak dependence on $p_T$ and strong dependence on $N_{\rm Part}$.

Figure~\ref{fig:RHICYnSRAANPart} shows
  the $\Upsilon$(nS) nuclear modification factor, $R_{AA}$, (left) as a function of $p_{T}$
  and (right) as a function of $N_{\rm Part}$ measured by the STAR experiment~\cite{Wang:2019vau} at $\sNN$ =0.2 TeV and
  CMS experiment~\cite{CMS:2018zza} at $\sNN$ =5.5 TeV.
The vertical bars denote statistical uncertainties, and the
rectangular boxes show the total systematic uncertainties.
One can note that the suppression of $\Upsilon$ states is slightly stronger at
LHC as compared to that at RHIC. This is an evidence of medium of increasing
temperature at increasing collision energy.

\begin{figure}
  \includegraphics[width=0.99\textwidth]{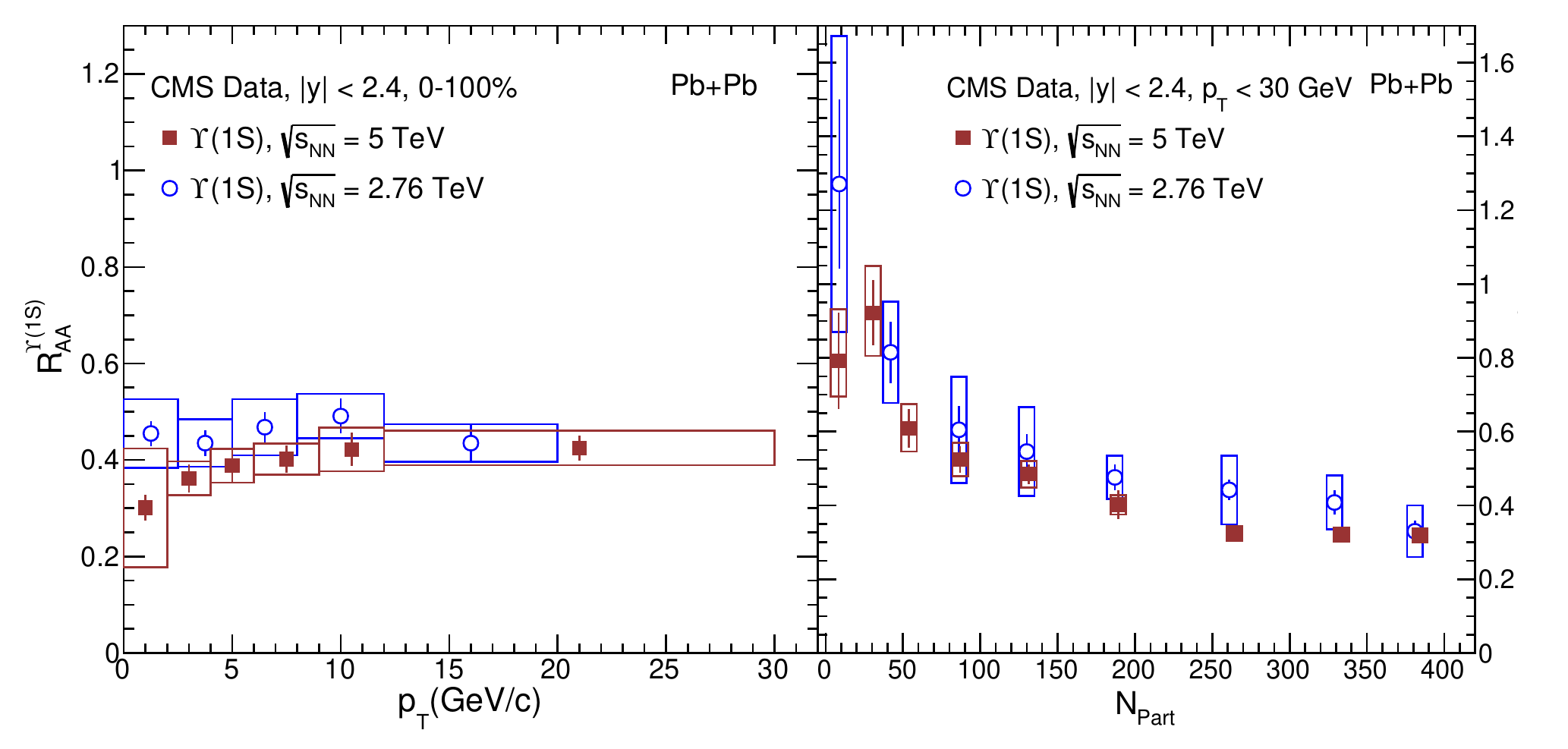}
  \caption{(Color online) The $\Upsilon$(nS) nuclear modification factor, $R_{AA}$ in
    Pb+Pb collisions,
    (left) as a function of $p_{T}$
    and (right) as a function $N_{\rm Part}$ measured by CMS 
    at 2.76~\cite{Khachatryan:2016xxp} and 5.02 TeV~\cite{CMS:2018zza}
  }
  \label{fig:LHCYnSRAAenergy}
\end{figure}

Figure~\ref{fig:LHCYnSRAAenergy} shows 
the $\Upsilon$(nS) nuclear modification factor, $R_{AA}$ in Pb+Pb collisions,
(left) as a function of  $p_{T}$
  and (right) as a function $N_{\rm Part}$ measured by CMS
    at 2.76~\cite{Khachatryan:2016xxp} and 5.02 TeV~\cite{CMS:2018zza}. 
 The CMS experiment measured slightly more amount of $\Upsilon$ suppression at
 $\sNN =$ 5.02 TeV than the suppression at $\sNN =$ 2.76 TeV.
 The ALICE experiment on the other hand observed less
suppression at $\sNN =$ 5.02 TeV than that at $\sNN =$ 2.76 TeV 
in the most central Pb+Pb collisions~\cite{ALICE:2018wzm,Abelev:2014nua}.

The overall conclusions from Figures~\ref{fig:RHICYnSRAANPart} and 
\ref{fig:LHCYnSRAAenergy} is that the suppression of $\Upsilon$ states
increases with collision energy albeit weakly. 
  
To summarize, LHC provided high statistics measurements of $R_{AA}$ in
Pb+Pb collisions for all three $\Upsilon$ states over wide kinematical ranges.
All $\Upsilon$ states are found to be suppressed in the Pb+Pb collisions,
the heavier states are more suppressed relative to the ground state.
The suppression of $\Upsilon$ states strongly depends on system size but
has a weak dependence on $p_T$ and rapidity. At high $p_T$, more precise
measurements are required to ascertain constancy of the suppression. 
Comparing the measurements at RHIC and at two energies of LHC, it can be
said that the suppression increases with energy albeit weakly.

\subsection{$\Upsilon$(nS) Azimuthal anisotropy}

In semi-central heavy-ion collisions,
the produced QGP has a lenticular shape in the transverse plane
which is reflected in the anisotropic
distribution of particles obtained using the magnitudes
of the Fourier co-efficients ($v_{n}$) of the azimuthal correlation of
particles~\cite{Voloshin:1994mz}. By studying the azimuthal distribution of
the produced quarkonia, it is possible to develop a more comprehensive understanding
of the dynamics of their production.

\begin{figure}
\includegraphics[width=0.99\textwidth]{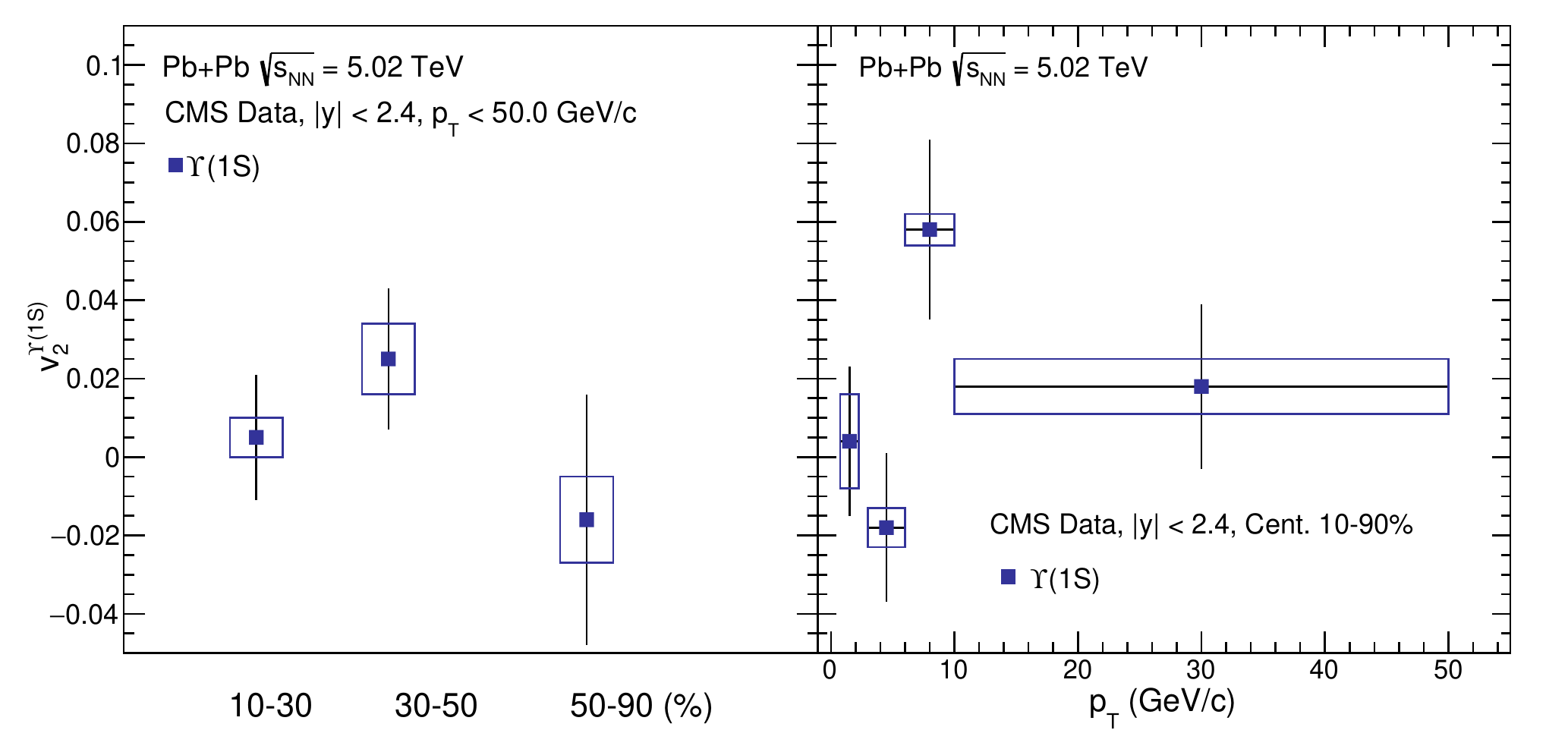}
\caption{(Color online) The $\Upsilon$(1S) azimuthal anisotropy ($v_{2}$) (left) as a
  function of collision centrality and (right) as a function of 
  $p_{T}$~\cite{CMS:2020efs}. The vertical bars denote statistical uncertainties,
  and the rectangular boxes show the total systematic uncertainties.
}
\label{fig:Upsilon1SV2CMS}
\end{figure}

The CMS experiment measured the second order Fourier coefficient $v_{2}$ for $\Upsilon$(1S) and $\Upsilon$(2S)
mesons in Pb+Pb collisions at $\sNN$ = 5.02 TeV.
Figure~\ref{fig:Upsilon1SV2CMS} shows the $\Upsilon$(1S) azimuthal
anisotropy ($v_{2}$) (left) as a function of collision centrality and (right) as a
function of $p_{T}$ measured by the CMS experiment at
LHC~\cite{CMS:2020efs}. The p$_{T}$ integrated results shown in
Fig.~\ref{fig:Upsilon1SV2CMS} (left) for three centrality intervals are consistent
with zero within the statistical uncertainties. The average $v_{2}$ values in the
10-90$\%$ centrality interval measured by the CMS the experiment are found to
be 0.007$\pm$0.011(stat)$\pm$0.005(syst) for $\Upsilon$(1S) and
-0.063$\pm$0.085(stat)$\pm$0.037(syst) for $\Upsilon$(2S).   
The p$_{T}$ dependence of  $v_{2}$ of $\Upsilon$(1S) meson is measured
for the 10-90$\%$ centrality interval. The values of $v_{2}$ are consistent with
zero in the measured p$_T$ range, except for the $6<p_T<10$ GeV/$c$ interval that
shows a 2.6 $\sigma$ deviation from zero~\cite{CMS:2020efs}. 

\begin{figure}
  \begin{center}
\includegraphics[width=0.6\textwidth]{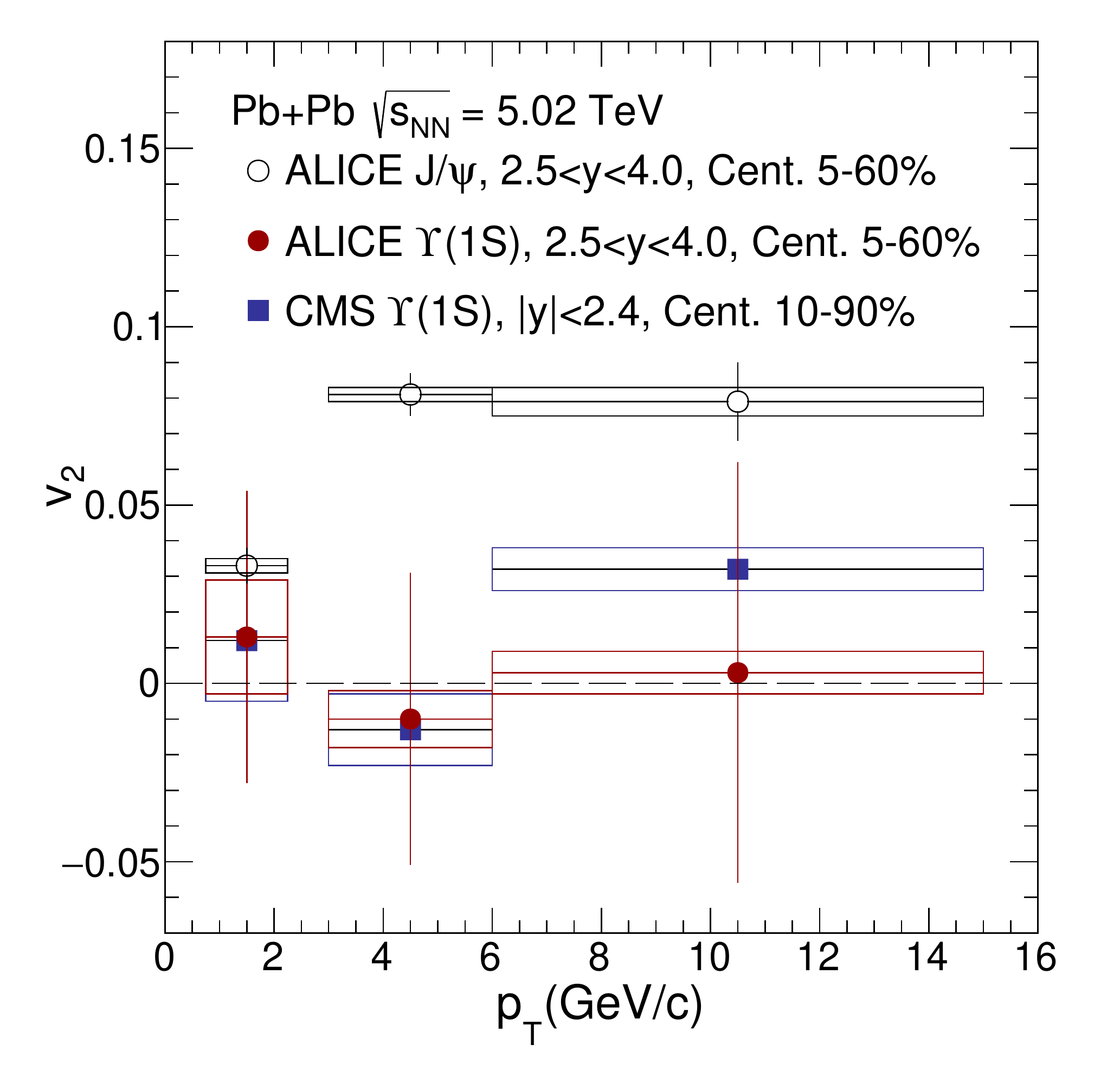}
\caption{(Color online) The $v_{2}$ for $\Upsilon$(1S) mesons as a function of p$_{T}$ in the
  rapidity range $|y|<2.4$ measured by
  CMS experiment~\cite{CMS:2020efs} compared with the ALICE results for $\Upsilon$(1S)
  and J/$\psi$ mesons measured in 2.5$<$y$<$4~\cite{ALICE:2019pox}.
  The vertical bars denote statistical uncertainties,
  and the rectangular boxes show the total systematic uncertainties.
}
\label{fig:Upsilon1SV2Compare}
\end{center}
\end{figure}

Figure~\ref{fig:Upsilon1SV2Compare} shows the p$_{T}$ differential results
for $v_{2}$ of $\Upsilon$(1S) mesons measured by CMS experiment along with the
measurements of $v_{2}$ for $\Upsilon$(1S) and J/$\psi$ from ALICE in the same
p$_{T}$ (0-15 GeV/$c$) and centrality (5-60$\%$) interval. The measurements from CMS
and ALICE are done in complementary rapidity ranges. The $\Upsilon$(1S) $v_{2}$ is
consistent with zero while the J/$\psi$ meson measured by ALICE 
has finite $v_{2}$.

Together, the CMS and ALICE
results indicate that the collective effects of the medium on the 
$\Upsilon$(1S) are small. This also indicates that the bottom quark is not
thermalized in the medium at LHC while the charm quark does thermalize. This has
implications for the recombination yield of bottomonia at LHC.

\subsection{$\Upsilon$(nS) in p+Pb collisions }
\label{sectionpA}

The nuclear modification factors and ratios of $\Upsilon$ states are measured by CMS
in p+Pb collisions as well as covering wide kinematic regions.

\begin{figure}
  \includegraphics[width=0.99\textwidth]{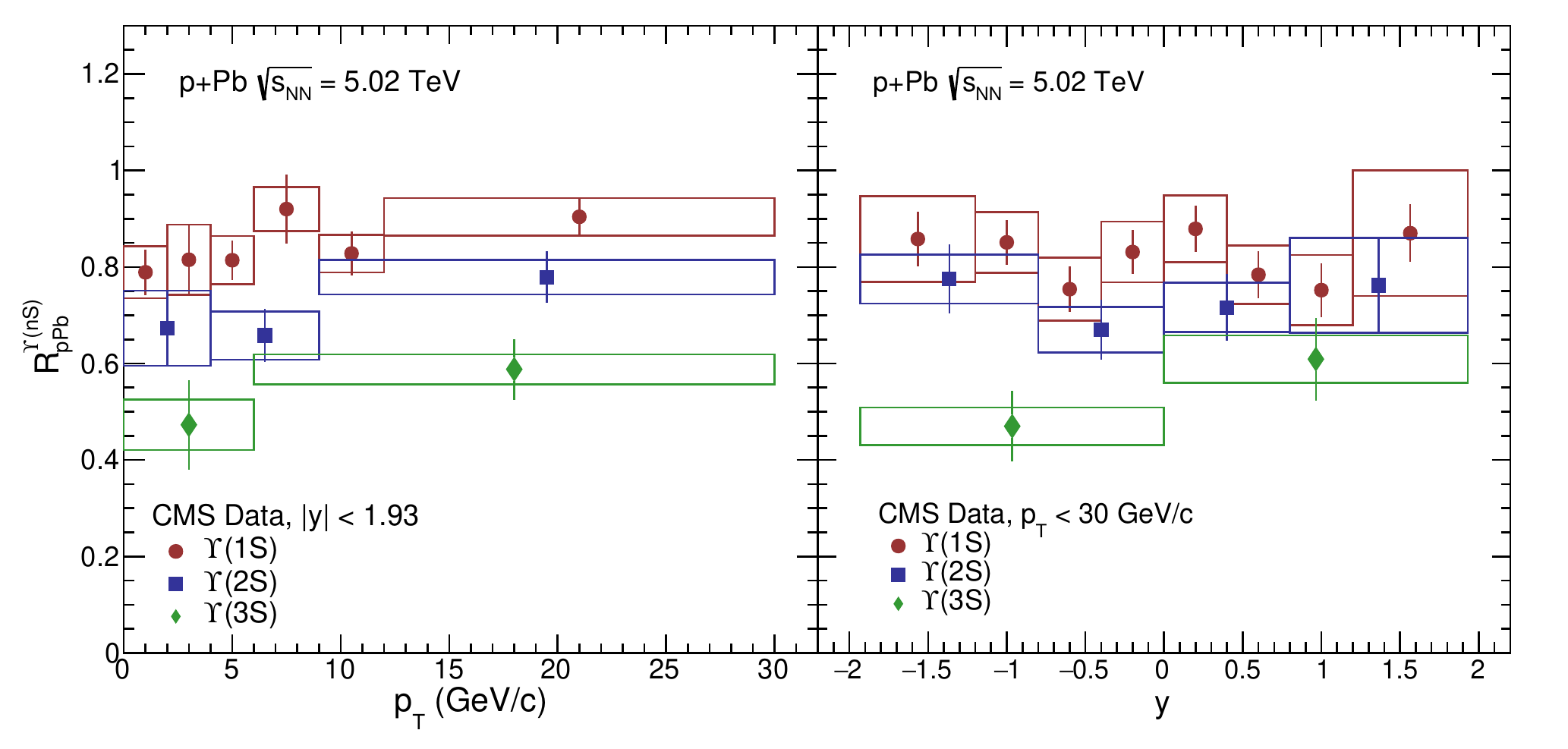}
  \caption{(Color online) The $\Upsilon$(nS) nuclear modification factor, $R_{\rm pPb}$,
    (left) as a function of $p_{T}$
    and (right) as a function of rapidity in p+Pb collisions at $\sNN =$ 5.02 TeV measured by CMS~\cite{CMS:2022wfi}
  }
  \label{fig:LHCpPb5}
\end{figure}

Figure~\ref{fig:LHCpPb5} shows the $\Upsilon$(nS) nuclear modification factor, $R_{\rm pPb}$,
(left) as a function of $p_{T}$
and (right) as a function of rapidity in p+Pb collisions at $\sNN =$ 5.02 TeV measured by CMS~\cite{CMS:2022wfi}.
It is observed that all three $\Upsilon$ states are suppressed in p+Pb collisions; while
$R_{\rm pPb}$ remains flat in the measured rapidity window, it shows an increasing trend with
increasing $p_{\rm T}$. The excited states are more suppressed as compared to the ground state.
Thus a sequential suppression pattern is observed in p+Pb collisions indicating final
state effects on bottomonia production. 

\begin{figure}
\centering
  \includegraphics[width=0.60\textwidth]{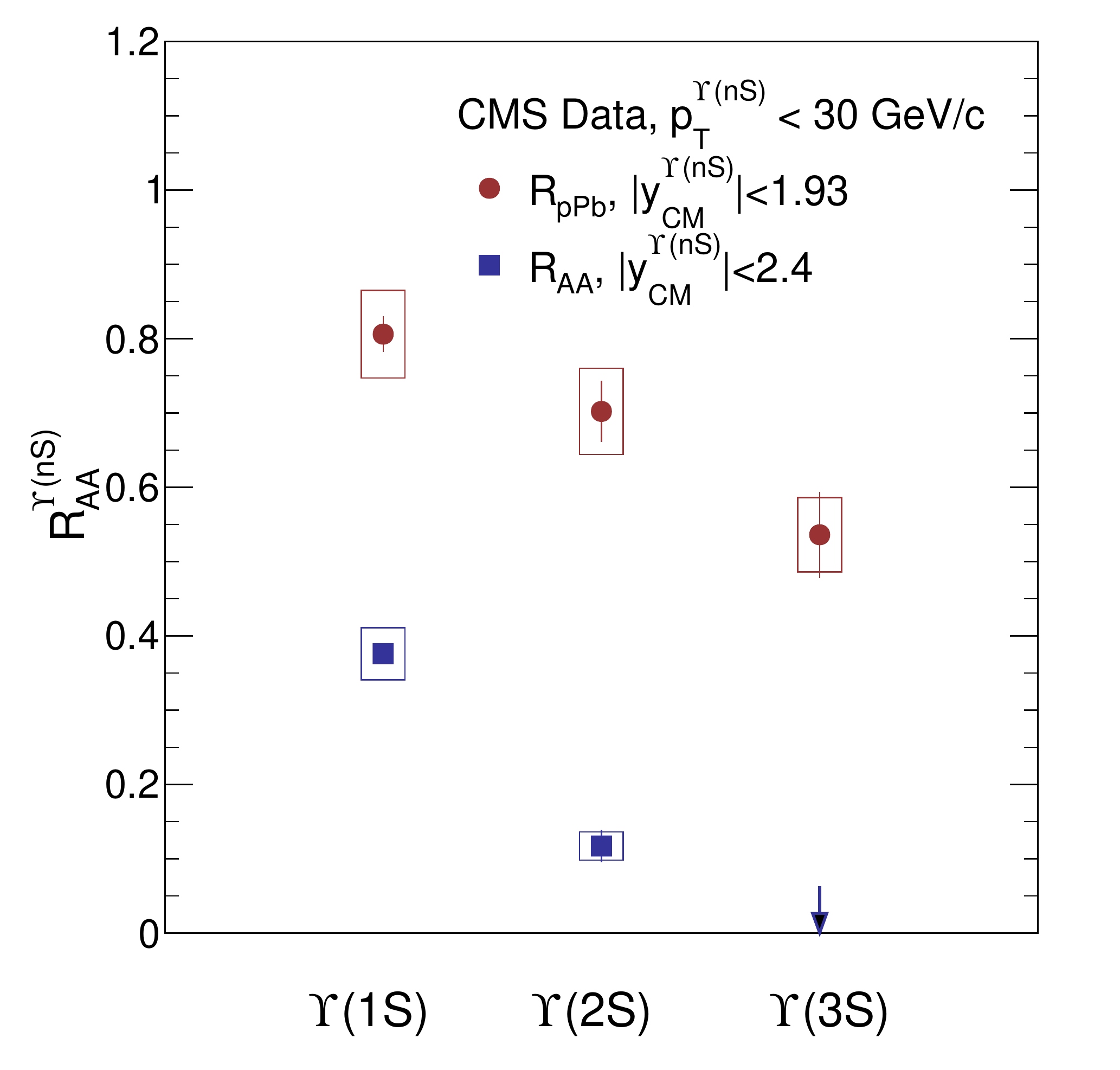}
  \caption{(Color online) The $\Upsilon$(nS) nuclear modification factors,
    $R_{pA}$~\cite{CMS:2022wfi} and $R_{AA}$~\cite{CMS:2018zza}
    at $\sNN =$ 5.02 TeV measured by CMS.
  }
 \label{fig:LHCpBPbPb}
\end{figure}

Figure~\ref{fig:LHCpBPbPb} shows the $\Upsilon$(nS) nuclear modification factors,
$R_{\rm pPb}$~\cite{CMS:2022wfi} and $R_{AA}$~\cite{CMS:2018zza}
at $\sNN =$ 5.02 TeV measured by CMS. It is observed that $\Upsilon$ states are suppressed
in both p+Pb and Pb+Pb collisions though the suppression is significantly stronger in Pb+Pb collisions.

The CMS experiment measured the $\Upsilon$ ratios as a function of event activity  
in p+Pb collisions at $\sNN$ = 5.02 TeV~\cite{CMS:2013jsu}.
The event activity in CMS is given by the number of tracks $N_{\rm tracks}^{|y|<2.4}$ within rapidity
range $|y|<2.4$.
The results were compared with p+p and Pb+Pb collisions at $\sNN$ = 2.76 TeV.
 The nuclear modification of all $\Upsilon$ states is also measured in p+Pb collisions
 at $\sNN$ = 5.02 TeV~\cite{CMS:2022wfi}.
 Recently, relative production of $\Upsilon$(nS) states are measured as a function of
 event activity in proton+proton collisions at $\sqrt{s}$ = 7 TeV~\cite{CMS:2020fae}.

\begin{figure}
  \begin{center}
    \includegraphics[width=0.60\textwidth]{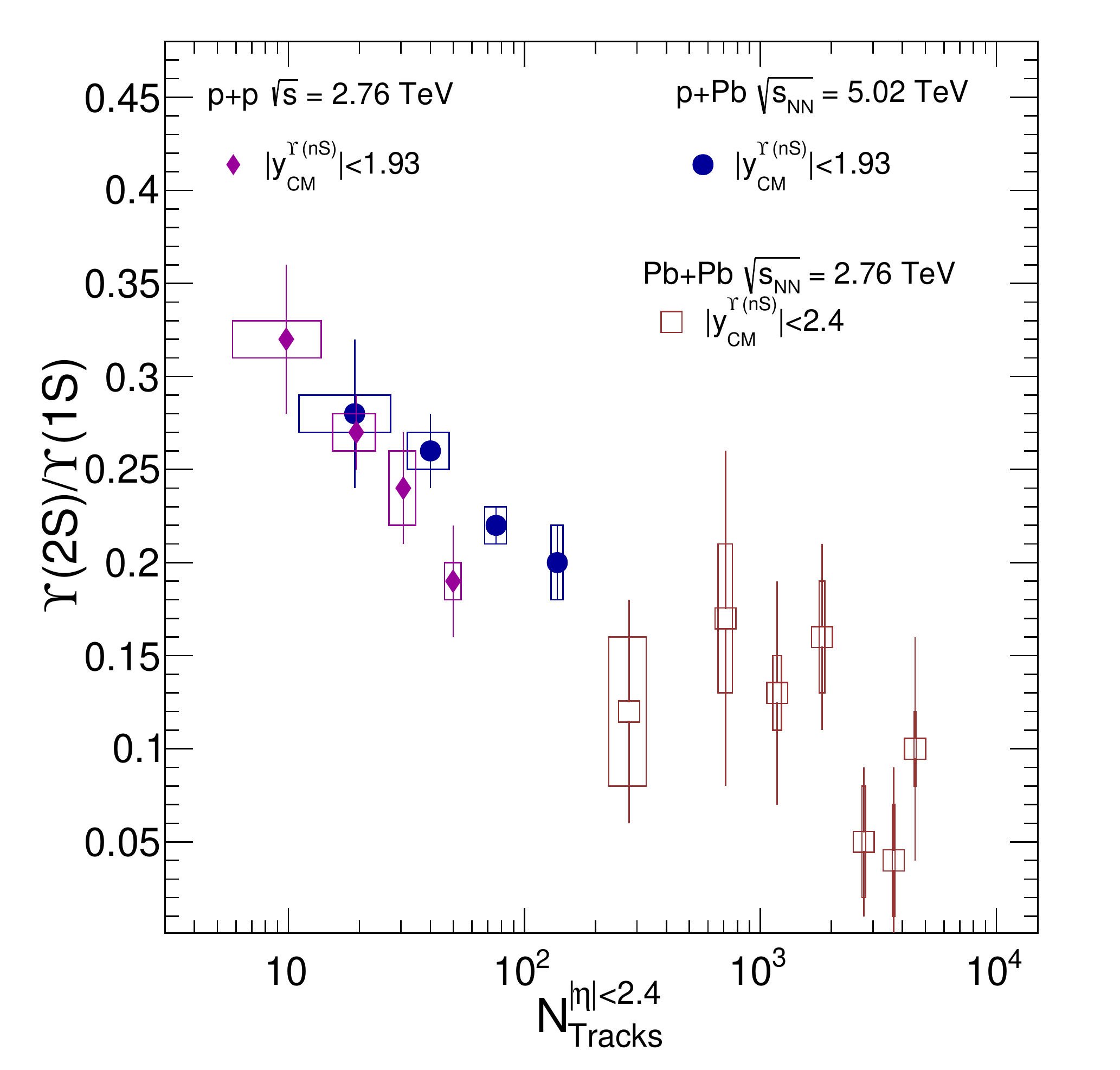}
\caption{(Color online)
   The ratio $\Upsilon$(2S)/$\Upsilon$(1S) as a function of event activity measured in 
$\sNN$ = 5.02 TeV p+Pb collisions~\cite{CMS:2013jsu} and compared with p+p
and Pb+Pb Collisions at $\sNN$ = 2.76 TeV.
}
\label{fig:UpsilonpPbA}
\end{center}
\end{figure}

\begin{figure}
  \begin{center}
    \includegraphics[width=0.60\textwidth]{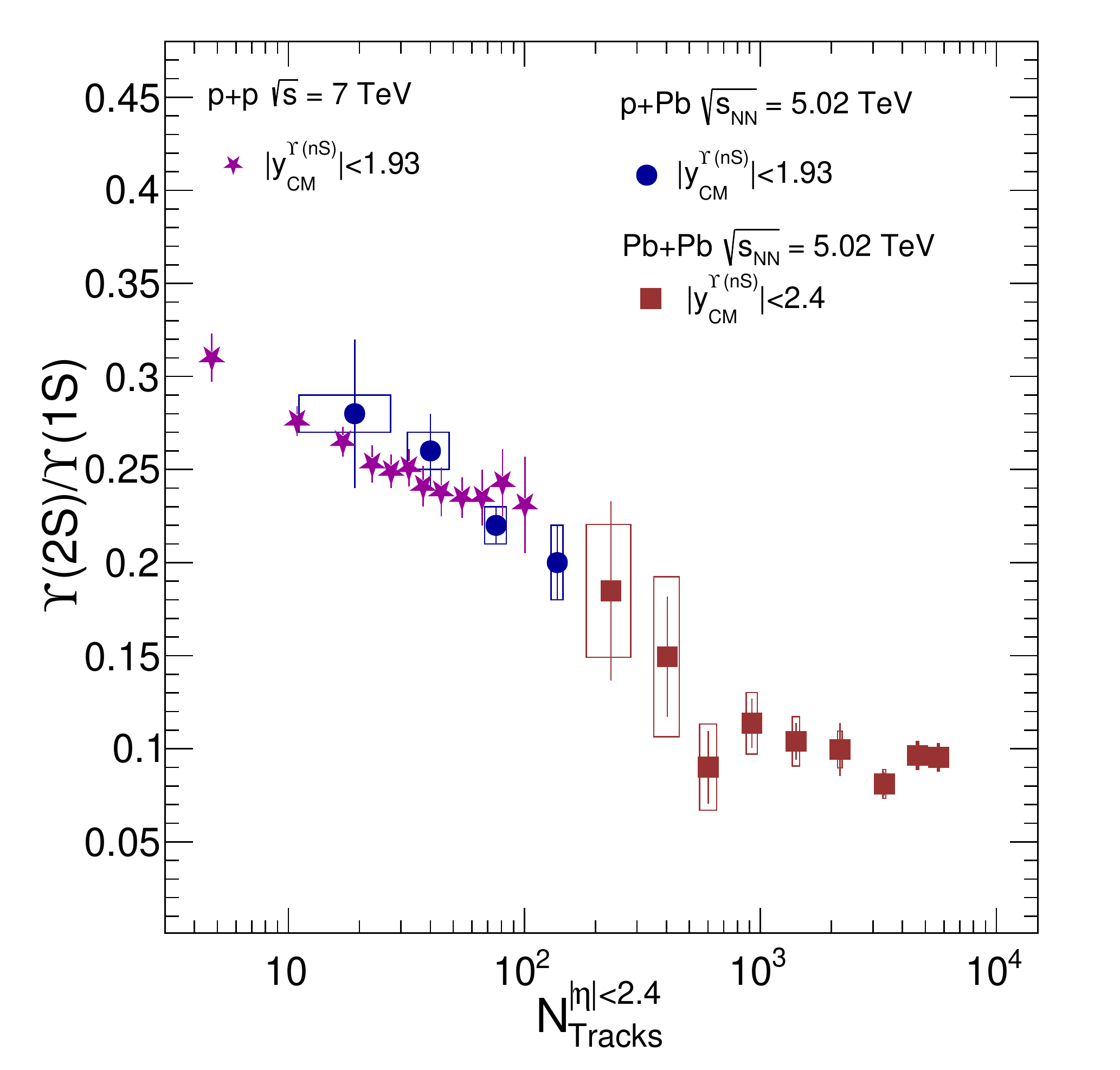}
\caption{(Color online)
 The ratio $\Upsilon$(2S)/$\Upsilon$(1S) as a function of event activity measured in 
$\sNN$ = 5.02 TeV p+Pb collisions~\cite{CMS:2013jsu} and is compared with p+p
collisions at $\sqrt{s}$ = 8 TeV \cite{CMS:2020fae}.
The ratio of $\Upsilon$(2S) and $\Upsilon$(1S) in Pb+Pb Collisions at
$\sNN$ = 5.02 TeV has been obtained using their $R_{AA}$ measured by CMS, 
the procedure explained in the text.
}
\label{fig:UpsilonpPb}
\end{center}
\end{figure}

Figure~\ref{fig:UpsilonpPbA} shows
the ratio $\Upsilon$(2S)/$\Upsilon$(1S) as a function of event activity measured in 
$\sNN$ = 5.02 TeV p+Pb collisions~\cite{CMS:2013jsu} and is compared with p+p
and Pb+Pb Collisions at $\sNN$ = 2.76 TeV.
The relative suppression of the excited state with the ground state indicates final
state effects in p+Pb collisions. 

Figure~\ref{fig:UpsilonpPbA} indicates that the relative suppression of
the two $\Upsilon$ states is falling steadily with the $N_{\rm tracks}$ and
there is no difference between different collision systems if
they are scaled with the event activity. However, because of large error bars
of data especially for Pb+Pb systems, this behavior can not be ascertained.
Moreover, the energies of p+Pb and Pb+Pb systems are different.

To get a clear picture, we obtain a new diagram using various CMS data sets.
Figure~\ref{fig:UpsilonpPb} shows the ratio $\Upsilon$(2S)/$\Upsilon$(1S) as a function of
event activity measured in $\sNN$ = 5.02 TeV p+Pb collisions~\cite{CMS:2013jsu}
and is compared with p+p collisions at $\sqrt{s}$ = 7 TeV \cite{CMS:2020fae} and
Pb+Pb Collisions at $\sNN$ = 5.02 TeV. One can observe from this new figure
that the ratio $\Upsilon$(2S)/$\Upsilon$(1S) decreases steadily for p+p and p+Pb systems
and the peripheral Pb+Pb data also follow this pattern. Then there is a step and most central Pb+Pb
data show a rather constant behavior as a function of event activity contrary to p+p and
p+Pb collisions which fall steadily with increasing $N_{\rm tracks}$.

Figure~\ref{fig:UpsilonpPb} shows the ratio of $\Upsilon$(2S)
and $\Upsilon$(1S) in Pb+Pb Collisions at
$\sNN$ = 5.02 TeV which has been obtained using their $R_{AA}$ measured by
CMS~\cite{CMS:2022wfi}. The procedure is explained in the following.
The $N_{\rm tracks}$ corresponding to $N_{\rm Part}$ at $\sNN =$ 5.02 TeV~\cite{CMS:2018zza}
can be obtained using the $N_{\rm tracks}$ for the same $N_{\rm Part}$ given
for 2.76 TeV~\cite{CMS:2013jsu} after scaling as

\begin{equation}
N_{\rm tracks}|_{5.02} =  N_{\rm tracks}|_{2.76} \times \frac{dN/d\eta |_{5.02}} {dN/d\eta|_{2.76}}.
\end{equation}
where $\frac{dN/d\eta |_{5.02}} {dN/d\eta|_{2.76}}=1.22$~\cite{CMS:2018zza,CMS:2013jsu}.
The ratio $\Upsilon$(2S)/$\Upsilon$(1S) at $\sNN$ = 5.02 can be obtained as 

\begin{equation}
\frac{\Upsilon(2S)}{\Upsilon(1S)} = \frac{R_{AA}^{2S}}{R_{AA}^{1S}} \times \frac{\sigma_{pp}^{1S}}{\sigma_{pp}^{2S}}.
\end{equation}
Here, $\sigma_{pp}^{1S}$ and $\sigma_{pp}^{2S}$ can be obtained by integrating the pp cross section
measured by CMS~\cite{CMS:2013jsu} at $\sNN =$ 5.02 TeV giving $\sigma_{pp}^{1S}/\sigma_{pp}^{2S}=0.26$.

To summarise, $\Upsilon$ states are suppressed in p+Pb collisions, although
the suppression is much smaller as compared to that in p+Pb collisions.
The relative suppression of the excited state with the ground state indicates final
state effects in p+Pb collisions. 
We have obtained a new figure for the ratio $\Upsilon$(2S)/$\Upsilon$(1S)
as a function of event activity measured in p+Pb and Pb+Pb collisions at
$\sNN$ = 5.02 TeV compared with the p+p collisions at $\sqrt{s}$ = 7 TeV.
In this figure, high statistics data Pb+Pb collisions at $\sNN$ = 5.02 TeV is used. 
 This study shows that the ratio $\Upsilon$(2S)/$\Upsilon$(1S) decreases steadily
with increasing $N_{\rm tracks}$ for
p+p and p+Pb systems and the peripheral Pb+Pb data also follow the same trend.
The most central Pb+Pb data show a rather constant behaviour as a function of
event activity contrary to p+p and p+Pb collisions which fall steadily with
increasing $N_{\rm tracks}$.
This shows that the final state suppression of $\Upsilon$ in p+Pb increases
as the number of particles increases which does not necessarily mean a
thermalized system.
The final state suppression (as given by the ratio of excited to ground state)
in Pb+Pb system remains similar for a range of event activity.
This could happen when a thermalized system is formed and the event
activity does not just mean the number of particles but gives the size of system.
Here the different sizes of the system may correspond to similar property like
temperature and hence the ratio  $\Upsilon$(2S)/$\Upsilon$(1S) remains similar
in Pb+Pb system for a large range of event activity.

%% file: BottAAthe.tex
 \section{Bottomonia production mechanism in heavy-ion collisions}
\label{sec:Bottomonia_hi}

The quarks and gluons which carry color quantum numbers are always confined inside 
hadrons. The quark gluon plasma (QGP) is state of matter where the color degrees of
freedom are at play over lengths much larger than the size of a typical hadron.  
Quarkonia are predicted to be suppressed in heavy-ion collisions
if a QGP is formed since the force among the quarks will be color screened
in the QGP phase~\cite{Matsui:1986dk}.
 However, very soon it was realized that the picture was not that simple.
There are many factors which affect the production of quarkonia in A+A collisions. 
In fact, the quarkonium suppression was observed in proton+nucleus (p+A)
collisions as well which is referred as cold-nuclear matter (CNM) effect.
Therefore it is necessary to disentangle hot 
and cold-matter effects. The CNM effects can arise at the initial
state and/or the final state. The initial state effect
arises due to modification of parton distribution functions (PDF) inside the nucleus
compared to the same inside the protons. The final state modification 
arises due to the  fact that the produced quarkonia would interact with the medium
leading to the breakup of the bound state.
Furthermore, the suppression of quarkonia is thought to be sequential in nature which
happens as a result of the differences in the  binding energy of different bound states. 
The tightly bound states, such as the $\upsa$ or the $\jpsi$,  melt at higher 
temperatures. On the other hand  more loosely bound states \psiP, \chic, \chib, 
\upsb or \upsc  melt at much lower temperatures. This property helps  
estimating the initial temperature achievable in 
the collisions~\cite{Digal:2001ue}. However, the prediction of a sequential 
suppression pattern gets complicated due to feed-down 
decays of higher-mass resonances. The production process is further 
enriched, in the high-energy scenario (like LHC), due to the recombination process. At very 
high energies, abundant production of $Q$ and $\bar Q$ may lead to new quarkonia production 
in the medium. The recombination process is more justified for the charmonia state and 
for the bottommonia states the contribution of this process is expected to be much
smaller since the bottom quark mass ($\sim$ 4.5 GeV) is three times more than the
charm quark and thus its thermalization at temperatures reached at LHC
($\sim$ 0.6 GeV) will be negligible.



\subsection{Cold-nuclear matter effects}

The baseline for quarkonium production and suppression in heavy-ion collisions 
are determined from studies of cold-nuclear matter (CNM) effects.
The most important CNM effect is due to the modifications of the parton distribution
functions (PDF) in the nucleus compared to that in the nucleon.
It depends mainly on two parameters, 
the momentum fraction of the parton $(x)$ and the scale of the parton-parton 
interaction $(Q^2)$. The nuclear density modified parton distribution
function is known as nPDF and the nPDF-to-PDF ratio,
$R_i(x,Q^2)=f_i^{p \epsilon A} (x, Q^2) /f_i^p  (x, Q^2)$
quantifies the modification due to nuclear effect. 
In the small $x$ regime $(x < 10^{-2})$, this ratio is less than unity
and is referred to as small-x shadowing. At intermediate  $x$ ($\sim 0.1$)
the ratio shows a hump-like structure, a phenomenon known as 
anti-shadowing. Around $x\approx 0.6$, one observes a dip which is known as EMC
effect. At very low $x$, the gluons density saturates as the gluon recombination
balances gluon splitting. Thus the dynamics of the system is described using
saturated gluon matter which is called the color-glass condensate \cite{Gelis_2010}.
In the final state, the quarkonia bound state scatters and re-scatters inelastically
while passing through the nucleus. This leads to 
the breakup or absorption of the bound state which is estimated by the
inelastic cross section of the quarkonia with the nucleon. 

The contributions to CNM effects look straightforward. However, there are several 
uncertainties associated with them. 
The nuclear modifications of the quark densities are relatively 
well-understood as they can be measured in  nuclear deep-inelastic scattering (nDIS).
On the other hand, the modifications of the gluon density are not directly measured.
The scaling violations in nDIS is one of the  
ways to constrain the nuclear gluon density. Another constraint is provided by 
overall momentum conservation.  However, more direct probes of the gluon 
density are needed. The shadowing parametrizations we have in hand are derived
from global fits to the nuclear parton densities.
This gives wide variations in the nuclear gluon 
density, from almost no effect to very large shadowing at low-$x$, 
compensated by strong antishadowing around $x \sim 0.1$.

The nuclear absorption survival probability depends on the absorption cross section 
of the quarkonium. There are even more inherent uncertainties 
in absorption than in the shadowing parametrization.
Typically an absorption cross section is extracted by a fit to the $A$ dependence 
of quarkonium production in p+A collision at a given energy. 
This is rather simplistic since it is not known whether the object traversing the
nucleus is a precursor color-octet state or a fully-formed color-singlet quarkonium state.
The \jpsi absorption cross section at $y \sim 0$ is observed to decrease with energy,
irrespective of which shadowing parametrizations are chosen~\cite{Lourenco:2008sk}. 

The analyses of $\jpsi$ production in fixed-target interactions~\cite{Lourenco:2008sk} 
show that the effective absorption
cross section depends on the energy of the initial beam and the rapidity or
$x_F$ of the observed $\jpsi$,  where $x_F$ is the $c{\bar c}$ longitudinal momentum fraction in the centre-of-mass frame of the
two colliding hadrons. One possible interpretation is that 
low-momentum color-singlet states can hadronize in the
target, resulting in larger effective absorption cross sections at lower
center-of-mass energies and backward $x_F$ (or center-of-mass rapidity).
At higher energies, the states traverse the target more rapidly so that
the $x_F$ values at which they can hadronize in the target move 
back from midrapidity toward more negative $x_F$.
Finally, at sufficiently high energies, the quarkonium states pass 
through the target before hadronizing, resulting in negligible absorption
effects.  Thus the {\it effective} absorption cross section decreases with 
the increasing center-of-mass energy.

This is a very simplistic picture. In practice, cold-nuclear matter effects 
(initial-state energy loss, shadowing, final-state breakup, {\it etc.}) 
depend differently on the quarkonium kinematic variables and the collision energy. 
Thus combining all these mechanisms into an {\it effective} absorption cross section,
as employed in the Glauber formalism is not adequate.
A better understanding of absorption requires more detailed knowledge of the 
production mechanisms which are not fully understood yet.

The nuclear modification factors and ratios of $\Upsilon$ states are measured by CMS
experiment covering wide kinematic regions.
In section~\ref{sectionpA}, Figure~\ref{fig:LHCpPb5} shows the $\Upsilon$(nS) nuclear
modification factor, $R_{pA}$,  as a function of transverse momentum $p_{T}$ 
and rapidity in p+Pb collisions at 5.02 TeV measured by CMS~\cite{CMS:2022wfi}.
It is observed that all three $\Upsilon$ states are suppressed in p+Pb collisions.
Moreover, it is noticed that the excited states are more suppressed as compared
to the ground state. Since the shadowing effects are expected to be similar
in all the three $\Upsilon$ states~\cite{Vogt:2015uba}
the measurements indicate final state effects on the $\Upsilon$ states which need
to be understood.

\subsection{Quarkonium in the hot medium}
\label{sec:media_sec3}

It has been argued that the color screening 
in a deconfined QCD medium will destroy $\QQbar$ bound states
at sufficiently high temperatures. If the size of the heavy
quark bound state is  much greater than the screening radius, then one heavy 
quark gets screened from the other and the pair is broken~\cite{Abdulsalam:2012bw}.
As the temperature 
increases, the screening radius becomes smaller and smaller compared to the 
size and the quarkonium states become more and more unstable. 
Although this idea was proposed long ago, the first principle QCD calculations, 
which go beyond qualitative arguments, have been performed quite recently. 
Such calculations include lattice QCD determinations of quarkonium 
correlators~\cite{Umeda:2002vr,Asakawa:2003re,Datta:2003ww,Jakovac:2006sf,Aarts:2007pk},
potential model calculations 
of the quarkonium spectral functions with potentials based on lattice 
QCD~\cite{Digal:2001ue,Wong:2004zr,Mocsy:2005qw,Mocsy:2004bv,Alberico:2006vw,Cabrera:2006wh,Mocsy:2007yj,Mocsy:2007jz},
also effective field theory
approaches~\cite{Laine:2007qy,Laine:2007gj,Laine:2008cf,Brambilla:2008cx}.  
Furthermore, better modeling of 
quarkonium production in the medium created by heavy-ion collisions has 
been achieved. These new advancements improve the understanding of 
the hot-medium effects on the quarkonium states which is crucial for the 
interpretation of heavy-ion data.

\begin{figure}[h]
   \begin{center}
      \includegraphics[width=0.6\textwidth]{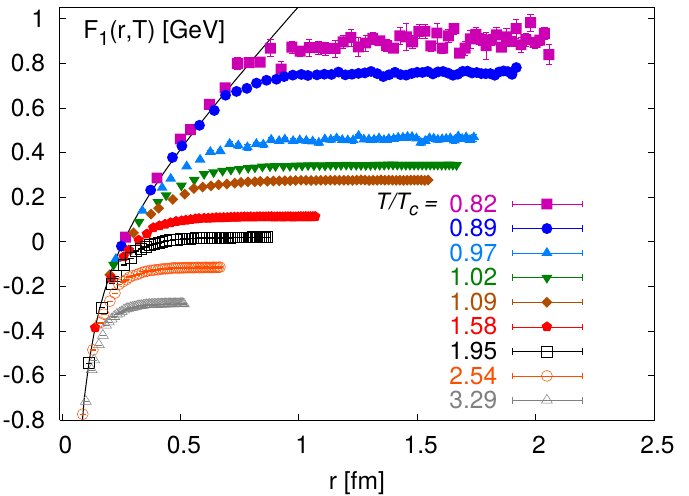}
      \caption{(Color online) The singlet free energy as a function of quark separation 
        calculated in (2+1) flavour QCD on $16^3 \times 4$ lattices at different 
             temperatures~\cite{Petreczky:2009ip,Petreczky:2010yn}.  
      }
      \label{Fig:LatticeSingEner}
   \end{center}
\end{figure}
In lattice gauge theory, color screening is studied  by 
calculating the spatial correlation function of a static quark and
antiquark in a color-singlet state which propagates in Euclidean time 
from $\tau=0$ to $\tau=1/T$, where $T$ is the temperature.
The result of such calculations on the lattice  with dynamical quarks have been
reported in Refs.~\cite{Petreczky:2009ip,Petreczky:2010yn,Kaczmarek:2002mc}.
The logarithm of the singlet correlation function, also called the singlet free energy,
is shown for (2+1) flavour in Fig.~\ref{Fig:LatticeSingEner} for
different temperatures.
As expected, in the zero-temperature limit,
the singlet free energy coincides with the zero-temperature potential. 
Also at sufficiently short distances, the singlet free energy is
temperature independent which is given by the zero-temperature potential. 
The range of interaction is shown to decrease with increasing temperature.
For temperatures just above the transition temperature, $T_c$, the heavy-quark 
interaction range becomes comparable to the charmonium radius. Based on 
this observation, it can be expected that the charmonium
states, as well as the excited bottomonium states, do not remain bound at
temperatures just above the deconfinement transition often referred to as 
dissociation or melting. 

\begin{figure}[]
   \begin{center}
      {\includegraphics[width=0.49\textwidth]{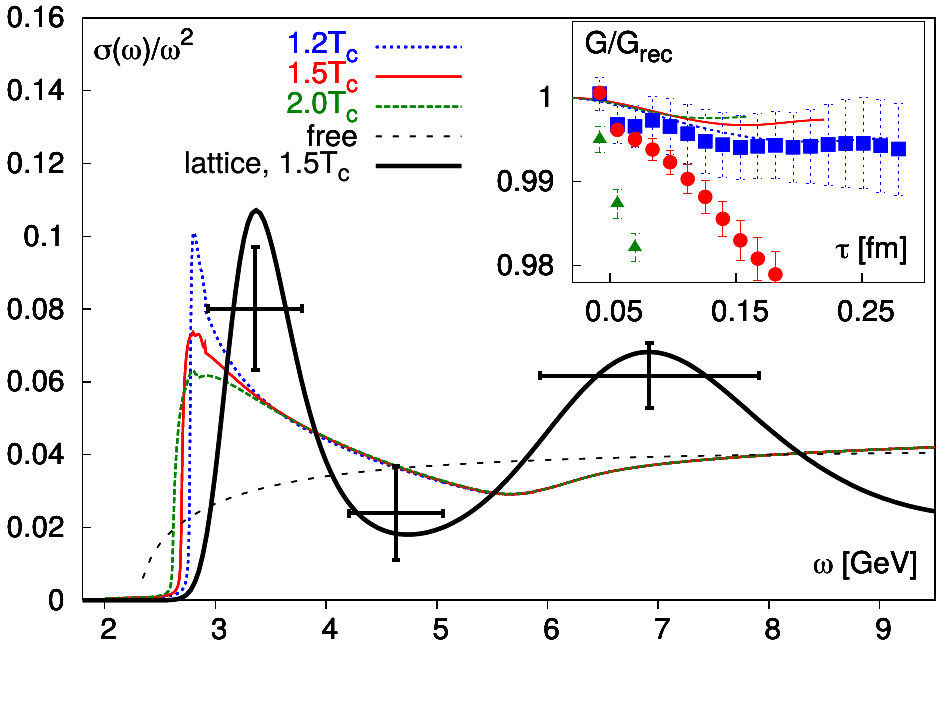}}
      {\includegraphics[width=0.49\textwidth]{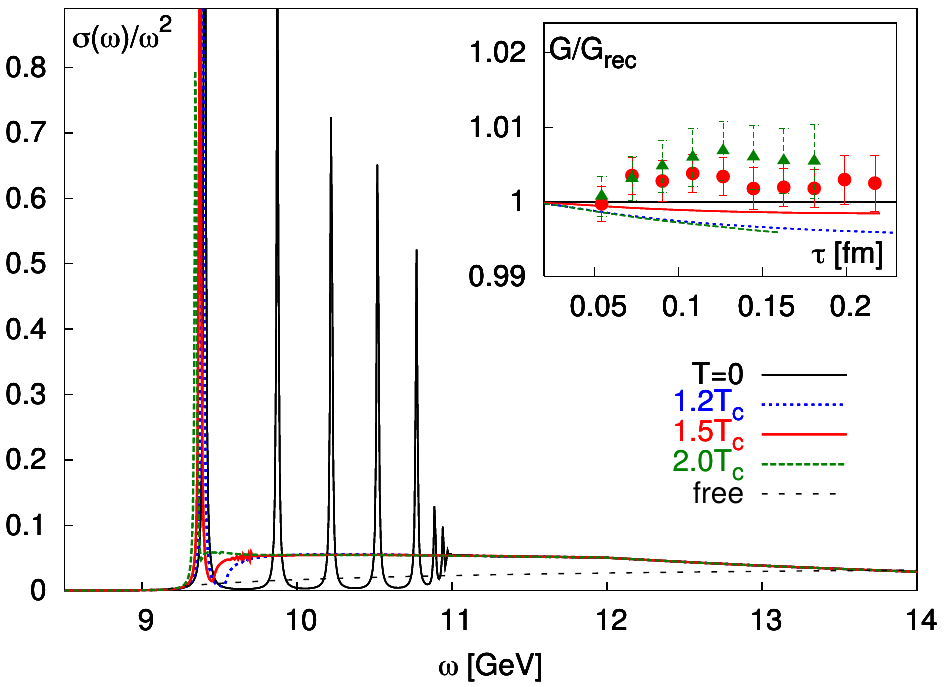}}
      \caption{(Color online) The S-wave charmonium (left) and 
        bottomonium (right) spectral functions calculated in potential 
        models. 
        Insets give correlators compared to lattice data.  
        The {\it dotted curves} are the
        free spectral functions. Figures are taken from Ref.~\cite{Mocsy:2007yj}.
      }
      \label{Fig:QuarkoniaSpecFuncLattice} 
   \end{center}
\end{figure}

In-medium quarkonium properties are encoded in the corresponding 
spectral functions, such as quarkonium dissociation
at high temperatures. Spectral functions are defined as
the imaginary part of the retarded correlation function of quarkonium
operators where the bound states appear as peaks.
The peaks broaden and eventually disappear with increasing temperature which
signals the melting of the given quarkonium state.
The quarkonium spectral functions can be calculated in potential models 
using the singlet free energy from Fig.~\ref{Fig:LatticeSingEner} or with different 
lattice-based potentials obtained using the singlet free energy
as an input~\cite{Mocsy:2007yj,Mocsy:2007jz}. 
The results for quenched QCD (without dynamical quarks) calculations
for S-wave charmonium  and bottomonium spectral functions~\cite{Mocsy:2007yj}
are shown in Fig.~\ref{Fig:QuarkoniaSpecFuncLattice}.
It shows that all charmonia states are dissolved in the deconfined phase above $T_c$ while the
bottomonium 1S state may persist up to $T \sim 2T_c$. An upper bound on the dissociation
temperature above which no bound state peaks can be seen in the spectral function can be
obtained from the analysis of the spectral functions.
Conservative upper limits on the dissociation temperatures for the different quarkonium
states obtained from a potential model which is based on a full QCD calculation (including dynamical quarks)~\cite{Mocsy:2007jz}
are given in Table~\ref{tab:LatticeDissTemp}.

\begin{table}[tb]
   \caption{Upper bounds on the dissociation 
             temperatures for different quarkonia states~\cite{Mocsy:2007jz}.
             }
   \label{tab:LatticeDissTemp}
   \setlength{\tabcolsep}{0.41pc}
   \begin{center}
      \begin{tabular}{ccccccc}
      \hline\hline
      State & $\chi_{cJ}(1P)$ & $\psi^{'}$ &J/$\psi$  &$\Upsilon(2S)$ & $\chi_{bJ}(1P)$ &$\Upsilon(1S)$ \\
      \hline 
      $T_{\rm diss}$ & $\le T_c$ & $\le T_c$ & $1.2T_c$ & $1.2T_c$ & $1.3T_c$ & $2T_c$\\ 
\hline\hline
\end{tabular}
\end{center}
\end{table}

Potential model calculations based on lattice QCD and resummed 
perturbative QCD calculations conclude that all charmonium states and the
excited bottomonium states dissolve in the deconfined medium.
This can account  
for the reduction of the quarkonium yields in heavy-ion collisions 
compared to the p+p collisions scaled by the number of binary collisions.
Recombination and edge effects, however, will produce a nonzero yield.

\subsection{Bottomonia suppression using lattice QCD inspired potential model rates}

Bottomonia suppression has been studied using first-principle
calculation of the thermal widths of the states and considering 
momentum anisotropy of the plasma~\cite{Strickland:2011aa,Krouppa:2016jcl,Krouppa:2018lkt}.
In this work, the phase-space distribution of gluons in the local
rest frame is taken to be 

\begin{equation} 
f({\bf x},{\bf p}) = f_{\rm iso}\left(\sqrt{{\bf p}^2+ \xi({\bf p}\cdot{\bf n})^2 }  / 
p_{\rm hard} \right).
\label{distribution}
\end{equation} 
In the above equation, $\xi$ is a measure of the degree of anisotropy of the plasma given as 
$\xi = \frac{1}{2} \langle 
{\bf p}_T^2\rangle/\langle p_z^2\rangle -1$
where $p_z$ and 
${\bf p}_T $ are the partonic longitudinal and transverse momenta in the local
rest frame, respectively. In equation (\ref{distribution}), $p_{\rm hard}$ is the momentum  
scale of the particles and can be identified with the temperature
in an isotropic plasma. 

An approximate form of the real perturbative heavy quark potential as a function of
$\xi$ can be written as~\cite{Dumitru:2007hy} (for $N_c=3$ and $N_f=2$). 
\begin{eqnarray}
Re[V_{\rm pert}] &=& - \alpha \exp(-\mu r)/r, \nonumber \\
\left(\frac{\mu}{m_D}\right)^{-4} &=&  
1 + \xi\left(1 + \frac{\sqrt{2}(1+\xi)^2\left(\cos(2\theta) - 1 \right)}{(2+ \xi)^{5/2}} \right),
\label{eq:muparam}
\end{eqnarray}
where $\alpha = 4\alpha_s/3$ and $m_D^2 = (1.4)^2 16 \pi \alpha_s  \, p_{\rm hard}^2/3$ gives
the isotropic Debye mass and $\theta$ is the angle with respect to the beamline.  
The factor of $(1.4)^2$ accounts for higher-order corrections to the isotropic Debye 
mass \cite{Kaczmarek:2004gv}.

This perturbative potential, given in equation (\ref{eq:muparam}) is modified to include
the non-perturbative (long-range) contributions. 
The modified real part of the potential is given as~\cite{Dumitru:2007hy} 

\begin{equation} 
\label{eq:repot}
Re[V] = -\frac{\alpha}{r} \left(1+\mu \, r\right) \exp\left( -\mu
\, r  \right) + \frac{2\sigma}{\mu}\left[1-\exp\left( -\mu
\, r  \right)\right] 
 - \sigma \,r\, \exp(-\mu\,r)- \frac{0.8 \, \sigma}{m_Q^2\, r} \, ,
\end{equation}
where the last term is a temperature- and spin-independent quark mass correction 
\cite{Bali:1997am} and $\sigma = 0.223$ GeV is the string tension.  Here  $\alpha$ 
is chosen to be  $0.385$ 
to match zero temperature
binding energy data for heavy quark states \cite{Dumitru:2007hy}.
The imaginary part of the potential is taken to be the same as the perturbative heavy quark
potential up to linear order in $\xi$ 
\begin{equation} 
Im[V_{\rm pert}] = -\alpha p_{\rm hard} \biggl\{ \phi(\hat{r}) - \xi \left[\psi_1(\hat{r},
\theta)+\psi_2(\hat{r}, \theta)\right]\biggr\} ,
\label{eq:impot}
\end{equation}
where $\hat{r}=m_D r$ and $\phi$, $\psi_1$, and $\psi_2$ are functions defined as~\cite{Krouppa:2016jcl}:  

\begin{equation}
\phi(\hat{r}) = 2\int_{0}^{\infty} dz \dfrac{z}{(z^2+1)^2}\left[1-\dfrac{\sin\left(z\hat{r}\right)}{\hat{r}}\right],
\end{equation}

\begin{equation}
 \psi_1(\hat{r}, \theta) = \int_0^{\infty} dz
 \frac{z}{(z^2+1)^2}\left(1-\frac{3}{2}
 \left[\sin^2\theta\frac{\sin(z\, \hat{r})}{z\, \hat{r}}
 +(1-3\cos^2\theta)G(\hat{r}, z)\right]\right),
 \end{equation}

 \begin{equation}
 \psi_2(\hat{r}, \theta) = - \int_0^{\infty} dz
\frac{\frac{4}{3}z}{(z^2+1)^3}\left(1-3 \left[
  \left(\frac{2}{3}-\cos^2\theta \right) \frac
 {\sin(z\, \hat{r})}{z\, \hat{r}}+(1-3\cos^2\theta)
 G(\hat{r},z)\right]\right).
\label{eq:psis}
\end{equation}
where $G(\hat r, z)$ is the Meijer G-function.

 The full model potential, given by $V = Re[V] + i Im[V]$, is used to 
solve the Schr\"odinger equation. 
The solution of the Schr\"odinger equation gives the real and imaginary parts of 
the binding energy of the states.  The imaginary part defines the instantaneous width of the state
$Im[E_{\rm bind}(p_{\rm hard},\xi)] \equiv -\Gamma_T(p_{\rm hard},\xi)/2$. 
The resulting width $\Gamma_T(\tau)$ implicitly depends on the initial temperature of the
system.

The following rate equation is used to account for in-medium bottomonia state decay,
\begin{equation} \label{eq:rate}
\frac{dn(\tau,{\bf x}_\perp,\varsigma)}{d\tau} = -\Gamma(\tau,{\bf x}_\perp,\varsigma)n(\tau,{\bf x}_\perp,\varsigma) ,
\end{equation}
where   $\tau = \sqrt{t^{2} - z^{2}}$ is the longitudinal proper time,  ${\bf x}_{\perp}$ is the transverse coordinate and 
 $\varsigma = {\rm arctanh}(z/t)$ is the spatial rapidity. The rate of decay is computed by~\cite{Strickland:2011aa}
\begin{eqnarray}
  \Gamma(\tau, {\bf x}_{\perp}, \varsigma)
  & =  2Im[E_{\text{bind}}(\tau, {\bf x}_{\perp}, \varsigma)] & \ \ Re[E_{\text{bind}}(\tau, {\bf x}_{\perp}, \varsigma)] > 0 \\ 
& =  \gamma_{\text{dis}} & \ \ Re[E_{\text{bind}}(\tau, {\bf x}_{\perp}, \varsigma)] \leq 0. 
\end{eqnarray}
The suppression factor $R_{AA}$ as a function of $p_T$ and centrality 
is obtained as follows
\begin{equation}
R_{AA}({\bf x}_\perp,p_T,\varsigma,b) =%
\exp\!\left(-\bar{\gamma}({\bf x}_\perp,p_T,\varsigma,b) \right),
\end{equation}
where
\begin{equation}
 \bar{\gamma}({\bf x}_\perp,p_T,
\varsigma,b) \equiv \Theta(\tau_f-\tau_{\rm form}(p_T)) \int_{{\rm max}(\tau_{\rm form}(p_T),\tau_0)}^{\tau_f} 
d\tau\,\Gamma_T(\tau,{\bf x}_\perp,\varsigma,b).
\end{equation}
  Here $\tau_{0}$ and $\tau_{f}$ are the initial and freeze-out times of the plasma and 
$\tau_{\rm form}$ is the formation time of the bottomonium state. 
Finally, one averages
over ${\bf x}_\perp$ to obtain 
\begin{equation}
\langle R_{AA}(p_T,\varsigma,b) \rangle \equiv 
\frac{\int_{{\bf x}_\perp} \! d{\bf x}_\perp \, T_{AA}({\bf x}_\perp)\,%
  R_{AA}({\bf x}_\perp,p_T,\varsigma,b)} 
{\int_{{\bf x}_\perp} \! d{\bf x}_\perp \, T_{AA}({\bf x}_\perp)}.
\end{equation} 

\begin{figure}[t]
\begin{center}
\includegraphics[width=0.5\textwidth]{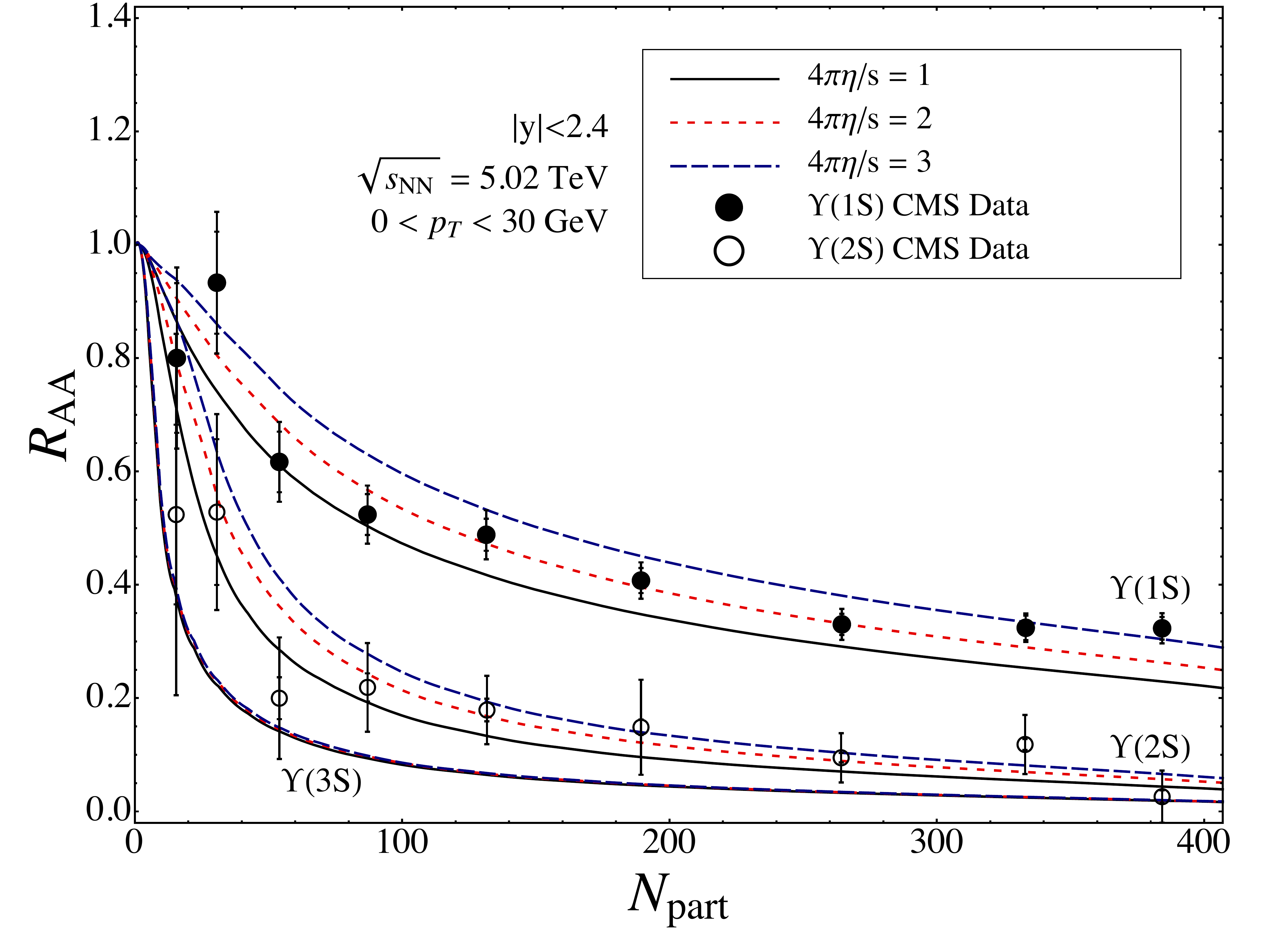}
\end{center}
\vspace{-7mm}
\caption{(Color online) Model calculations \cite{Krouppa:2018lkt} of the $R_{AA}$
  of $\Upsilon$(1S) and $\Upsilon$(2S) as a function of the $N_{\text{part}}$
  in Pb+Pb collisions at $\sNN$ = 5.02 TeV.   
  A comparison is made to the data from the CMS experiment~\cite{CMS:2018zza} at
  the LHC.}
\label{fig:raasep}
\end{figure}

Figure~\ref{fig:raasep} shows the model calculations~\cite{Krouppa:2018lkt}
of the $R_{AA}$ of $\Upsilon$(1S) and $\Upsilon$(2S) as a function of
$N_{\text{part}}$  in Pb+Pb collisions at $\sNN$ = 5.02 TeV for three
values of shear viscosity to entropy ratio ($4\pi\eta/s$).
A comparison is made to the data from CMS experiment \cite{CMS:2018zza} at
  the LHC. It is shown that there is substantial 
suppression  of $\Upsilon$(1S) and $\Upsilon$(2S) which are attributed to the in-medium decay.  
A  similar suppression pattern is observed for $\chi_{b1}$ which 
may be attributed to the finite formation time of the $\chi_{b1}$.

\subsection{Gluon dissociation of quarkonia in a dynamical medium}

{\color{black}
  The quarkonia can undergo both dissociation and recombination during the
  evolution of medium.
 The evolution of quarkonia population $N_{Q}$ with proper time $\tau$ can be studied via
a kinetic equation~\cite{Thews:2000rj}
  \begin{equation}\label{eqkin}
    {dN_{Q} \over d\tau}  =  - \lambda_D  \rho_g N_{Q} + \lambda_F {N_{q \bar{q}}^{2} \over V(\tau)},
  \end{equation}
  where $V(\tau)$ is the volume of the deconfined spatial region.
  The $\lambda_{D}$ is the dissociation rate obtained by the dissociation cross section
averaged over the momentum distribution of gluons $\rho_g$ and $\lambda_{F}$ is the formation
rate obtained by the formation cross section 
averaged over the momentum distribution of heavy quark pair $q$ and $\bar{q}$. 
$N_{q \bar{q}}$ is the number of initial heavy quark pairs produced per event which depends
on the centrality defined by the number of participants.
  The number of quarkonia at freeze-out time $\tau_f$ is given by the solution of Eq.~(\ref{eqkin}),
  \begin{equation}
    N_{Q}(p_T) = S(p_T) \,N_{Q}^{\rm PbPb}(p_T)+N_{Q}^F(p_T).
    \label{eqbeta}
  \end{equation}
  Here $N_{Q}^{\rm PbPb}(p_T)$ is the number of initially-produced quarkonia (including shadowing)
  as a function of $p_T$ and $S(p_T)=S(\tau_f, p_T)$ is their survival probability from gluon collisions at
  freeze-out given by 
  \begin{equation}
    S(\tau_f, p_T) = \exp \left( {-\int_{\tau_0}^{\tau_f}f(\tau) \lambda_{\rm D}(T,p_T)\,\rho_g(T)\,d\tau} \right).
  \end{equation}
  The temperature $T(\tau)$ and the QGP fraction $f(\tau)$ evolve from initial time $\tau_0$ 
  to freeze-out time $\tau_f$ due to expansion of the QGP. The initial temperature and the 
  evolution are dependent on the collision centrality $N_{\rm part}$.
  $N_{Q}^F(p_T)$ is the number of regenerated quarkonia per event,
  \begin{equation}
    N_{Q}^F(p_T)=S(\tau_f, p_T)N_{q \bar{q}}^{2} \int_{\tau_0}^{\tau_f}{{\lambda_{\mathrm{F}}(T,p_T) \over V(\tau)\,S(\tau,p_T)} d\tau}.
  \end{equation}
  The nuclear modification factor ($R_{AA}$) then can simply be written as~\cite{Kumar:2014kfa, Kumar:2019xdj}
  \begin{equation}
    R_{AA}(p_T)=S(p_T) \, R(p_T) + \frac{N_{Q}^F(p_T)}{N_{Q}^{pp}(p_T)}.
    \label{raa}
  \end{equation}
  Here $R(p_T)$ is the shadowing factor.

  The gluon dissociation rate can be obtained in the color dipole
approximation~\cite{Bhanot:1979vb} as a function of gluon energy, $q^0$ as
 \begin{equation}
    \sigma_{D}(q^{0}) = {8\pi \over 3} \, {16^2 \over 3^2} {a_0 \over m_q}  \frac{(q^0/\epsilon_0 - 1)^{3/2}} {(q^0/\epsilon_0)^5},
 \end{equation}
  where $\epsilon_0$ is the quarkonia binding energy and $m_q$ is the charm/bottom quark mass 
  and $a_0=1/\sqrt{m_q\epsilon_0}$.
  The value of $\epsilon_0$ is equal to $1.10$ GeV for $\Upsilon$(1S) \cite{Karsch:1987pv}. 
For the first excited state of bottomonia, $\Upsilon$(2S), the dissociation
 cross section is given in Ref.~\cite{Arleo:2001mp}.

  \begin{figure}
    \begin{center}
    \includegraphics[width=0.50\textwidth]{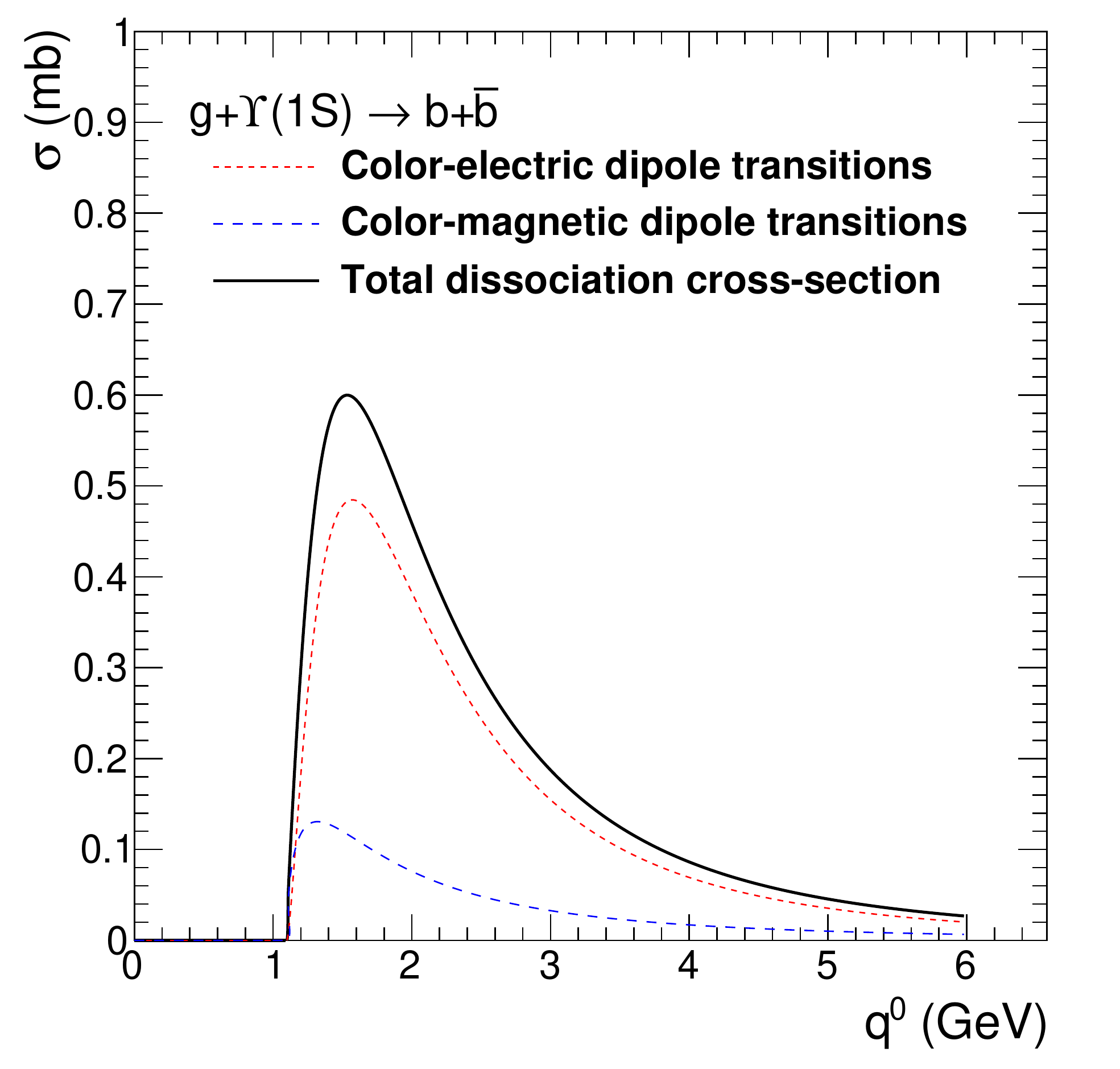}
    \caption{(Color online) Gluon dissociation cross section of $\Upsilon$(1S) as a
      function of gluon energy ($q^{0}$) in the $\Upsilon$(1S) rest frame.}
    \label{fig:SigmaDQ0}
    \end{center}
  \end{figure}

  Figure \ref{fig:SigmaDQ0} shows the gluon dissociation cross sections of
$\Upsilon$(1S) as a function of gluon energy. The dissociation cross section
is zero when the gluon energy is less than the binding energy of the quarkonia.
It increases with gluon energy and reaches a maximum at 1.5 GeV for 
$\Upsilon$(1S). At higher gluon energies, the interaction
probability decreases. The dissociation rate as a function of quarkonium
momentum can be obtained by integrating the dissociation cross section over thermal gluon momentum 
distribution.

 The formation cross section can be obtained from the dissociation cross section using
detailed balance~\cite{Thews:2000rj,Thews:2005vj},
  \begin{equation}
    \sigma_{F} = \frac{48}{36}\,\sigma_{D}(q^0)\frac{(s-M_{Q}^2)^{2}}{s(s-4m_q^{2})}.
  \end{equation}
  The formation rate of quarkonium as a function of with momentum can be obtained using
  thermal distribution functions of  $q/\bar{q}$.

\begin{figure}
\includegraphics[width=0.49\textwidth]{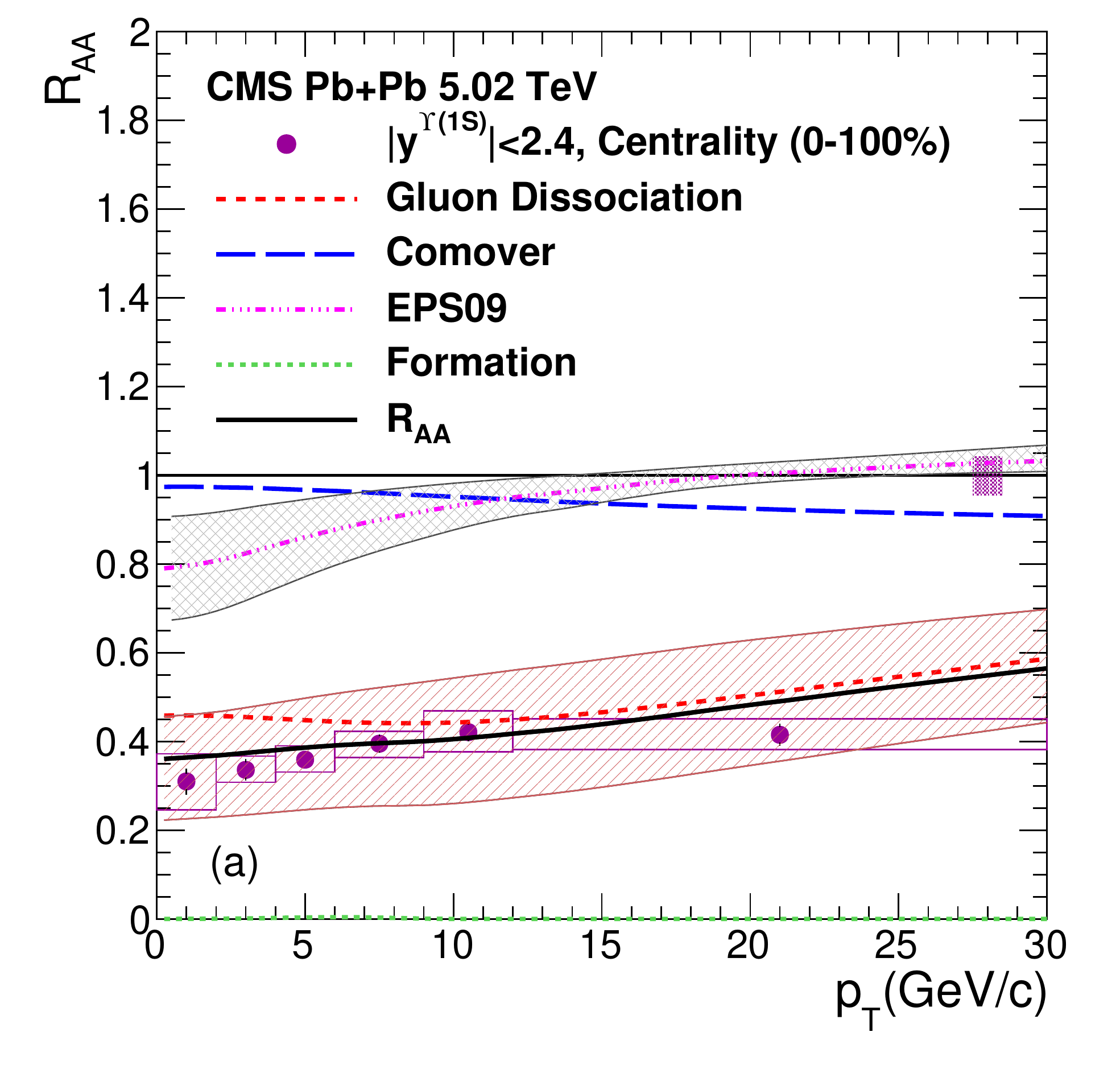}
\includegraphics[width=0.49\textwidth]{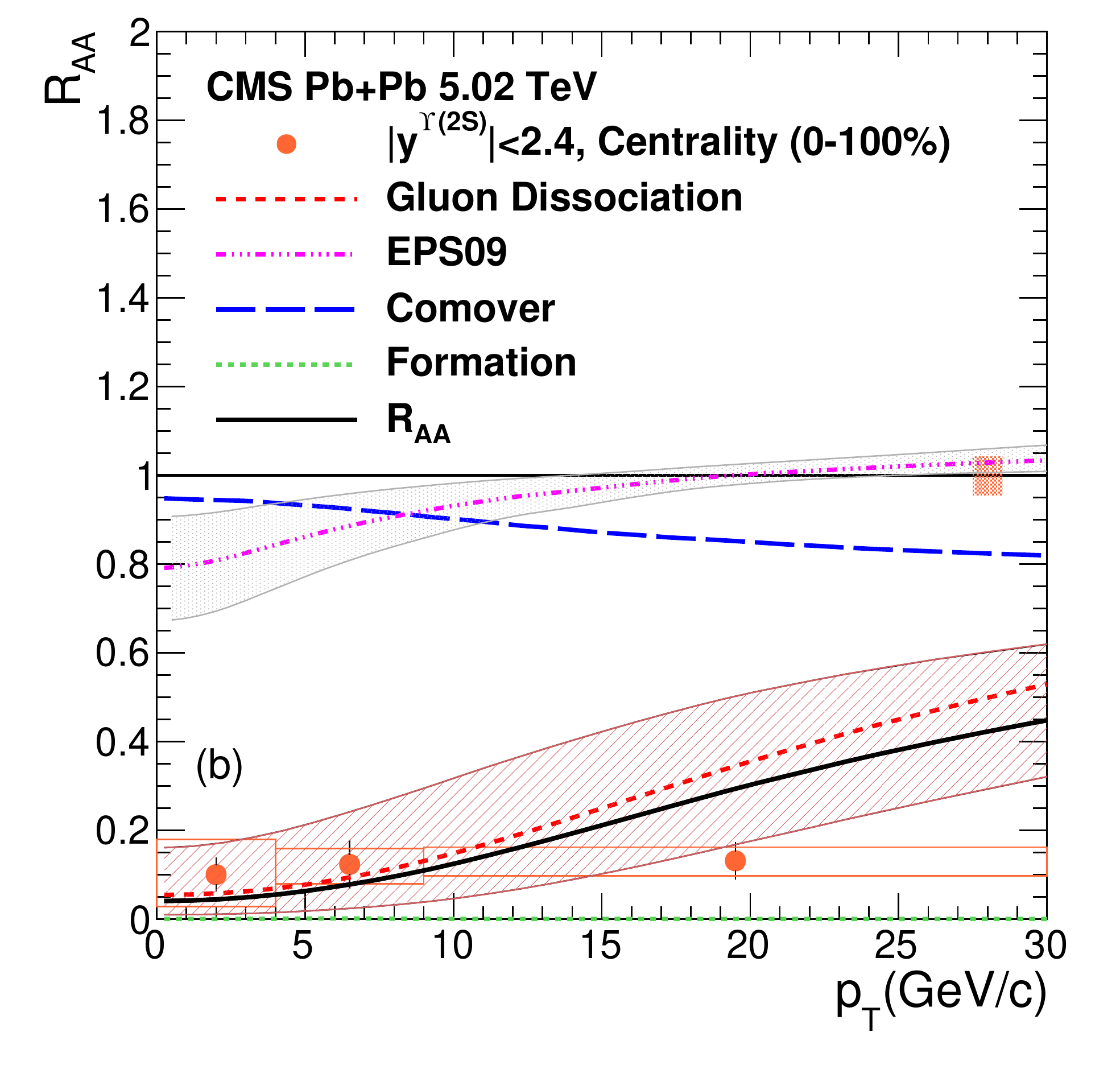}
\caption{(Color online) Calculated nuclear modification factor ($R_{AA}$) \cite{Kumar:2019xdj}
  of (left) $\Upsilon$(1S) and 
  (right) $\Upsilon$(2S) as a function of $p_{T}$ 
  compared with CMS measurements~\cite{CMS:2018zza}.
The global uncertainty in $R_{AA}$ is shown as a band around the line at 1.
}
\label{fig:UpsilonRaaPtCMS}
\end{figure}

\begin{figure}
\includegraphics[width=0.49\textwidth]{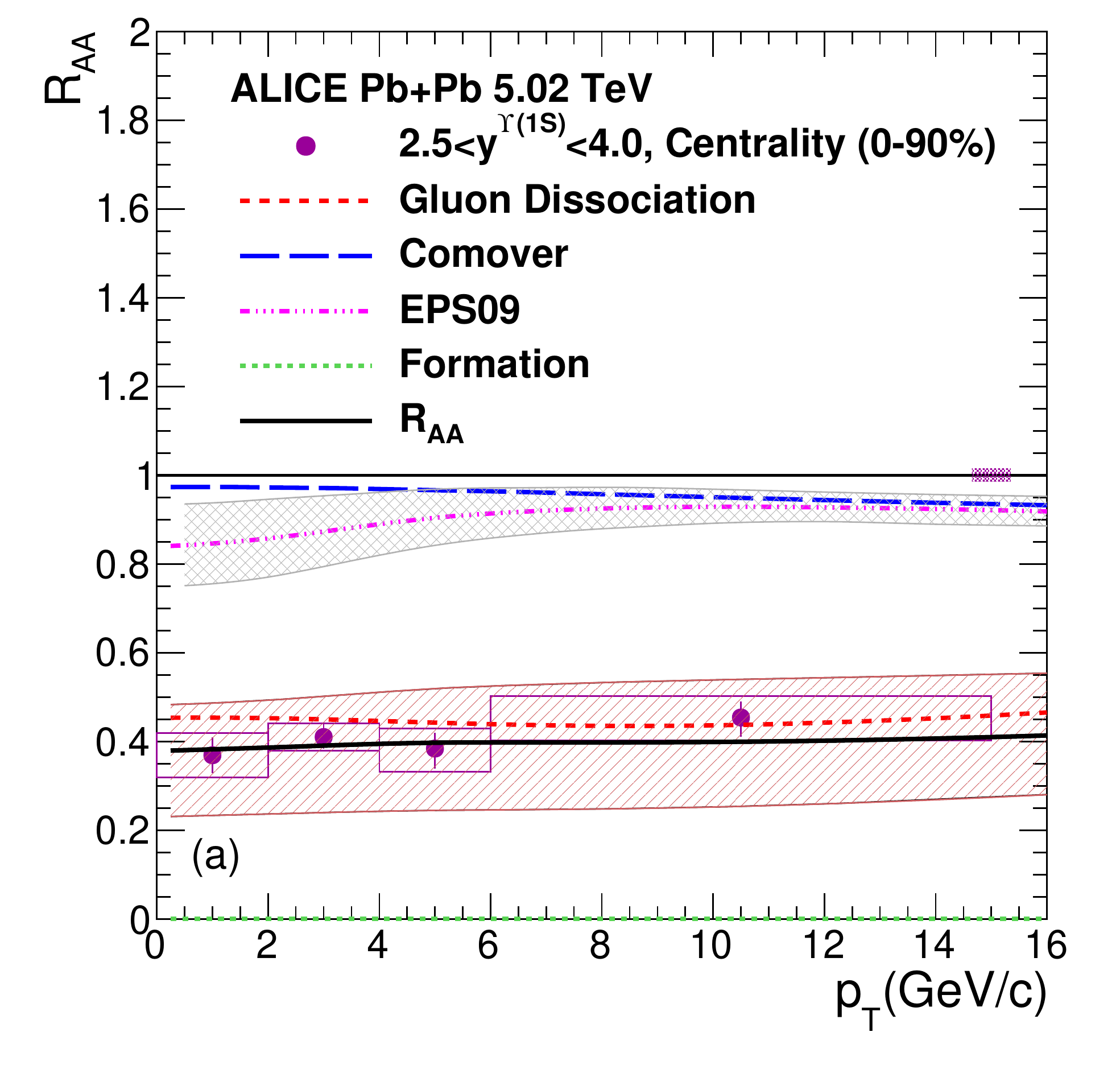}
\includegraphics[width=0.49\textwidth]{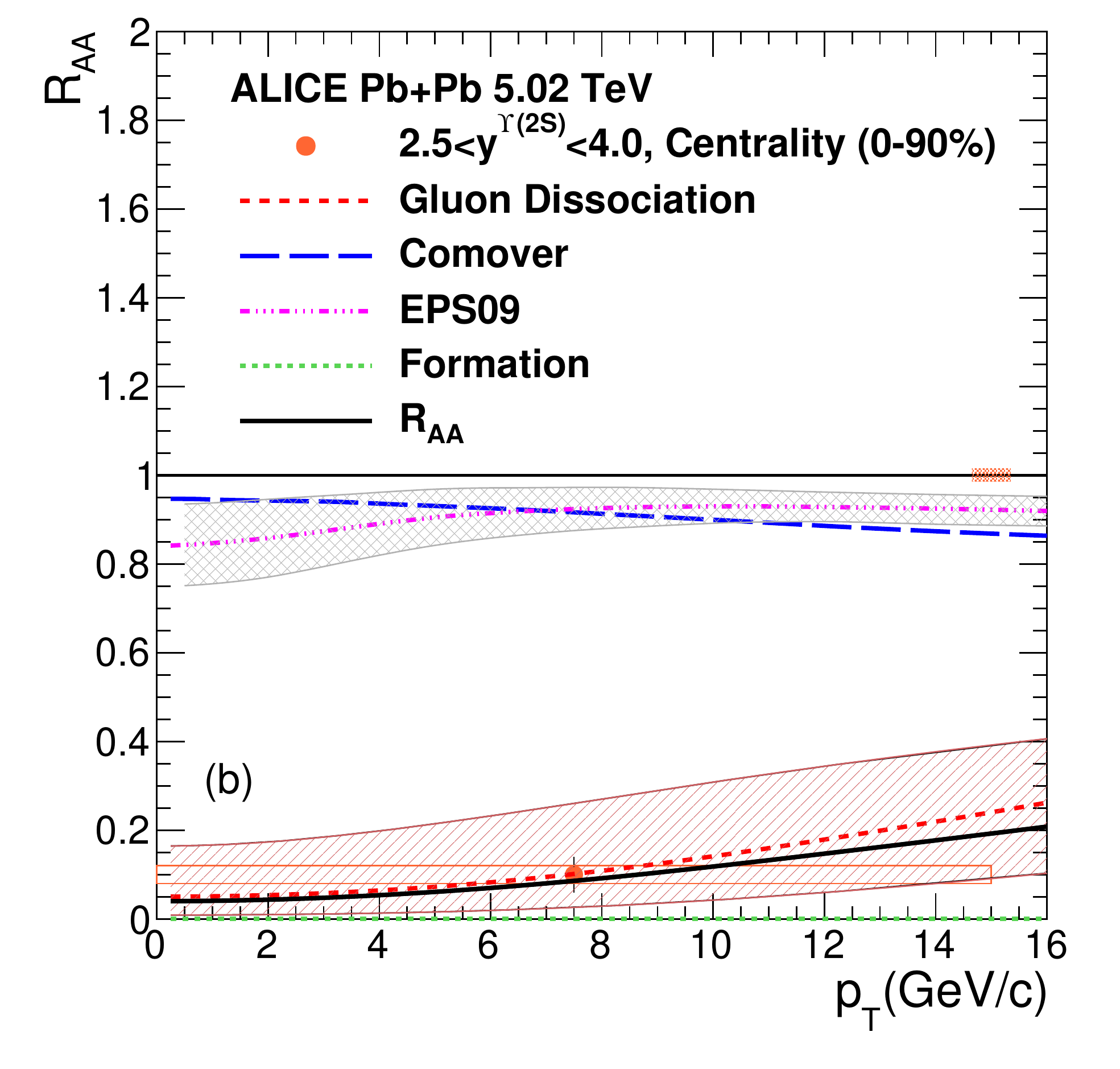}
\caption{(Color online) Calculated nuclear modification factor ($R_{AA}$) \cite{Kumar:2019xdj}
  of (left) $\Upsilon$(1S) and 
  (right) $\Upsilon$(2S) as a function of $p_{T}$ in the kinematic range of ALICE detector
at LHC ~\cite{ALICE:2020wwx}. The global uncertainty in $R_{AA}$ is shown as a band
around the line at 1.
} 
\label{fig:UpsilonRaaPtALICE}
\end{figure}

\begin{figure}
\includegraphics[width=0.49\textwidth]{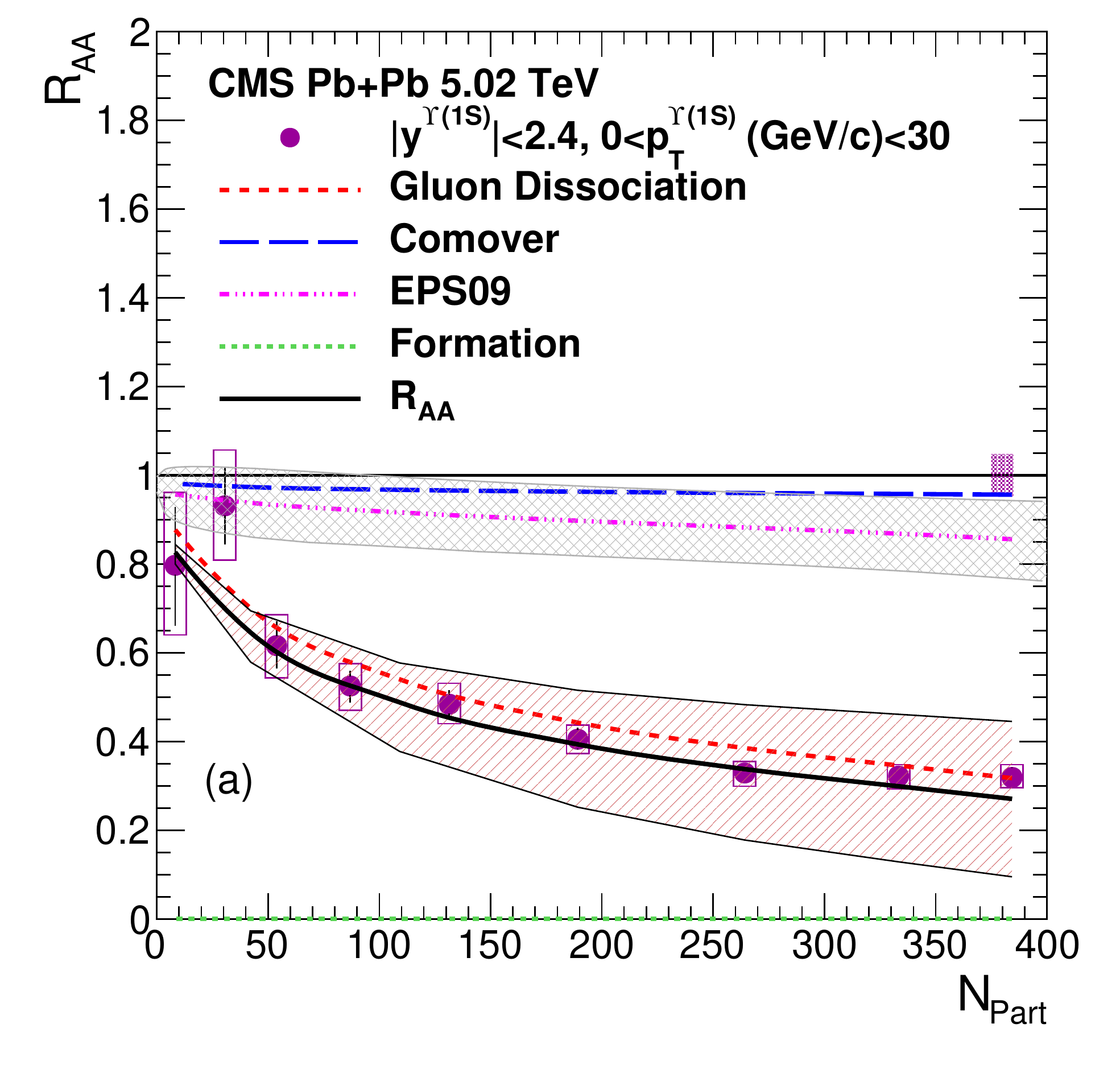}
\includegraphics[width=0.49\textwidth]{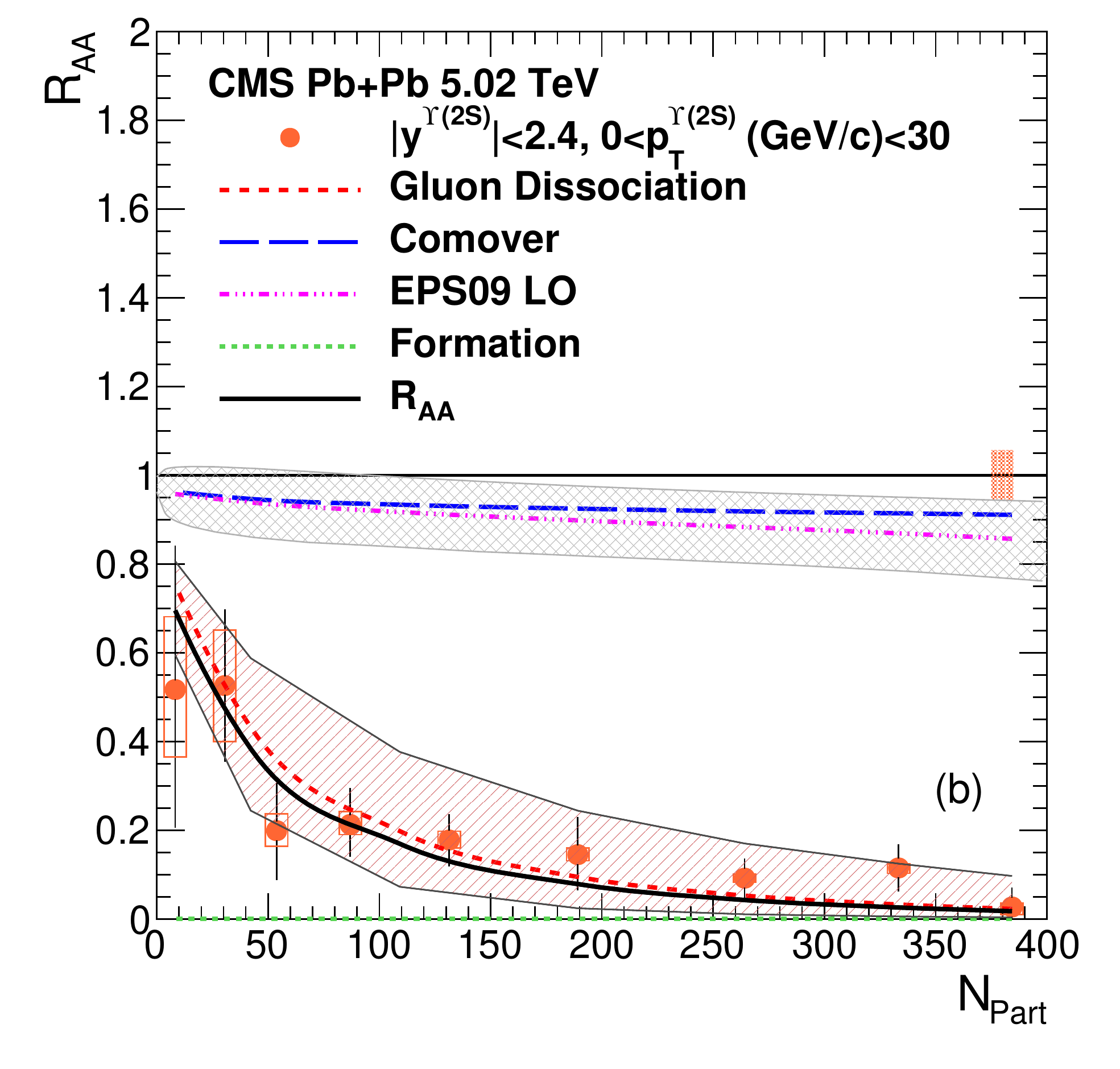}
\caption{(Color online) Calculated nuclear modification factor ($R_{AA}$) \cite{Kumar:2019xdj} of 
  (left) $\Upsilon$(1S) and (right) $\Upsilon$(2S) as a function of the centrality of the 
  collisions compared with the CMS measurements~\cite{CMS:2018zza}. The global uncertainty
  in $R_{AA}$ is shown as a band around the line at 1.
}
\label{fig:UpsilonRaaNPartCMS}
\end{figure}

Figure~\ref{fig:UpsilonRaaPtCMS}(left) and (right) show the calculations~\cite{Kumar:2019xdj}
of various contributions to
the nuclear modification factor, $R_{AA}$, for the $\Upsilon$(1S) and $\Upsilon$(2S)
respectively as a function of $p_T$ compared with the mid rapidity measurements from
CMS~\cite{CMS:2018zza}.  
The gluon dissociation mechanism combined with the pion dissociation and shadowing
corrections gives a good description of data in $p_{T}$ range ($p_{T}\approx$ 0-15 GeV/c)
for both $\Upsilon$(1S) and $\Upsilon$(2S).
The contribution from the regenerated $\Upsilon$s is negligible even at LHC energies.
The calculations under-predict the suppression observed at the highest measured
$p_{T}$ for $\Upsilon$(1S) and $\Upsilon$(2S) which is similar to the case
of J/$\psi$.

The feeddown corrections in the states $\Upsilon$(1S) and $\Upsilon$(2S) 
from decays of higher b$\bar{\rm b}$ bound states are obtained as
  \begin{equation}
    R_{AA}^{\Upsilon(2S)} = f_1 R_{AA}^{\Upsilon(2S)} +  f_2 R_{AA}^{\Upsilon(3S)},  
  \end{equation}
   \begin{equation}
    R_{AA}^{\Upsilon(1S)} = g_1 R_{AA}^{\Upsilon(1S)} +  g_2 R_{AA}^{\chi_b(1P)} + g_3 R_{AA}^{\Upsilon(2S)} + g_4 R_{AA}^{\Upsilon(3S)}. 
  \end{equation}
The $f$ and $g$ factors are obtained from the CDF measurement~\cite{Affolder:1999wm}.
The values of $g_1$, $g_2$, $g_3$ and $g_4$ are 0.509, 0.27, 0.107
and 0.113 respectively. Here $g_4$ is assumed to be the combined fraction of 
$\Upsilon$(3S) and $\chi_b$(2P).
The values of $f_1$ and $f_2$ are taken as 0.50~\cite{Strickland:2011aa}.

Figure~\ref{fig:UpsilonRaaPtALICE} (left) and (right) show the model 
calculation~\cite{Kumar:2019xdj} of the nuclear modification factor, $R_{AA}$, for the $\Upsilon$(1S)
and $\Upsilon$(2S) respectively as a function of $p_T$ in the kinematic range
covered by the ALICE detector. The ALICE data~\cite{ALICE:2020wwx} is well described by the model.

Figure~\ref{fig:UpsilonRaaNPartCMS} (left) depicts the calculated \cite{Kumar:2019xdj}
centrality dependence of the $\Upsilon$(1S) nuclear
modification factor, along with the midrapidity data from CMS~\cite{CMS:2018zza}.
The calculations combined with the pion dissociation and shadowing corrections 
gives a very good description of the measured data. Figure~\ref{fig:UpsilonRaaNPartCMS}(right)
shows the same for the $\Upsilon$(2S) along with the midrapidity
CMS measurements. The suppression of the excited $\Upsilon$(2S) states 
is also well described by the model. As stated earlier, the effect of regeneration is
negligible for $\Upsilon$ states. 

To summarise, the gluon dissociation mechanism combined with the shadowing
corrections gives a very good description of the available data in the mid $p_{T}$ range ($p_{T}\approx$ 5-10 GeV/c)
for both $\Upsilon$(1S) and $\Upsilon$(2S).
The contribution from the regenerated $\Upsilon$s is negligible even at LHC energies.
The calculations under-predict the suppression observed at the highest measured
$p_{T}$ for $\Upsilon$(1S) and $\Upsilon$(2S) which is similar to the case
of J/$\psi$.

The suppression of quarkonia by comoving pions can be calculated by folding the quarkonium-pion
dissociation cross section $\sigma_{\pi Q}$ over thermal pion distributions \cite{Vogt:1988fj}. 
It is expected  that at LHC energies, the comover cross section will be small~\cite{Lourenco:2008sk}.
{\color{black}
The pion-quarkonia cross section is calculated by convoluting the gluon-quarkonia cross section $\sigma_D$
over the gluon distribution inside the pion~\cite{Arleo:2001mp},
\begin{equation}
\sigma_{\pi Q} (p_{\pi}) = {p_+^2 \over 2(p_\pi^2 - m_\pi^2)} \int_0^1 \, dx \, G(x) \, \sigma_D(xp_+/\sqrt {2}),
\end{equation}
where $p_+ = (p_\pi + \sqrt{p_\pi^2-m_\pi^2})/\sqrt{2}$. The gluon distribution, $G(x)$, inside a pion is 
given by the GRV (Gluck-Reya-Vogt) parameterization~\cite{Gluck:1991ey}. 
The dissociation rate $\lambda_{D_{\pi}}$  can be obtained using the 
thermal pion distribution.

\subsection{Transport approach for bottomonia in the medium}
 The studies in Refs.~\cite{Grandchamp:2005yw,Rapp:2017chc} use 
transport approach for the bottomonia production in the medium~\cite{Grandchamp:2005yw,Rapp:2017chc}.
The rate equation for bottomonium evolution in the medium's rest frame
can be written as,
\begin{equation}
\frac{\mathrm{d} N_Y(\tau)}{\mathrm{d}\tau} =
-\Gamma_Y(T)\left[N_Y(\tau)-N^{\rm eq}_Y(T)\right] \ .
\end{equation}
Here $\Gamma_Y$, is the inelastic reaction rate  and $N^{\rm eq}_Y(T)$ is the thermal
equilibrium limit  for each state $\Upsilon$(1S), $\Upsilon$(2S), $\chi_b$.
In the reaction rates  both gluon-dissociation and quasi-free mechanisms have
been incorporated.  An important ingredient in this calculation is the bottomonium
binding energies. The thermal-equilibrium limit is evaluated from the statistical
model with bottom quarks~\cite{Grandchamp:2002wp}. 
The initial conditions are obtained from the p+p collision data. With these inputs,
the study is carried out in a hydrodynamically expanding scenario.

\begin{figure}[t]
\includegraphics[width=0.49\textwidth]{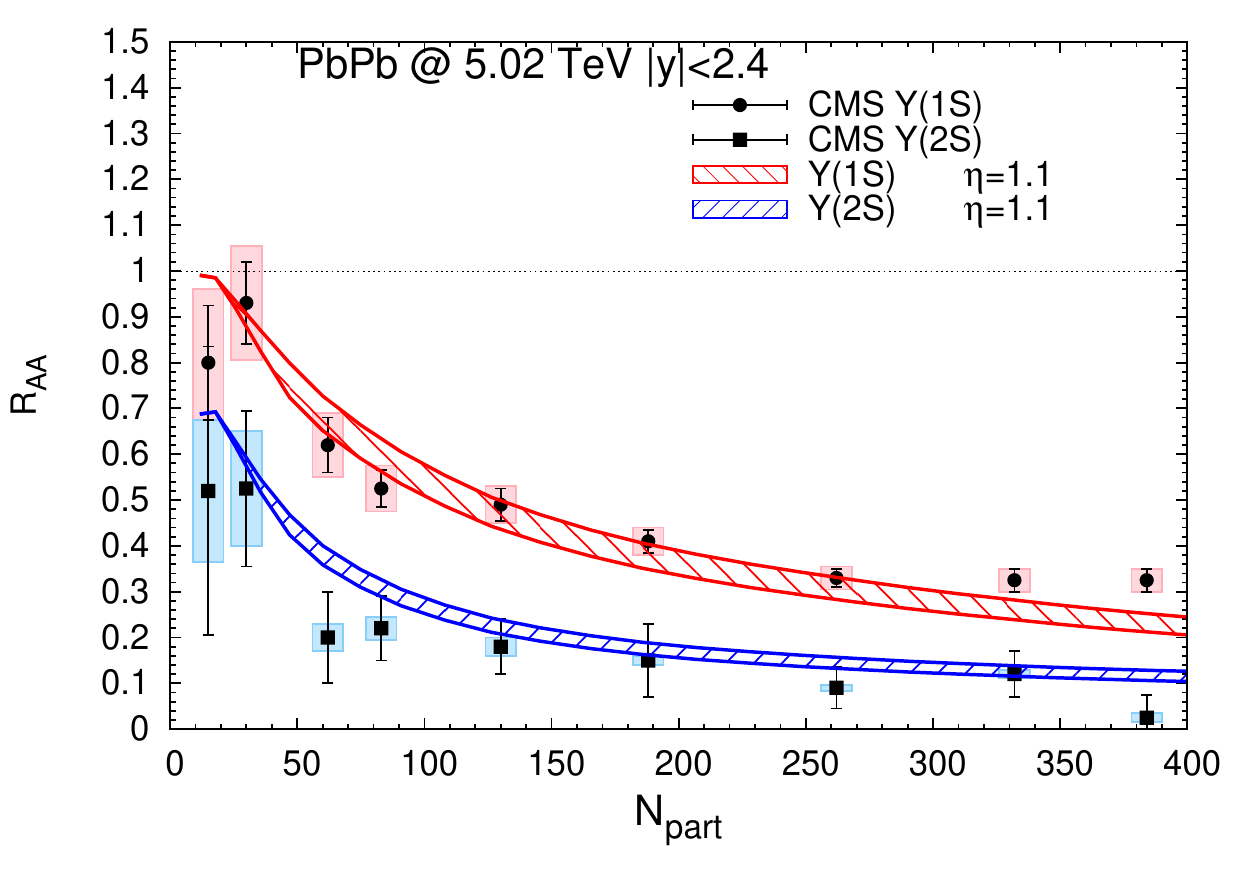}
\includegraphics[width=0.49\textwidth]{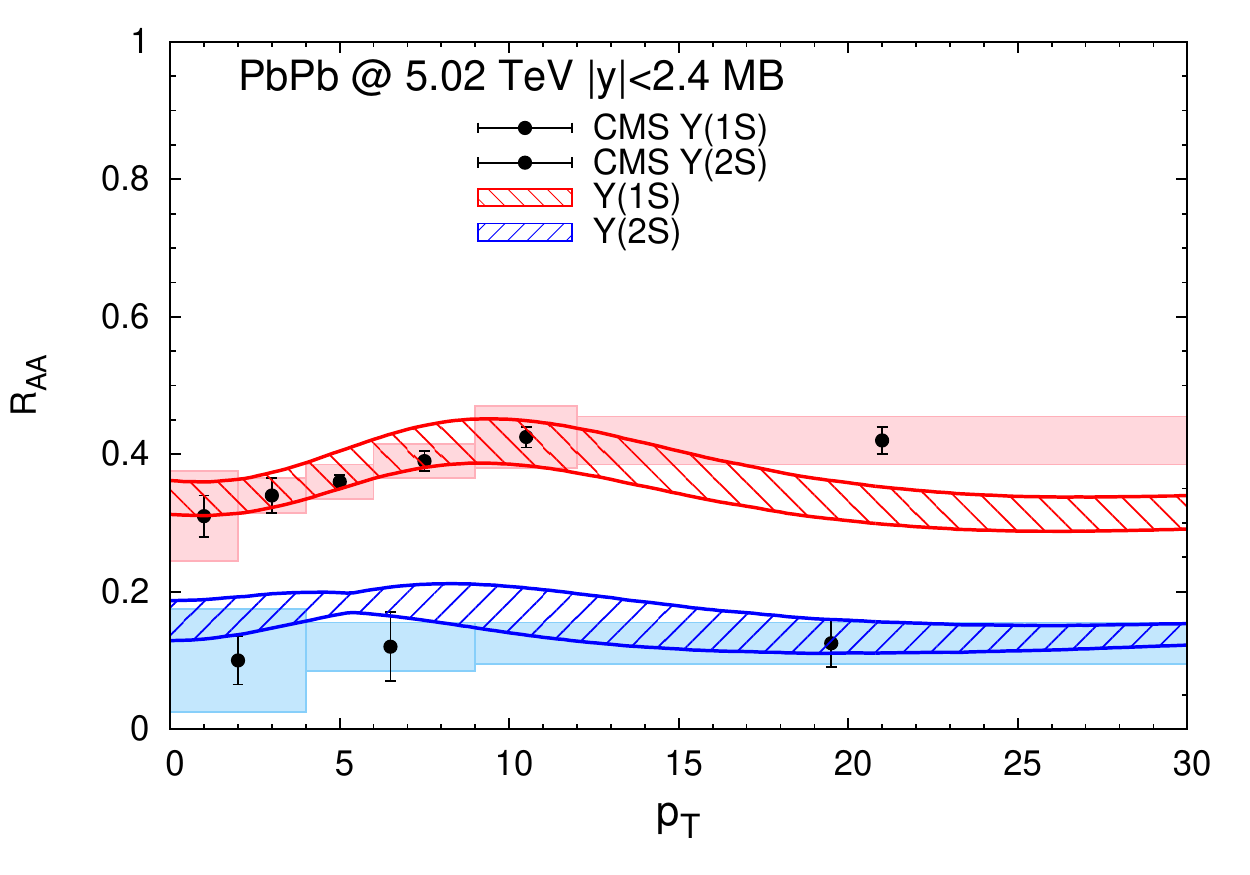}
\caption{(Color online) Centrality (left) and transverse-momentum (right) dependence of the $R_{AA}$~\cite{Rapp:2017chc,rappqm17}
  for $\Upsilon$(1S) and $\Upsilon$(2S) in 5.02\,TeV Pb+Pb collisions at the LHC,
  compared to CMS data~\cite{CMS:2018zza}.
  The bands represent a 0-15\,\% shadowing~\cite{Eskola:2009uj} on open-bottom and bottomonia.}
\label{fig_cms}
\end{figure}

Figure \ref{fig_cms} shows the centrality (left) and transverse-momentum (right)
dependence of the $R_{AA}$ calculated by model in Ref~\cite{Rapp:2017chc,rappqm17}
for $\Upsilon$(1S) and $\Upsilon$(2S) in 5.02\,TeV Pb+Pb collisions at the LHC,
compared to CMS data~\cite{CMS:2018zza}.
The authors of this model found a reasonable agreement  with experimental data
for the transverse momentum and centrality dependence of both $\Upsilon$(1S) and $\Upsilon$(2S)
at measured collision energies.


%% file: Conclusions.tex
\section{Summary and Conclusions}
\label{sec:conclusions}

In this writeup, we have reviewed the field of bottomonia production in
p+p, p+A and A+A collisions. With immense experimental and theoretical activities
especially due to LHC measurements, many features of the bottomonia production
and their behavior in the medium are well understood.

In section~\ref{Sec:BottPP}, we have reviewed the experimental status of the bottomonia production
in p+p and in p+$\overline{\rm p}$ collisions. The measurements at Tevatron, by CDF and D0 collaborations,
have been discussed. The measurements at LHC, by CMS and ATLAS, at $\sqrt s$ = 7 TeV and
13 TeV have been reviewed. There have been some measurements 
of $\Upsilon$ polarization by CDF collaboration.
The CMS and LHCb data, on $\Upsilon$ polarizability, confirms 
negligible polarization. 
The measurements of the cross sections and polarizations have shed light on the
$\Upsilon$(1S, 2S, 3S) production mechanisms in p+p collisions.
LHC data has substantially extended the reach of the kinematics to test the
Non-Relativistic QCD (NRQCD) and other models with higher-order corrections which become more
distinguishable at higher transverse momenta.

In section~\ref{sec:Bottomonia_pp_th}, we have discussed theoretical models of the bottomonia production
mechanism in p+p collisions. The bottomonia study in p+p involves heavy quark
pair production treatable by perturbative process and the formation of bottomonia 
which is a non-perturbative process.
For the later,  one has to take recourse to some effective models. We have discussed the 
color singlet model, the color evaporation model and the NRQCD factorisation approach.
In the color singlet model, it is assumed that the $Q\bar Q$ pair that evolves into
the quarkonium is in a color-singlet state. 
On the other hand, in the color evaporation model,
it is assumed that every produced $\QQbar$ pair can evolve into a quarkonium,
if it has an invariant mass that is less than the threshold for
producing a pair of open-flavor heavy mesons.
The probability factor of a pair evolving into quarkonium is obtained by fitting 
the experimental data which is supposed to be independent of collision energy. 
The Improved CEM reproduces the transverse momentum dependence of the
quarkonium cross section at CDF and LHC energies.  
The NRQCD formalism, along with the color singlet state, includes the color octet state.
In this formalism the evolution probability of $Q\bar{Q}$
pair into a state of quarkonium is expressed as matrix elements of NRQCD operators
expanded in terms of heavy-quark velocity $v$.
The work using NLO cross sections are discussed and LO calculations have been
reproduced for $\Upsilon$(nS) production in p+p collision at $\sqrt s$ = 7 and 13 TeV.

In section~\ref{Sec:BottAAexp}, we have presented an experimental overview of the bottomonia
results in p+A and A+A collisions at RHIC and LHC. We have looked into the
$R_{AA}$ for $\Upsilon$(nS) as a function of kinematic variables 
$p_{\rm T}$, rapidity and $N_{\rm Part}$ at different energies and by different experiments. 
We have also studied $v_2$ for these states with centrality and $p_{\rm T}$.
LHC has provided high-statistics measurements of $R_{AA}$ for
Pb+Pb collisions for all three $\Upsilon$ states over wide kinematical ranges.
All $\Upsilon$ states are found to be suppressed in the Pb+Pb collisions,
the heavier states are more suppressed relative to the ground state.
The suppression of $\Upsilon$ states strongly depends on system size but
has a weak dependence on $p_T$ and rapidity. At high $p_T$, more precise
measurements are required to ascertain flatness in the suppression. 
Comparing the measurements at RHIC and at two energies of LHC, it can be
said that the suppression increases with energy albeit weakly.

All three $\Upsilon$ states are suppressed in p+Pb collisions as well
and the excited states are more suppressed than the ground state indicating
final state effects.
We have obtained a new figure for the ratio $\Upsilon$(2S)/$\Upsilon$(1S)
as a function of event activity measured in p+Pb and Pb+Pb collisions at
$\sNN$ = 5.02 TeV compared with the 
p+p collisions at $\sqrt s$ = 7 TeV.
In this figure, high statistics data Pb+Pb collisions at $\sNN$ = 5.02 TeV is used. 
 This study shows that the ratio $\Upsilon$(2S)/$\Upsilon$(1S) decreases steadily
with increasing $N_{\rm tracks}$ for
p+p and p+Pb systems and the peripheral Pb+Pb data also follow the same trend.
The most central Pb+Pb data show a rather constant behaviour as a function of
event activity contrary to p+p and p+Pb collisions which fall steadily with
increasing $N_{\rm tracks}$.
This shows that the final state suppression of $\Upsilon$ in p+Pb increases
as the number of particles increases which does not necessarily mean a
thermalized system.
The final state suppression (as given by the ratio of excited to ground state)
in Pb+Pb system remains similar for a range of event activity.
This could happen when a thermalized system is formed and the event
activity does not just mean the number of particles but gives the size of system.
Here the different sizes of the system may correspond to similar property like
temperature and hence the ratio  $\Upsilon$(2S)/$\Upsilon$(1S) remains similar
in Pb+Pb system for a large range of event activity.

No significant $v_2$ is found for $\Upsilon$(1S) measured by the CMS
experiment. This shows that the bottom quark is not thermalized in the medium.
The recombination yield of bottomonia calculated using our model
is very small.

In section~\ref{sec:Bottomonia_hi}, we have discussed the mechanisms for the modification of 
bottomonia yields in heavy-ion collisions. Starting with the idea of color screening we have
discussed the more recent ideas like the modification of spectral functions of quarkonia states as
a function of temperature.
The cold-nuclear matter effects have been reviewed in a certain amount of detail.
The excited bottomonia states are more suppressed as compared to the ground state,
an effect that can not be understood by the shadowing of nPDFs alone. The final state
effects in p+A collisions require a better theoretical understanding. 
We have discussed the current state-of-the-art theoretical models
treating both
the quarkonia dissociation and recombination in the dynamical medium.
The comparison of the theoretical results with the experiments shows that
the bottomonia production can be understood in terms of color screening
or gluon dissociation. There is no significant recombination needed,
a picture that is also consistent with the small values of $v_2$ measured
by the experiments.